\documentclass[]{emulateapj}
\usepackage{natbib,amsmath}

\begin{document}

\def \nar  {NAR} 
\def \jcap  {JCAP}
\def \na  {Nature}
\def \npairs  {650} 
\def \nesi  {12}
\def \nmgii  {3?}
\def \nmagell  {13}
\def \nsys  {572} 
\def \mlox {\ell(X)}
\def \fnx {\ell(X)}
\def \lox {$\mlox$}
\def \mwstack {W_{\rm Ly\alpha}^{\rm stack}}
\def \wstack {$\mwstack$}
\def \mncom{n^{\rm com}}
\def \ncom{$\mncom$}
\def \maeff{A^{\rm eff}}
\def \aeff{$\maeff$}
\def \mfc {f_C}
\def \fc {$\mfc$}
\def \mslls {\sigma_{\rm gal}^{\rm LLS}}
\def \slls {$\mslls$}
\def \mdrr {\delta \rho/\rho}
\def \drr {$\mdrr$}
\def \mscgm {\sigma_{\rm gal}^{\rm CGM}}
\def \scgm {$\mscgm$}
\def \mngcom {n_{\rm gal}^{\rm com}}
\def \ngcom {$\mngcom$}
\def \msigf {\sigma(\mavgf^{2000})}
\def \sigf {$\msigf$}
\def \mdeltf {\delta_{\mavgf}}
\def \deltf {$\mdeltf$}
\def \msnlya {{\rm S/N}_{\rm Ly\alpha}}
\def \snlya {$\msnlya$}
\def \sncut {5.5}
\def \mllya {\lambda_{\rm Ly\alpha}^{\rm fg}}
\def \llya {$\mllya$}
\def \zem {$z_{\rm em}$}
\def \zbg {$z_{\rm bg}$}
\def \zfg {$z_{\rm fg}$}
\def \mzfg {z_{\rm fg}}
\def \zlya {$z_{\rm Ly\alpha}$}
\def \mkafr {\langle F(r) \rangle}
\def \kafr {$\mkafr$}
\def \mavgf {\langle F \rangle}
\def \avgf {$\mavgf^{\Delta v}$}
\def \mmnaf {\overline{\mavgf}}
\def \mnaf {$\mmnaf$}
\def \wlya {$W_{\rm Ly\alpha}$}
\def \mwstack {W_{\rm Ly\alpha}^{\rm stack}}
\def \wstack {$W_{\rm Ly\alpha}^{\rm stack}$}
\def \wsubj {$W_{\rm Ly\alpha}^{\rm line}$}
\def \mdvline {\delta v^{\rm line}}
\def \dvline {$\mdvline$}
\def \mwsubj {W_{\rm Ly\alpha}^{\rm line}}
\def \mwlya {W_{\rm Ly\alpha}}
\def \rcom {$R_{\rm com}$}
\def \mrcom {R_{\rm com}}
\def \rphys {$R_\perp$}
\def \mrphys {R_\perp}
\def \mrcom {R_\perp^{\rm com}}
\def \rcom {$\mrcom$}
\def \lbol {$L_{\rm Bol}$}
\def \guv {$g_{\rm UV}$}
\def \nhi  {$N_{\rm HI}$}
\def \mnhi  {N_{\rm HI}}
\def \kms  {\, km~s$^{-1}$}
\def \mkms  {{\rm km~s^{-1}}}
\def \lya  {Ly$\alpha$}
\def \mlya  {{\rm Ly\alpha}}
\def \lyb  {Ly$\beta$}
\def \hMpc      {h^{-1}{\rm\ Mpc}}
\def \mcMpc      {h^{-1}{\rm\ Mpc}}
\def \cMpc      {$\mcMpc$}
\def \msol      {{\rm\ M}_\odot}
\def\cm#1{\, {\rm cm^{#1}}}
\def \rAA  {{\rm \AA}}
\def \cgsflux   {{\rm erg\ s^{-1}\ cm^{-2}}}
\def \cgssflux   {{\rm erg\ s^{-1}\ Hz^{-1} cm^{-2}}}
\def\sci#1{{\; \times \; 10^{#1}}}
\def\ltk{\left [ \,}
\def\ltp{\left ( \,}
\def\ltb{\left \{ \,}
\def\rtk{\, \right  ] }
\def\rtp{\, \right  ) }
\def\rtb{\, \right \} }
\def\perd{\;\;\; .}
\def\smm{\sum\limits}
\def\intl{\int\limits}
\def \mfnx {f(\mnhi,X)}
\def \fnx {$\mfnx$}
\def \mrzkc {r_{KC}}
\def \rzkc {$\mrzkc$}
\def \mgkc {\gamma_{KC}}
\def \gkc {$\mgkc$} 
\def \sixrdla {3.9 \pm 2.3 \mcMpc}
\def \sixrslls {15.3 \pm 1.5 \mcMpc}
\def \sixrlls {14.6 \pm 0.8 \mcMpc}
\def \rslls {14.0^{+7.6}_{-2.7} \mcMpc}
\def \glls {1.68^{+0.14}_{-0.30}}
\def \rlls {12.5^{+2.7}_{-1.4} \mcMpc}
\def \gslls {1.68^{+0.06}_{-0.13}}
\def \mrvir {r_{\rm vir}}
\def \rvir {$\mrvir$}
\def \mmmin {M_{\rm halo}^{\rm min}}
\def \mmin {$\mmmin$}

\title{Quasars Probing Quasars VI. Excess \ion{H}{1} Absorption within 
One Proper Mpc of $z\sim 2$ Quasars}

\author{
J. Xavier Prochaska\altaffilmark{1,2},
Joseph F. Hennawi\altaffilmark{2},
Khee-Gan Lee\altaffilmark{2},
Sebastiano Cantalupo\altaffilmark{1},
Jo Bovy\altaffilmark{3},
S.G. Djorgovski\altaffilmark{4},
Sara L. Ellison\altaffilmark{5},
Marie Wingyee Lau\altaffilmark{1},
Crystal L. Martin\altaffilmark{6},
Adam Myers\altaffilmark{7,2},
Kate H.R. Rubin\altaffilmark{2},
Robert A. Simcoe\altaffilmark{8}
}
\altaffiltext{1}{Department of Astronomy and Astrophysics, UCO/Lick
  Observatory, University of California, 1156 High Street, Santa Cruz,
  CA 95064}
\altaffiltext{2}{Max-Planck-Institut f\"ur Astronomie, K\"onigstuhl 17,
  D-69115 Heidelberg, Germany} 
\altaffiltext{3}{Institute for Advanced Study, Einstein Drive,
  Princeton, NJ 08540, USA; Hubble Fellow}
\altaffiltext{4}{California Institute of Technology, Pasadena, CA
  91125}
\altaffiltext{5}{Department of Physics \& Astronomy, University
of Victoria, Finnerty Road, Victoria, British Columbia, V8P 1A1,
Canada}
\altaffiltext{6}{Department of Physics, University of California, Santa Barbara, Santa
Barbara, CA 93106}
\altaffiltext{7}{Department of Physics and Astronomy, University of
  Wyoming, Laramie, WY82072, USA}
\altaffiltext{8}{MIT-Kavli Institute for Astrophysics and Space
  Research, Massachusetts Institute of Technology, Cambridge, MA
  02139, USA}

\begin{abstract}


With close pairs of quasars at different redshifts, a background
quasar sightline can be used to study a foreground quasar's
environment in absorption. We use a sample of \npairs\ projected
quasar pairs to study the \ion{H}{1} \lya\ absorption transverse to
luminous, $z \sim 2$ quasars at proper separations of 
$30\,{\rm kpc} < \mrphys < 1$\,Mpc. In contrast to measurements along the line-of-sight, regions
transverse to quasars exhibit enhanced \ion{H}{1} \lya\ absorption and
a larger variance than the ambient intergalactic medium, with
increasing absorption and variance toward smaller scales.  Analysis of
composite spectra reveals excess absorption characterized by a
\lya\ equivalent width profile $W = 2.3 \, {\rm \AA} (\mrphys/{\rm
  100\,kpc})^{-0.46}$.  We also observe a high ($\simeq 60\%$)
covering factor of strong, optically thick \ion{H}{1} absorbers
(\ion{H}{1} column $\mnhi > 10^{17.3} \cm{-2}$) at separations
$\mrphys < 200\,{\rm kpc}$, which decreases to $\sim 20\%$ at $\mrphys
\simeq 1\,{\rm Mpc}$, but still represents a significant excess over
the cosmic average. This excess of optically thick absorption can be
described by a quasar-absorber cross-correlation function $\xi_{\rm
  QA}(r) = (r/r_0)^{\gamma}$ with a large correlation length $r_0 =
\rlls$ (comoving) and $\gamma = \glls$.  
The \ion{H}{1} absorption measured
around quasars exceeds that of any previously studied
population, consistent with quasars being hosted by massive dark
matter halos $M_{\rm halo} \approx 10^{12.5} \msol$ at $z \sim
2.5$. The environments of these massive halos are highly biased
towards producing optically thick gas, and may even dominate
the cosmic abundance of Lyman limit systems and hence the
intergalactic opacity to ionizing photons at $z\sim 2.5$. 
The anisotropic absorption around quasars implies the
transverse direction is much less likely to be illuminated by ionizing
radiation than the line-of-sight, which we interpret in terms of the
same obscuration effects frequently invoked in unified models of
active galactic nuclei.

\end{abstract}


\section{Introduction}


In cold dark matter (CDM) cosmology, galaxies form within the
potential wells of virialized dark matter halos where the overdensity
relative to the cosmic mean exceeds $\delta \rho/\rho \gg 100$, driving gravitational collapse 
of gas into these halos and the subsequent formation of stars.
These models also predict that the most massive galaxies
arise in the highest-mass dark matter halos which, in the early
Universe, trace the rarest density fluctuations.  In addition, such
`peaks' in the density field typically occur on top of larger-scale
overdensities that extend well beyond the virial radius of the collapsed halo, implying 
$\mdrr \gtrsim 1$ to many Mpc.  This large-scale structure is
arranged in a network of filaments, sheets, clusters, etc.\ making up the 
the so-called cosmic web.

At modest overdensities $\delta\rho/\rho \lesssim 10$, the Universe's
baryons are predicted to closely track the dark matter density field.
Therefore, in the vicinity of high-$z$ galaxies, the intergalactic
medium (IGM) -- revealed by \ion{H}{1} \lya\ absorption -- should
trace the corresponding large-scale matter distribution \citep[e.g.][]{mco+96,kim08}.  
%
%
%
Although experiments to test this paradigm are difficult to perform 
because
surveys of high-$z$ galaxies are observationally expensive,  
there has been progress in the past decade.
\cite{adel03} examined the mean transmission of \ion{H}{1} \lya\
flux through gas in the environments of the $z \sim 2-3$, star-forming
Lyman break galaxies (LBGs).  Aside from a peculiar behavior on the
smallest scales (not confirmed by subsequent studies), 
they found excess \ion{H}{1}
absorption associated with LBGs on scales of a few Mpc.
\cite{crighton+11} extended this experiment to a larger dataset of
quasar sightlines and LBGs;  their results confirm reduced \ion{H}{1}
\lya\ flux on scales of $2 -7 h^{-1}$\,Mpc.  \cite{rakic12} studied
the \ion{H}{1} \lya\ opacity towards 15 quasars probing 679 
LBGs.  They reported
an excess of \ion{H}{1} \lya\ absorption to proper impact
parameters $\mrphys = 2$\,Mpc around this galactic population.
Furthermore, the opacity increases with decreasing 
\rphys\ down to the survey limit of $\approx 50$\,kpc, where the
sightlines are believed to intersect the so-called circumgalactic
medium (CGM).  On these scales, non-linear and complex astrophysical
processes related to galaxy formation may dominate the baryonic
density field and the physical state of the gas
\citep[e.g.][]{simcoe02,kkw+05,fumagalli11a,smg+13}.

Complimentary work on the LBG-IGM connection has studied the
association of individual absorption systems to these galaxies.
\cite{ass+05} assessed the cross-correlation of \ion{C}{4} absorbers
to LBGs and measured a clustering amplitude $r_0 \approx 4 \mcMpc$
indicating a physical association between the metal-enriched IGM and
galaxies \citep[see also][]{crighton+11}.  
\cite{rudie12} examined the incidence of \ion{H}{1}
absorbers at $\mrphys < 2$\,Mpc from LBGs and found systematically
higher \ion{H}{1} column densities \nhi.  They also concluded that the 
majority of $\mnhi > 10^{15} \cm{-2}$ absorbers 
may arise in the CGM of these galaxies.
Altogether, these results confirm that the IGM traces the
overdensities marked by luminous, star-forming galaxies at $z \sim
2-3$, supporting the concept of a cosmic web permeating the
Universe between galaxies.

Galaxy formation models built on the $\Lambda$CDM hierarchical structure
formation paradigm predict that more massive 
galaxies should occupy higher mass halos, exhibiting larger
overdensities that extend to greater distances.  
This enhancement around massive galaxies should be reflected as
signatures in the IGM absorption.  In this manuscript, we test this
hypothesis by focusing on the dark matter halos of galaxies hosting
luminous quasars at $z \sim 2-3$.  Measurements of the quasar-quasar
autocorrelation function yields a correlation length of $r_0 = 8.4
\hMpc$ for a projected correlation function with slope $= -1$
\citep[][see also \citet{pmn04,myers07b,Shen-th09}]{white12}.  
For a $\Lambda$CDM cosmology,
this large correlation length implies a bias factor $b \approx 3.5$ and one
infers that $z>2$ quasars are hosted by dark matter halos with typical
mass $M_{\rm DM} \approx 10^{12.5} \msol$.  
This correlation
length and associated mass significantly exceed that measured for
luminous LBGs, the best-studied, coeval galaxy population
\citep[$r_0^{\rm LBG} = 4.0 \hMpc$, $M_{\rm DM}^{LBG} \approx
  10^{11.5}-10^{12} \msol$;][]{adel05,cwg+06-2,conroy+08,bielby11}.
Therefore, one predicts that the environments of massive galaxies
hosting $z\sim 2$ quasars will exhibit stronger \ion{H}{1}
\lya\ absorption \citep{kim08}.

On the other hand, a variety of astrophysical processes may alter this
simple picture, especially on scales influenced by the galaxy and/or
its neighbors (i.e.\ in the CGM).  For example, the gas may shock to the virial
temperature of the dark matter halo (i.e.\ $T > 10^6$\,K) which would
substantially reduce the hydrogen neutral fraction.   On the other
hand, the galactic winds of
star-forming galaxies drive a non-negligible fraction of gas and dust
from their interstellar medium \citep[ISM;
e.g.][]{rvs05b,shapley03,wcp+09,martin+12,rubin+13} 
and may therefore raise the surface density of \ion{H}{1} gas at
distances $R \gg 1$\,kpc.  
Similarly, quasar driven outflows may inject energy and material on
galactic scales, via radiative pressure and/or kinetic feedback
\citep[e.g.][]{moe+09,QPQ3}. 
As a third example, the massive stars 
in the galaxy and the quasar may produce a significant flux
of ionizing photons that would photoionize the surrounding gas on
scales of at least tens kpc \citep[e.g.][]{schaye06,cmb+08,QPQ2}.   
This proximity effect would suppress \ion{H}{1} absorption
\citep{bdo88} but may yield a greater abundance of highly ionized
gas (e.g.\ \ion{N}{5}).  For luminous quasars, such effects could extend
to proper distances $R \gtrsim 1$\,Mpc \citep{QPQ2}.
On these scales, 
therefore, one may be more sensitive to the astrophysics of galaxy
formation rather than the (simpler) physics of structure formation.

In this manuscript, we explore several of these processes and predictions
through the analysis of \ion{H}{1} absorption in the $\mrphys \le
1$\,Mpc (proper, i.e.\ $\approx 3 \mcMpc$ comoving)
environments surrounding the massive galaxies tagged by $z\sim
2$ luminous quasars.
This marks the sixth paper in our quasars probing quasars series,
which we refer to as QPQ6.  Previous work in this series introduced the novel technique
of using projected quasar pairs to study quasar environments, \citep[][QPQ1]{QPQ1}, 
measured the anisotropic clustering of strong \ion{H}{1} systems around
quasars \citep[][QPQ2]{QPQ2}, 
studied the physical conditions in the gas at $\approx 100$\,kpc from
a quasar \citep[][QPQ3]{QPQ3}, 
searched for fluorescent \lya\ emission from optically thick absorbers illuminated
by the foreground quasars \citep[][QPQ4]{QPQ4},  
and characterized the circumgalactic medium of the massive galaxies hosting
quasars \citep[][QPQ5]{QPQ5}.  
In the latter manuscript,
we reported on strong \ion{H}{1} absorption to $\mrphys =
300$\,kpc and a high covering fraction to optically thick gas (see
also QPQ1 and QPQ2).  This gas also shows significant enrichment of
heavy elements, suggesting a gas metallicity in excess of 1/10 solar
abundance (see also QPQ3).  This implies a massive, enriched and cool
($T\sim 10^4$K) circumgalactic medium surrounding these massive
galaxies, despite the presence of a luminous quasar whose ionizing
flux is sufficient to severely reduce the local \ion{H}{1} content.

Indeed, this cool CGM gas is generally not apparent along
the illuminated line-of-sight.  In QPQ2, we measured the incidence of
strong \ion{H}{1} absorbers in $\Delta v = \pm 1500\mkms$ windows
centered on the f/g quasar redshift to measure the clustering of such
gas to quasars.  We then used this clustering signal to predict the
incidence along the quasar sightline and found it greatly exceeds the
observed incidence, i.e.\ there is an anisotropic clustering of 
strong \ion{H}{1} systems around quasars.
Taken together with the general absence of fluorescent \lya\ emission
(QPQ4) from these absorbers, these observations imply that the
surrounding gas observed in background sightlines is not illuminated
by the foreground quasar.  
Such anisotropic emission follows naturally from unification models of
AGN where the black hole is obscured by a torus of 
dust and gas \citep[e.g.][]{Anton93,elvis00}.

Based on the methods we have presented in the QPQ series, there is now a
growing literature on the analysis of quasar pair spectroscopy to
examine gas in the environments of quasar hosts.  \cite{bhm+06}
searched for strong \ion{Mg}{2} absorption at small scales ($\mrphys <
100$\,kpc) from a sample of 4 quasars at $z\sim 1$ and found a
surprising 100\%\ detection rate.  \cite{farina13} expanded the search
for \ion{Mg}{2} absorption on small scales, also finding a high
detection rate (7 of 10).  They also reported on the detection of more
highly ionized gas traced by the \ion{C}{4} doublet.  These results
lend further support to the concept of a cool, enriched CGM
surrounding quasars.  On much larger scales ($\gg 1$\,Mpc),
spectroscopy of quasar pairs has been analyzed to measure
quasar-absorber clustering.  \cite{wkw+08} measured the large-scale
($>5\mcMpc$), transverse clustering of \ion{Mg}{2} and \ion{C}{4}
absorbers with quasars at $z \sim 1$ and 2 respectively.  The
clustering amplitudes ($r_0 \approx 5 \mcMpc$) were used to infer that
quasars are hosted by halos with masses $M > 10^{12.5} \msol$ at $z
\sim 2$ and over $10^{13} \msol$ at $z \sim 1$.  These inferences
assume, however, that the absorbers are unbiased tracers of the
underlying dark matter density field.  Most recently, \cite{font13}
have assessed the correlation of \ion{H}{1} \lya\ opacity with quasars
on scales of $\approx 5-50 \mcMpc$.  Their cross-correlation
measurements confirm the results from the quasar auto-correlation
function that quasars inhabit massive, dark matter halos.

At the heart of our project is a large sample of quasar pairs
\citep{thesis,BINARY,hennawi10} 
drawn predominantly but not exclusively from the Sloan
Digital Sky Survey \citep{sdssdr7} and ongoing BOSS experiment
\citep{boss_dr9}.  
We focus on projected pairs of quasars, which 
are physically unassociated, but  project to small angular separations
on the sky. In these unique sightlines, spectra 
of the b/g quasar are imprinted with absorption line signatures of the gas associated
with the foreground (f/g) quasar.  With sufficient signal-to-noise
(S/N) and spectral resolution, one is sensitive to the full suite of 
ultraviolet diagnostics traditionally used to study the intergalactic
medium: 
(1) \ion{H}{1} Lyman series absorption to assess neutral
hydrogen gas and by extension the underlying density field;
(2) low-ion transitions of Si, C, O that track cool and metal-enriched
gas;
(3) high-ion transitions of C, O, and N that may trace ionized or shock-heated
material associated with photoionization, virialization and/or feedback processes.

Here, we focus exclusively on \ion{H}{1} \lya\ absorption and defer
metal-line analysis for future papers (see also QPQ3, QPQ5).  As
described above, our principal motivation is to trace the density
field surrounding massive galaxies at $z \sim 2$ to scales of one
proper and projected Mpc.  
The decision to cut the sample at 1\,Mpc was somewhat arbitrary;  we
aimed to extend the analysis beyond the halo hosting the quasar but
still focus on the neighboring environment.
On these scales, our analysis offer constraints on the physical
processes that drive the accretion of gas into dark matter halos and
onto galaxies \citep[e.g.][]{barkana04,fg11,fumagalli11a}.
Models of these processes are still in a formative stage and exploring
trends with mass and redshift offer valuable insight.

Our experiment uses luminous quasars as a signposts for distant
massive dark matter halos. Because quasar activity represents a brief
energetic phase of galaxy evolution, our results could show peculiarities
related to quasar activity, which are not representative of the massive halo
population as a whole. 
Ionizing radiation from the quasar, for example, may photoionize gas
in the surrounding environment on scales to 1\,Mpc and beyond,
imposing a so-called transverse proximity effect (TPE). 
Work to date, however, has not shown strong evidence for such an
effect \citep{croft04,KT08};  in fact (as noted above), we have
identified excess \ion{H}{1} absorption on scales of the CGM
(QPQ2, QPQ5).
Quasars may also drive outflows, frequently invoked to suppress
star-formation and/or remove the cold ISM of massive galaxies,
which would inject energy and material into the surrounding medium. 
Indeed, a high incidence of metal-line absorption is observed in the CGM
of quasars \citep[QPQ3,QPQ5][]{bhm+06,farina13}.
In these respects, therefore, 
our experiment also offers insight into processes of 
quasar feedback on scales of tens kpc to 1\,Mpc.

This paper also describes
the methodology, sample selection, data
collection, reduction, and 
quasar redshift and continua measurements of our ongoing program.
As a result, this is a lengthy manuscript intended to provide a
nearly complete description of the methodology and our 
program's assessment of \ion{H}{1} gas on
1\,Mpc scales.  The casual reader, therefore, may wish to focus
his/her attention on $\S$~\ref{sec:discuss} which discusses the key
results and their implications.  The full paper is organized as
follows:
In $\S$~\ref{sec:data}, we describe detail the spectral datasets 
that comprise QPQ6 including data reduction, continuum normalization,
and quasar redshift measurements.
Non-parametric measurements of the \ion{H}{1} absorption are presented
in $\S$~\ref{sec:wlya}.   Measurements of the equivalent width and
\ion{H}{1} column densities are given in $\S$~\ref{sec:indiv}.
We generate and analyze composite spectra at \ion{H}{1} \lya\ in
$\S$~\ref{sec:stacks}.  In $\S$~\ref{sec:discuss}, we discuss the
main results and draw inferences.  We conclude with
a summary of the main findings in $\S$~\ref{sec:summary}.
Throughout this manuscript, we adopt a $\Lambda$CDM cosmology with
$\Omega_M = 0.26, \Omega_\Lambda = 0.74$, and $H_0 = 70 {\rm km \,
  s^{-1} \, Mpc}$.  In general, we refer to proper 
distances in units of Mpc.   The primary exception
is in the clustering analysis of \S~\ref{sec:clustering} where we employ 
comoving distances in units of \cMpc\ for consistency with the
conventions used in clustering.

\section{Data and Preparation}
\label{sec:data}

In this section, we discuss the criteria that define the QPQ6 sample
and the corresponding, diverse spectroscopic dataset that
forms the basis of analysis for this QPQ6 manuscript.  
We also describe several procedures required to prepare the data for
absorption-line analysis.

\subsection{Experimental Design and the QPQ6 Sample}

The primary goal of this paper is to explore the \ion{H}{1} \lya\
absorption of the environment surrounding $z \sim 2.5$ quasars on
proper scales of 10\,kpc to 1\,Mpc.  
To accomplish this goal, we utilize projected quasar pairs.  Analysis
of the
absorption-line spectroscopy for the background (b/g) quasar diagnoses
the gas (in projection) associated to a foreground (f/g) quasar.
To effectively probe a wide
dynamic range in projected radii, we have leveraged several large spectroscopic survey datasets
and have performed
dedicated follow-up observations on a number of large-aperture telescopes.
For the former, we use the spectroscopic quasar databases of the
Sloan Digital Sky Survey \citep[SDSS;][]{sdssdr7} and the recently
released Baryonic Oscillation Spectroscopic Survey data release 9 (DR9)
\citep[BOSS][]{boss_dr9}. For the latter, we have collected follow-up observations
of quasar pairs from the Keck, Magellan, Gemini, and the Large Binocular
Telescope. 

The starting point of our experiment is to discover projected quasar
pairs with angular separation $\theta$ corresponding to proper
separations of $\mrphys < 1$\,Mpc.  Modern spectroscopic surveys select \emph{against}
close pairs of quasars because of fiber collisions.  For the SDSS and BOSS
surveys, the 
finite size of their optical fibers preclude discovery of pairs with
separation $<55\arcsec$ and $<62\arcsec$, corresponding respectively to 
 to 414\,kpc and 467\,kpc at $z=2$.
At $z = 2$, however,
our target separation of $< 1$\,Mpc corresponds to $\theta < 2.1'$, which 
exceeds the fiber collision scale of these surveys. Therefore, these survey  
datasets provide a large sample of quasar pairs for $\mrphys \gtrsim
500$\,kpc, but relatively few at smaller separations. 
In the regions where spectroscopic plates overlap, this fiber collision limit
can be circumvented. However, presently only $\approx 30\%$ of the SDSS spectroscopic footprint 
and $\approx 40\%$ of the BOSS footprint are in overlap regions. 
Unfortunately, small separation quasar pairs are rare, 
and only a small fraction of the of SDSS/BOSS spectra from overlapping plates have 
sufficient data quality to meet our analysis criteria. 

To better sample the gas surrounding quasars at small \rphys, we have
been conducting a comprehensive spectroscopic survey to discover
additional close quasar pairs and to follow-up the best examples for our
scientific interests. Close quasar pair candidates are selected from 
a photometric quasar catalog \citep{bovy11,bovy12}, and are confirmed
via spectroscopy on 4m class telescopes including: the 3.5m telescope
at Apache Point Observatory (APO), the Mayall 4m telescope at Kitt
Peak National Observatory (KPNO), the Multiple Mirror 6.5m Telescope,
and the Calar Alto Observatory (CAHA) 3.5m telescope. Our continuing
effort to discover quasar pairs is described in \citet{thesis},
\citet{BINARY}, and \citet{hennawi10}. 
Projected pair
sightlines were then observed with 8m-class telescopes at the 
Keck, Gemini, MMT, Magellan, and LBT observatories to obtain
science-grade, absorption-line spectra.
Over the years, we have had various scientific goals when conducting
the follow-up spectroscopy.  
This includes 
measuring the small-scale clustering of quasars
\citep{BINARY,HIZBIN,hennawi10}, 
exploring correlations in the IGM along close-separation sightlines
\citep{ehm+07,martin+10},
analyzing small-scale transverse Ly$\alpha$ forest correlations
\citep{rorai+13}, 
characterizing the
transverse proximity effect (Hennawi et al.\, in prep), and using the b/g sightline to
characterize the circumgalactic medium of the f/g quasar (our QPQ
series).  Only the latter effort is relevant to this paper.

We emphasize that the nature of previously measured
absorption at \ion{H}{1} \lya\ for the
f/g quasar very rarely\footnote{The only significant exceptions are
  the MagE observations which often targeted systems with known,
  strong absorption at the f/g quasar.  However, all of these pairs
  were first surveyed at a lower dispersion with an 8m class telescope
  and would have been included in the QPQ6 sample even without the
  MagE observations.}  influenced the target selection.  Therefore,
the overall dataset has no explicit observational bias regarding
associated \ion{H}{1} absorption\footnote{It is conceivable that
  subtle effects, e.g.\ dust obscuration, could affect target
  selection, but at present we consider this highly improbable.}.  
The diverse nature of our
scientific programs and evolving telescope access, however, has led to
a follow-up dataset which is heterogeneous in terms of spectral
resolution ($R \approx 1000-40,000$), S/N, and wavelength coverage.
From the master dataset, we generated the QPQ6 sample as described
below.  A discussion of the spectroscopic observations and data
reduction procedures are given in the following sub-section.

The parent sample
is all unique, projected quasar pair sightlines which
have a proper transverse separation of $\mrphys < 1\,{\rm Mpc}$
at the redshift of the f/g quasar. The initial list of potential quasar pair
members includes any known systems, irrespective of the survey design
or spectroscopic characteristics.  This includes 
all of the quasars in the SDSS \citep{sdss_qso_dr7}, 2QZ
\citep{croom04} and BOSS \citep{paris+12} samples, 
all the quasars confirmed from our 4m telescope follow-up targeting quasar
pairs, and any quasars discovered during our observations on
large-aperture telescopes. 
For the following analysis, we have further restricted to pairs where
each member has
either an SDSS, BOSS, or large-aperture telescope science
spectrum of the b/g quasar.

An initial cut on velocity difference between
the redshifts of the two quasars of $\Delta v > 2000\mkms$ 
was made to minimize confusion between physically unassociated projected pairs
and physically associated binary quasars. For physical binaries, it is impossible to 
distinguish absorption intrinsic to the background quasar from absorption
associated with the foreground quasar.  
Strong broad absorption line
(BAL) quasars with large \ion{C}{4} equivalent widths (EWs) were
excluded from the analyses, if apparent in either the f/g or b/g
quasar.   Mild BALs were excluded if BAL
absorption in the b/g quasar coincided with \lya\ of the f/g quasar
redshift or if BAL absorption precluded a precise redshift
estimate of the f/g quasar.  
This yielded a parent sample of over 2000 quasars pairs with proper
separation at the f/g quasar of $\mrphys \le 1$\,Mpc.

We further require that the wavelength of the f/g quasar's \lya\ line,

\begin{equation}
\mllya = (1+\mzfg) \, 1215.6701 \rAA \;\;\; ,
\end{equation}
lie within the wavelength coverage of the b/g
spectrum, and at a velocity corresponding to 
500\kms\ {\it blueward} of the b/g quasar's \lyb\ emission line. 
The latter criterion is a first, lenient cut to avoid
\lya\ absorption being confused with higher order Lyman series lines. 
We impose a stricter cut after re-measuring the f/g quasar
redshift (see below).
In the b/g quasar spectrum, we measured the
average signal-to-noise per rest-frame \AA\ 
in a $\pm 500\mkms$ window centered on \llya, \snlya. 
This \snlya\ ratio is measured from an estimated
continuum for the b/g quasar not the absorbed flux.  
If no continuum had been generated in
the pursuit of previous analyses (e.g.\ QPQ5), we used the
algorithm developed by \cite{lee+12} to measure quasar continua 
for SDSS and BOSS spectra (see $\S$~\ref{sec:continuum} for details).
All pairs with $\msnlya > 4.5$ were passed through for further
consideration. We also
visually inspected the spectra with $\msnlya < 4.5$
and passed through those cases where the automated algorithm had failed to
generate a sensible continuum. 

\begin{figure*}
\includegraphics[width=5in,angle=90]{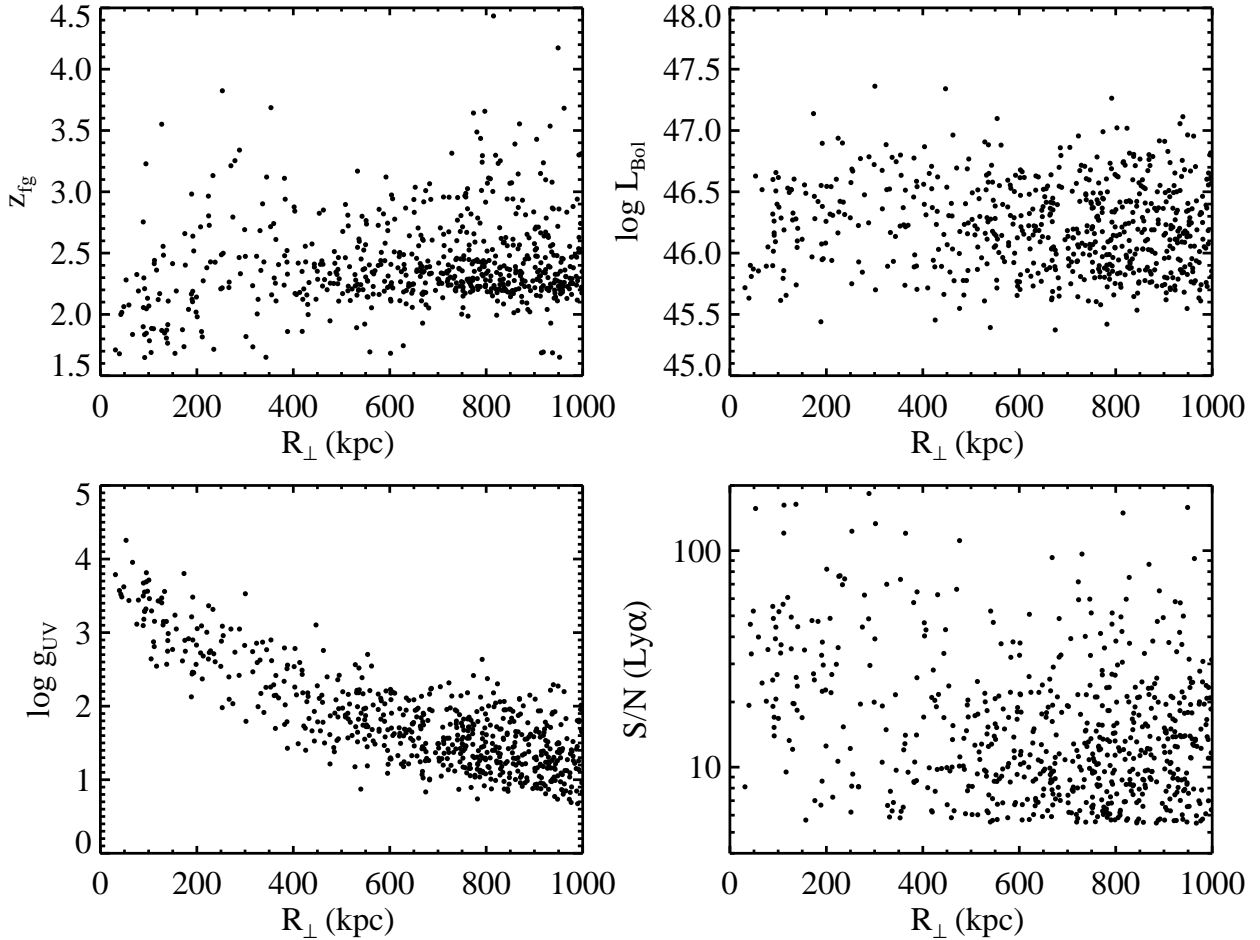}
\caption{These panels summarize key properties of the QPQ6 quasar pair
  dataset as a function of pair separation \rphys\ (projected and
  proper, measured at the f/g quasar redshift).  
  The f/g quasars have redshifts predominantly at $\mzfg =
  2-3$ and Bolometric luminosities $L_{\rm Bol} = 10^{46} -
  10^{47} \, {\rm erg \, s^{-1}}$.  Assuming isotropic emission and 
  a distance equal to the impact parameter, we estimate enhancements in the UV flux
  relative to the EUVB of $g_{\rm UV} = 10-10,000$ scaling inversely with
  $\mrphys^2$.   Lastly, the signal-to-noise of the b/g spectrum at
  \lya\ of the f/g quasar (\llya) ranges from 5.5 (the QPQ6 cutoff) to
  over 100 per rest-frame \AA.
}
\label{fig:demograph}
\end{figure*}

For the pairs that survived these cuts, we re-analyzed the f/g quasar
spectrum to measure a more precise emission redshift \zfg\ (see
$\S$~\ref{sec:redshifts} for details).  
Note that this is the only analysis performed on the f/g quasar
spectrum in this manuscript.
This cut on redshift quality 
eliminated $\approx 25\%$ of the pairs.  Using these
revised redshifts, we recorded the velocity offset between \llya\ 
and \lyb\ of the b/g quasar and demanded a separation of 
$1500\mkms$ (to the red).  
Similarly, we further restrict the sample to pairs where the velocity
difference between the new f/g quasar redshifts and the b/g redshift exceeds
4000\kms.  This insures the quasars are projected and should minimize
the impact from the proximity region of the b/g quasar.
Next, we generated a continuum for any b/g quasar spectrum 
without one or with
a poor estimation from the automated algorithm
($\S$~\ref{sec:continuum}).
Lastly, we re-measured \snlya\ 
and required that it exceed \sncut\ per rest-frame \AA.  
This criterion is a compromise between maximizing sample size versus
maintaining a high-level of data quality on the individual
sightlines.  We adopt even stricter criteria on \snlya\ for several of
the following analyses.
Lastly,
we identified a small set ($\approx 30$) of pairs
where the b/g quasar spectrum was compromised by insufficient
wavelength coverage, a detector gap, or previously unidentified BAL 
features.  These quasars were eliminated from any
further consideration.  
 
\begin{deluxetable*}{lcccccccccc}
\tablewidth{0pc}
\tablecaption{QPQ6 Survey\label{tab:qpq6_survey}}
\tabletypesize{\footnotesize}
\tablehead{\colhead{f/g Quasar} & \colhead{$\mzfg^a$} & 
\colhead{$L_{\rm 912}^b$} & \colhead{\lbol$^b$} & \colhead{b/g Quasar} & \colhead{\zbg} & \colhead{\rphys} & \colhead{\guv$^b$}
\\
&& (cgs) & (cgs) &&& (kpc) }
\startdata
J000211.76$-05$2908.4&2.8190&30.46&46.49&J000216.66$-05$3007.6&3.147&768&   65&\\
J000426.43$+00$5703.5&2.8123&29.96&46.00&J000432.76$+00$5612.5&2.920&882&   17&\\
J000536.29$+00$0922.7&2.5224&30.38&46.42&J000531.32$+00$0838.9&2.848&725&   57&\\
J000553.32$-03$1200.3&2.5468&29.87&45.91&J000551.25$-03$1104.7&3.058&533&   33&\\
J000629.92$-00$1559.1&2.3327&29.79&45.85&J000633.35$-00$1453.3&2.882&711&   16&\\
J000839.31$-00$5336.7&2.6271&30.70&46.71&J000838.30$-00$5156.7&2.887&841&   89&\\
J001028.78$-00$5155.7&2.4268&29.87&45.89&J001025.73$-00$5155.3&2.800&387&   61&\\
J001247.12$+00$1239.4&2.1571&30.59&46.64&J001250.49$+00$1204.0&2.203&532&  164&\\
J001351.21$+01$2717.9&2.2280&... & ...&J001357.14$+01$2739.2&2.303&784&...&\\
J001605.88$+00$5654.2&2.4021&30.15&46.28&J001607.27$+00$5653.0&2.598&176&  558&\\
\enddata
\tablenotetext{a}{Redshifts measured as described in $\S$~\ref{sec:redshifts} and listed in Table~\ref{tab:qpq6_zsys}.}
\tablenotetext{b}{See text and QPQ2 for a description of the derivation of these quantities.  Pairs without a value do not have a reliable photometric measurement.}
\tablecomments{[The complete version of this table is in the electronic edition of the Journal.  The printed edition contains only a sample.]}
\end{deluxetable*}

The final QPQ6 sample comprises \npairs\ pairs at
$\mzfg >1.6$ with $\mrphys < 1$\,Mpc.
Figure~\ref{fig:demograph} presents a series of
plots summarizing the demographics of the f/g quasars and
the spectral quality;
Table~\ref{tab:qpq6_survey} lists these properties.
From $\mrphys \approx 30 - 500$\,kpc we have a fairly uniform sampling 
of impact parameters.  Beyond 500\,kpc, where we are no longer limited
by fiber collisions, the sample is dominated by
BOSS spectroscopy and the f/g quasars tend toward higher redshift and
the number of pairs per \rphys\ interval increases with separation. 
Nevertheless, there is no strong dependence on the bolometric
luminosity with \rphys.  The values range from
$10^{45.5-47}$\,erg\,s$^{-1}$, where we have combined 
SDSS $i$-band photometry and the \cite{mclure04} relation to convert magnitudes
to bolometric luminosities.
If these sources are shining at near the Eddington limit (we adopt
10\%\ as a fiducial value), they
correspond to black holes with masses of $\gtrsim 10^{9} \msol$
\citep{shen11}.

The specific luminosities at 1\,Ryd ($L_{912}$) were estimated from the quasar
redshift and the SDSS photometry \citep{QPQ1}, and they
range from $10^{29.5} - 10^{31.2} \cgssflux$. This implies
an enhancement in the radiation field by the quasar 
relative to the extragalactic UV background at 
100\,kpc (1\,Mpc) of $g_{\rm UV} \approx 1000 (10)$ (see QPQ1 for how
$g_{\rm UV}$ is computed). 

The quasar pair sample presented in QPQ5 is a subset of the QPQ6
dataset, restricted to have $\mrphys \le 300$\,kpc to isolate the CGM,
$\msnlya > 9.5$, and further restricted to SDSS, BOSS-DR9, Keck/LRIS,
Gemini/GMOS data taken prior to 2011, and any of our Magellan
observations.

\subsection{Spectroscopic Observations}
\label{sec:spec}

Our analysis draws on several datasets to explore the \ion{H}{1}
absorption associated with $z\sim 2$ quasars.
In practice, we have utilized at least two spectra per pair: one to
measure the redshift of the f/g quasar and
another to gauge the \ion{H}{1} \lya\ absorption in the spectrum of a b/g
quasar.  A majority of the sources
rely on spectra from the SDSS DR7 \citep{sdssdr7} and BOSS DR9 \citep{boss_dr9}
surveys, which have spectral resolution $R \approx 2000$ 
and wavelength coverage from $\lambda \approx 3800-9000$\AA\ 
and $\lambda \approx 3600-10000$\AA\, respectively.  
We refer interested readers to the survey
papers for further details.

\begin{deluxetable}{lcccccccccc}
\tablewidth{0pc}
\tablecaption{Journal of Observations\label{tab:qpq6_obs}}
\tabletypesize{\footnotesize}
\tablehead{\colhead{b/g Quasar$^a$} & \colhead{Obs.} & \colhead{Instr.$^b$} & 
\colhead{Date$^c$} & \colhead{Exp. (s)$^d$} }
\startdata
J002802.60$-10$4936.0&Keck&ESI&2008-07-04&3600\\
J022519.50$+00$4823.7&Keck&ESI&2005-11-28&7100\\
J022519.50$+00$4823.7&Keck&ESI&2005-11-28&7100\\
J081806.87$+07$1920.2&Keck&ESI&2006-11-18&1800\\
J102616.11$+46$1420.8&Keck&ESI&2008-01-04&3600\\
J103900.01$+50$2652.8&Keck&ESI&2008-01-04&2000\\
J121533.54$-03$0925.1&Keck&ESI&2007-04-12&10800\\
J131428.97$+28$1840.2&Keck&ESI&2008-07-04&2400\\
J154225.81$+17$3322.9&Keck&ESI&2008-06-05&3000\\
J155952.67$+19$2310.5&Keck&ESI&2008-07-04&1800\\
\enddata
\tablenotetext{a}{The majority of the f/g quasars were observed at the same time and for the same exposure.}
\tablenotetext{b}{See the text for details on the instrument configuration.}
\tablenotetext{c}{UT date of the first night this object was observed by our program with this instrument.}
\tablenotetext{d}{Total exposure time for the spectrum covering \lya\ using this instrument.}
\tablecomments{[The complete version of this table is in the electronic edition of the Journal.  The printed edition contains only a sample.]}
\end{deluxetable}

Our QPQ survey has been gathering follow-up optical spectra on
large-aperture telescopes using spectrometers with a diverse range of
capabilities. 
This includes data from the Echellette Spectrometer and Imager
\citep[ESI;][]{ESI}, the Low Resolution Imaging
Spectrograph \citep[LRIS;][]{LRIS}, and the 
High Resolution Echelle Spectrometer \citep[HIRES;][]{vogt94} on the
twin 10m Keck telescopes, 
the Gemini Multi-Object Spectrograph \citep[GMOS;]{gmos} on the 8m
Gemini North and South telescopes, 
the Magellan Echellette Spectrograph \citep[MagE;][]{mage} 
and the Magellan Inamori Kyocera Echelle \citep[MIKE;][]{bernstein03} 
spectrometers on the 6m Magellan telescopes, and the 
Multi-Object Double Spectrograph \citep[MODS;][]{mods} on the Large
Binocular Telescope (LBT).  A summary of all these observations is
provided in Table~\ref{tab:qpq6_obs}.

At the W.M.~Keck Observatory, we have exploited three optical
spectrometers to obtain spectra of quasar pairs.
For the Keck/LRIS observations, we generally used the 
multi-slit mode with custom
designed slitmasks that enabled the placement of slits on other known
quasars or quasar candidates in the field.  LRIS is a double
spectrograph with two arms giving simultaneous coverage of the near-UV
and red.  We used the D460 dichroic with the $1200$ lines mm$^{-1}$
grism blazed at $3400$\,\AA\ on the blue side, resulting in wavelength
coverage of $\approx 3300-4200$\,\AA, a dispersion of 
$0.50$~\AA\ per pixel,  and the $1''$ slits give a FWHM resolution 
of $125\mkms$.
These data provide the coverage of \lya\ at $z_{\rm fg}$. 
On the red side we typically used
the R600/7500 or R600/10000 gratings with a tilt chosen to cover the
\ion{Mg}{2} emission line at the f/g quasar redshift, useful for
determining accurate systemic redshifts of the quasars  (see \S~\ref{sec:redshifts}).  Occasionally
the R1200/5000 grating was also used to give additional bluer
wavelength coverage.  The higher dispersion, better sensitivity, and
extended coverage in the red provided high signal-to-noise ratio
spectra of the \ion{Mg}{2} emission line and also enabled a more
sensitive search for metal-line absorption in the b/g quasar (see
Prochaska et al.\ in prep.).  Some of our older data also used the
lower-resolution 300/5000 grating on the red-side covering the
wavelength range $4700-10,000$~\AA. About half of our LRIS
observations were taken after the atmospheric dispersion corrector was installed, 
which reduced slit-losses (for point sources) in the UV.
The Keck/LRIS
observations took place in a series of runs from 2004-2008. 
Keck/HIRES observations ($R \approx 35,000$) were taken for one pair
in the sample;  these observations and data reduction were described
QPQ3.
Keck/ESI observations ($R\approx 5,000$) were obtained for \nesi\
pairs covering \lya\ in the QPQ6 sample.  These data have been
previously analyzed for \ion{C}{4} correlations between neighboring
sightlines \citep{martin+10} and for the analysis of an intriguing
triplet of strong absorption systems \citep{ehm+07}.  We refer the
reader to those papers for a full description of the data acquisition
and reduction.

The Gemini data were taken with the GMOS on the Gemini North and South
facilities. We used the B$1200\_G5301$ grating which has 1200 lines
mm$^{-1}$ and is blazed at 5300~\AA. The detector was binned in the
spectral direction resulting in a pixel size of 0.47\,\AA, and the
$1\arcsec$ slit corresponds to a FWHM~$\simeq 125\mkms$.  The slit was
rotated so that both quasars in a pair could be observed
simultaneously.  The wavelength center depended on the redshift of the
quasar pair being observed. We typically observed $z \sim 2.3$ quasars
with the grating centered at 4500\,\AA, giving coverage from
$3750-5225$\,\AA, and higher redshift $z\sim 3$ pairs centered at
4500~\AA, covering $4000-5250$\,\AA. The Gemini CCD has two gaps in
the spectral direction, corresponding to 9\,\AA\ at our
resolution. The wavelength center was thus dithered by
15-50\AA\ between exposures to obtain full wavelength coverage in the
gaps. The Gemini North observations were conducted over three
classical runs during UT 2004 April 21-23, UT 2004 November 16-18, and
UT 2005 March 13-16 (GN-2004A-C-5, GN-2004B-C-4, GN-2005A-C-9,
GN-2005A-DD-4).  We are also pursuing a new project to study the CGM
of damped \lya\ systems (unassociated with the f/g quasar) which began
in Semester~2012A on Gemini South (GN-2012A-Q-12), and is continuing 
on Gemini North and South (GN-2012-B-Q-12 and GS-2012-B-Q-20). 
These data were taken with a
$0.5''$ wide longslit using the B600\_G5307 grating yielding a
FWHM~$\approx 125 \mkms$ spectral resolution.  We employed two central
wavelengths covering $3600-6270$\AA\ and $4350-7150$\AA respectively.
The data presented here were taken prior to August 20, 2012, and are thus
restricted to the GN-2012A-Q-12 program only.

Observations of \nmagell\ pairs were obtained with the Magellan telescopes
using the MagE and MIKE spectrometers.   These data have spectral
resolution $R = 4,000$ for MagE and $R=28,000$ ($R=22,000$) for blue (red) 
side of MIKE. The wavelength coverage of the MagE instrument is fixed at
$\lambda \approx 3050 - 10300$\AA\ and we observed with MIKE in its
standard configuration giving $\lambda \approx 3300 - 9150$\AA.
MagE data were obtained on the nights of UT 2008 January 07-08, UT 2008 April 5-7, and 
UT 2009 March 22-26, whereas MIKE data was obtained only on the latter
observing run.

We have a complementary program to study the CGM of DLAs using the MODS
spectrometer on the LBT.  Our current sample includes 5 pairs from that
survey.  Each was observed with a $0.6''$ longslit oriented to include
each member of the pair.  The blue camera was configured with the
G400L grating giving a FWHM~$\approx 200\mkms$ spectral resolution and the
red camera used the G670L grating giving FWHM~$\approx 200 \mkms$.
Together the data span from $\lambda \approx 3000 - 10000$\AA.

\begin{figure}
\includegraphics[width=3.7in]{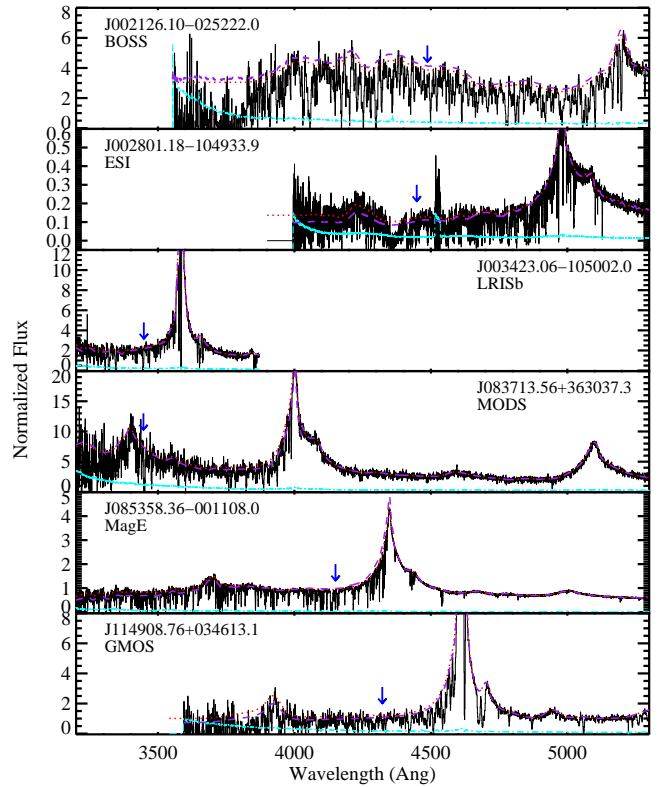}
\caption{A set of representative spectra of the b/g quasars in the
  QPQ6 dataset.   
The data are plotted in the observer reference frame
  and the blue arrows indicate the position of \lya\ for the f/g
  quasar, \llya.   
 The cyan dash-dotted line traces the
  $1 \sigma$ error array.  The dotted red line traces the original
  continuum, while the dashed purple
  line shows the mean-flux regulated (MFR) continuum 
  (see $\S$~\ref{sec:continuum} for definitions). 
}
\label{fig:rep_spec}
\end{figure}

All of the follow-up spectra that our team acquired were reduced
with custom IDL data reduction pipelines (DRPs)
developed primarily by 
J. Hennawi  and
J.X. Prochaska 
and are publicly available and distributed within the XIDL software
package\footnote{http:www.ucolick.org/$\sim$xavier/XIDL}.  
We refer the interested reader to the paper describing the MagE
pipeline \citep{mage_pipeline} which summarizes the key algorithms
employed in all of the DRPs.  In short, the spectral images are bias
subtracted, flat-fielded, and wavelength calibrated, and the codes
optimally extract the data producing a calibrated (often fluxed) 1D
spectrum.  
We estimate a $1\sigma$ uncertainty vector for each co-added 
spectrum based on
the detector characteristics, sky spectrum, and the measured RMS in
multiple exposures. 
Wavelength calibration was always performed using calibration arc
lamps and frequenly corrected for instrument flexure using sky
emission lines.  Uncertainties in this calibration are less than
one-half binned pixel, i.e.\ less than 35\kms\ for all of the
spectra.  Such error does not contribute to uncertainty in any o the
analysis that follows. 
A set of representative spectra are shown in
Figure~\ref{fig:rep_spec}.

\subsection{Preparation for the \ion{H}{1} \lya\ Analysis}

\subsubsection{Continuum Normalization}
\label{sec:continuum}

An assessment of the \ion{H}{1} \lya\ absorption requires
normalization of the b/g quasar spectrum at \llya.
In addition, control measures of \ion{H}{1}
absorption are performed on `random' spectral regions throughout the
data; this also requires normalized spectra.  Therefore, we have
estimated the quasar continuum at all wavelengths blueward of the
\ion{N}{5} emission line 
for each of the b/g quasar spectra in the QPQ6 dataset.

We have generated two estimates of the
continuum for each b/g quasar spectrum in QPQ6 with the following recipe.
First, for each spectrum which covered both the \lya\ forest
and the \ion{C}{4} emission-line of the b/g quasar, 
we used 
the Principal Component Analysis (PCA) 
algorithm developed by K.G. Lee \citep{lee+12} to generate a continuum.
The Lee algorithm generates a quasar SED based on a
PCA analysis\footnote{We employed the DR7 templates provided in the
  algorithm which we found gave better results.} 
of the data redward of \lya\ and then modulates this SED
by a power-law so that the transmitted flux at
rest-frame wavelengths
$\lambda = 1040-1180$\AA\ best matches 
the mean transmission of the IGM measured by \cite{fpl+08}.    
This latter treatment is referred to as mean-flux regulation.
In the version of the code available to the lead author at the time,
there was no masking of strong absorption lines (e.g.\ DLAs) along the
sightlines.   Therefore, this first estimate often showed regions of
the continuum that were biased too low.
Similarly, stochasticity in the \lya\ forest means that some spectra have
significantly lower/higher \lya\ absorption than the average,  which
biases the continuum for that individual quasar.  To mitigate
these effects, one of us (JXP) visually inspected each
continuum and re-normalized the estimate when it was obviously 
required (e.g.\ the continuum lies well below the observed flux).  
In some cases an
entirely new continuum was required, and in all cases the continuum
was extended to longer and shorter wavelengths than the region
fitted by the Lee algorithm. 
These modifications to the continuum were done by-hand, 
using a spline algorithm.  

A spline algorithm was also adopted for the spectra where
the Lee algorithm could not be applied or where it failed.
One of us (JXP) generated a spline function
by-hand that traces the obvious undulations and emission features of
the b/g quasar.  
These features, of course, are more easily discerned
in the higher $S/N$ spectra.  This led, in part, to the imposed
\snlya\ criterion of the QPQ6 sample.  
At the typical spectral resolution of our QPQ6 sample, one
generally expects the normalized flux to lie below unity (in the
absence of noise) owing to integrated \lya\ opacity from the IGM.  We
took this into full consideration when generating the spline continuum
and also allowed for the expected increase in \lya\ opacity with increasing
redshift.

At the end of this first stage, we had generated a continuum for every
spectrum of the b/g quasars in the QPQ6 sample.  In the following, we
refer to this set of continua as the `Original' continua.  These
were applied to the data to estimate S/N and to perform the
line-analysis in $\S$~\ref{sec:indiv}.   For all other analyses, we
modulated these original continua as follows.  
First, any BOSS spectrum was renormalized by the Balmer flux
correction recommended for these data \citep{lee+13}.  
Second, we mean-flux regulated
every original continuum following the \cite{lee+12} prescription,
i.e. by solving for the power-law
that best matches the mean flux of the
IGM estimated by \cite{fpl+08}:
\begin{equation}
<F>_{\rm IGM} = \exp(-\tau_{\rm IGM}) \;\; {\rm with} \;
\tau_{\rm IGM} = 0.001845 (1+z)^{3.924}  \;\;\; .
\label{eqn:FIGM}
\end{equation}
In this analysis, we have masked the $\pm 50$\AA\ region surrounding 
\lya\ of the f/g quasar and $\pm 100$\AA\ around strong \lya\ absorbers
(primarily DLAs) given by \cite{lee+13} and our
own visual inspection (Rubin et al.\, in prep.).
The derived power-law was then applied to the original continuum at
all wavelengths $\lambda_r < 1220$\AA\ in the rest-frame of the b/g
quasar. 
We refer to this second set as the ``Mean-Flux-Regulated (MFR)'' continua.  

Figure~\ref{fig:rep_spec} shows the original and MFR continua 
on the sample of representative data.
We estimate the average uncertainty in the continuum to be 
$\approx 5-10\%$ (dependent on S/N), improving to a few percent outside the
\lya\ forest.  The majority of this error is systematic (e.g.\ poorly
modeled fluctuations in the quasar emission lines), 
but such errors should be uncorrelated with properties
of the f/g quasars.  And, we reemphasize that we have masked the
spectral region near the f/g quasar when performing the mean-flux
regulation to avoid it from influencing the result. 

\begin{deluxetable}{lcccccccccc}[hb]
\tablewidth{0pc}
\tablecaption{QPQ6 Quasar Redshifts\label{tab:qpq6_zsys}}
\tabletypesize{\footnotesize}
\tablehead{\colhead{f/g Quasar} & 
\colhead{Spect} &
\colhead{Lines} &
\colhead{\zem} &
\colhead{$\sigma_z$} \\ &&&& (\kms)}
\startdata
J000211.76$-05$2908.4&SDSS&SiIV,CIV, [CIII]&2.8190& 520&\\
J000426.43$+00$5703.5&BOSS&SiIV&2.8123& 792&\\
J000536.29$+00$0922.7&BOSS&MgII&2.5224& 272&\\
J000553.32$-03$1200.3&BOSS&SiIV, CIV&2.5468& 714&\\
J000629.92$-00$1559.1&BOSS&CIV&2.3327& 794&\\
J000839.31$-00$5336.7&BOSS&SiIV,CIV, [CIII]&2.6271& 520&\\
J001028.78$-00$5155.7&BOSS&CIV&2.4268& 794&\\
J001247.12$+00$1239.4&SDSS&SiIV,CIV, [CIII]&2.1571& 520&\\
J001351.21$+01$2717.9&BOSS&CIV&2.2280& 794&\\
J001605.88$+00$5654.2&BOSS&CIV, [CIII]&2.4021& 653&\\
\enddata
\tablecomments{[The complete version of this table is in the electronic edition of the Journal.  The printed edition contains only a sample.]}
\end{deluxetable}

\subsubsection{Redshift Analysis}
\label{sec:redshifts}

The redshifts used for the initial selection of quasar pairs 
were taken from the SDSS or BOSS catalogs.  
The methodology used by those
projects, detailed in \cite{sdss_qso_dr7,paris+12}, 
is to fit a quasar template to
the observed spectrum and solve for the emission redshift \zem.  
It is now well-recognized, however, that these redshifts are not
optimal and may even have a significant and systematic offset 
from the systemic redshift of the quasar's host galaxy
\citep{shen07,hw11,font13}.

For our analysis of quasar pairs, the results are most sensitive to
the systemic redshift adopted for the f/g quasar.  We
aim to associate the source with \ion{H}{1} absorption which shows
significant variations on scales of $100\mkms$ in the $z \sim 2$ IGM.
Therefore, we have refined the SDSS/BOSS redshift measurements as
follows. Our methodology uses the custom line-centering algorithm described 
in QPQ1 to determine the line-center of one or more
far-UV emission lines (\ion{Mg}{2}, [\ion{C}{3}], \ion{Si}{4},
\ion{C}{4}). We then use the recipe in \cite{shen07} for combining these 
measurements from different emission lines.
We center all emission lines with $S/N > 5$ covered by our
spectroscopic dataset.  These data, of course, are distinct from the
spectra of the b/g quasars and may not even include the f/g quasar's
\lya\ line.  The specific spectrum and emission-lines analyzed are
listed in Table~\ref{tab:qpq6_zsys}.  When it is available, we adopt
the redshift estimated from \ion{Mg}{2} alone because its offset from
systemic is the smallest and it also exhibits the smallest scatter
about systemic \citep[after applying the offset;][]{richards02}.  We
cover \ion{Mg}{2} emission for many of the f/g quasars in the
QPQ6 sample having $\mzfg < 2.4$ ($\approx 40\%$), 
and we assume a $1\sigma$ uncertainty of $272 \mkms$ following
\cite{richards02}.

In lieu of \ion{Mg}{2}, we analyze one or more of the remaining
emission lines depending on the wavelength coverage and $S/N$ of the
spectra.  If none of the lines could be analyzed, the pair has not
been included here.  The precision assumed for \zem\ depends on which
lines were analyzed \citep[Table~\ref{tab:qpq6_zsys};][]{richards02}
and is in the range of $\sigma(z_{\rm em}) \simeq 270-770\mkms$ 
All of the automated fits
were inspected by-eye and minor modifications were occasionally
imposed (e.g.\ eliminating a blended or highly-absorbed line from the
analysis).

We measure an average offset 
between \zem\ and $z_{\rm orig}$ 
of $\Delta v \approx +390 \mkms$, which is due to a systematic offset
results from the redshift determination algorithm used by the SDSS
survey survey \citep{richards02,hw11}.  
\cite{font13} report a similar offset based on their analysis of
quasar clustering with the \lya\ forest.
There is no strong redshift
dependence for $\Delta v$ aside from larger uncertainties at higher
$z$ where the spectra no longer cover the \ion{Mg}{2} emission line.
We proceed with the analysis using these new estimates for the
redshifts (tabulated in Table~\ref{tab:qpq6_zsys}).  For the systems
with \zem\ values derived from \ion{Mg}{2} emission, the precision
($\approx 270\mkms$) is comparable to the peculiar velocities expected
in the dark matter halos hosting our luminous f/g quasars.  The
uncertainties for the remainder of the sample, however, likely exceed
these motions and result in a significant source of uncertainty in our associations
of quasars with IGM absorption. We are performing a survey of near-IR quasar spectroscopy that includes
$\approx 100$ QPQ6 members, to establish more precise redshifts from
[\ion{O}{3}], H$\alpha$, and/or H$\beta$ emission.


\section{\ion{H}{1} Absorption in Fixed Velocity Windows around \zfg}
\label{sec:wlya}

\subsection{Definitions and Tests}
\label{sec:tests}

In previous papers in the in the QPQ series (QPQ1,QPQ2,QPQ4,QPQ5)
we have demonstrated that the CGM surrounding
quasars exhibits significant \ion{H}{1} absorption relative to the
average opacity of the $z \sim 2$ \lya\ forest on scales of $\mrphys <
300$\,kpc. In QPQ5 this result was recovered from the analysis of composite spectra, which
collapses the distribution of \ion{H}{1} absorption along many sightlines
to a single measure.  A large fraction of our
spectra, however, are of sufficient quality to perform a
pair-by-pair analysis, subject to the uncertainties of the quasar
redshifts and continuum placement.  
In this section, we measure the \ion{H}{1}
absorption for individual sightlines and explore the results as a
function of the impact parameter and quasar properties.
In a later section ($\S$~\ref{sec:stacks}), 
we return to stacking, extending
the QPQ5 measurement to 1\,Mpc and explore correlations and systematic
uncertainties. 

We quantify the strength of \ion{H}{1} absorption in two steps:
(1) associate regions in the b/g quasar spectrum with the \lya\ `location'
of the f/g quasar; and
(2) assess the \ion{H}{1} absorption.  
For the latter, there are several standard measures -- 
(i) the average normalized flux \avgf, measured over a specified
velocity window $\Delta v$;
(ii) the rest-frame equivalent width of \lya, \wlya; and 
(iii) the physical column density of \ion{H}{1} gas, \nhi.  
The first two quantities are relatively straightforward to measure with spectra of
the quality that we have obtained.  If measured in the same velocity interval
$\Delta v = c \Delta \lambda / \lambda$, then
\avgf\ and \wlya\ are essentially interchangeable:
$\mwlya = (1-\mavgf^{\Delta v}) \, \Delta\lambda$. 
In the following, we treat these as equivalent measures of the
\ion{H}{1} absorption strength.
We examine the \nhi\ values in Section~\ref{sec:NHI}, which
includes the more subjective association of individual absorption
lines to the f/g quasar and the challenges of determining
column densities from low-resolution spectra.
For all of the analyses in this section, we adopt the MFR continua
($\S$~\ref{sec:continuum}).  

Ideally, one might measure \avgf\ or
\wlya\ in a spectral region centered on the
quasar redshift and encompassing only the interval physically
associated to the host galaxy's environment.  In practice, this
analysis is challenged by several issues.
First, as discussed in $\S$\ref{sec:redshifts}, the f/g quasars
comprising QPQ6 have $1\sigma$ uncertainties for their emission
redshifts of at least $272\,\mkms$ and  frequently as large as
$770 \mkms$.   The latter corresponds to many \AA\ in the observer frame.
Second, the dark matter halos hosting quasars are estimated  to have
masses of $M_{\rm DM} \sim 10^{12.5} \msol$ at $z \approx 2.5$. 
The characteristic velocity\footnote{The line-of-sight velocity
  dispersion will be even larger (e.g.\ QPQ3).} 
of such halos is $v_c \approx 330\,\mkms
\, (M_{\rm DM}/10^{12.5}\msol)^{1/3}$, i.e.\ gas associated with such structures
should have peculiar velocities of several hundreds \kms\ (analogous to
individual galaxies in a cluster).
Therefore, even when the quasar's redshift is precisely constrained (e.g.\ 
via [OIII] emission), one must still analyze a relatively large velocity
window. 

The negative consequence of adopting a large velocity window
is that the intergalactic medium at $z>2$
exhibits a thicket of \ion{H}{1} absorption at nearly all wavelengths.
Even within a spectral window of 100\kms, one is likely 
to find strong absorption related to the \lya\ forest. 
It is only at $z<1$, where \lya\ absorption is rare,
that one can confidently associate the observed 
\ion{H}{1} absorption with the environment of a given galaxy
\citep[e.g.][]{pwc+11,tumlinson+13}.  
For $z>2$, a non-zero \wlya\ value is nearly guaranteed.
Interpretation of the observed distribution therefore requires
comparison to control distributions measured from random regions of
the Universe.

\begin{figure}
\includegraphics[width=3.7in]{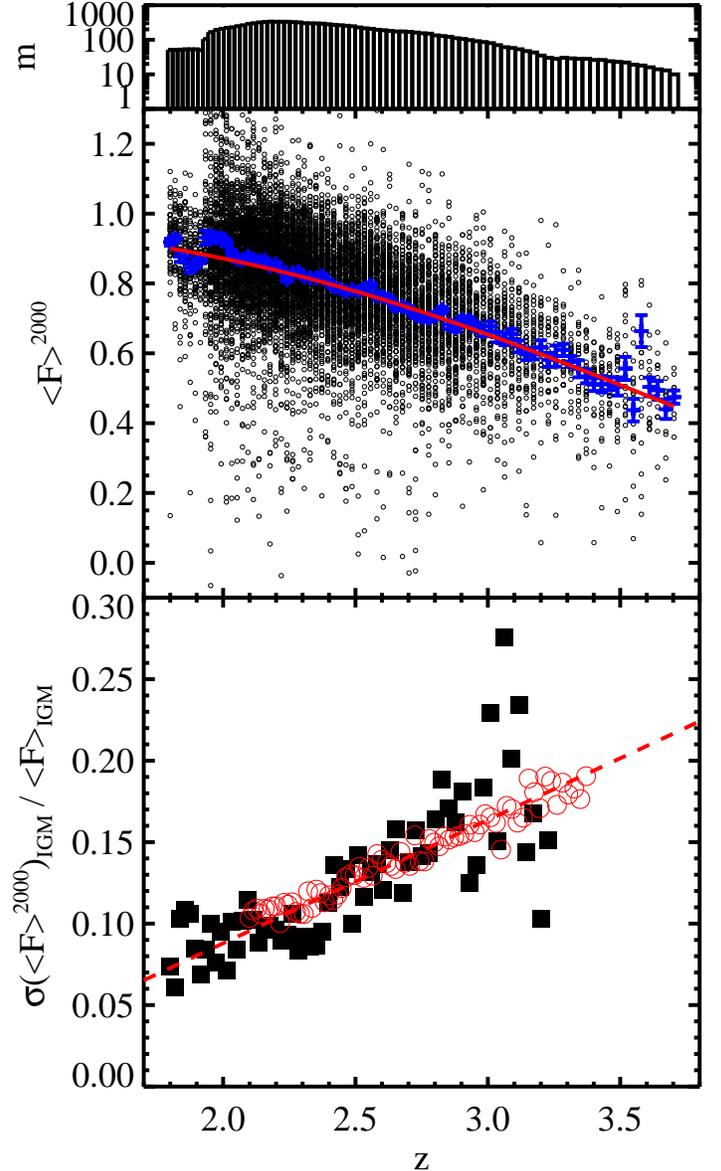}
\caption{({\it upper two panels}): Each circle in the middle panel shows a single flux
  measurement in a 2000\kms\ window at the given redshift, with the
  number of unique measurements $m$ described by the histogram in the
  top panel.  The blue points indicate the average of the
  $\mavgf^{2000}$ measurements and the standard deviation of the
  mean.  The solid red line shows the mean flux
  (Equation~\ref{eqn:FIGM}), which was used to modulate the MFR continua
  of our quasar spectra ($\S$~\ref{sec:continuum}).  There is
  excellent agreement at all redshifts except at $z \approx 2$ which
  may be attributed to the poorer data quality of BOSS spectra at
  $\lambda < 3700$\AA\ and to greater uncertainty in our approach to
  mean flux regulate these data \citep[see also][]{lee+13}.
  ({\it lower panel}): Black squares show the scatter in the measured
  $\mavgf^{2000}$ values from the QPQ b/g quasar sightlines,
  restricted to $\msnlya > 15$ and normalized by the mean flux of the
  IGM.  Given the high signal-to-noise ratio, the variance is dominated by intrinsic fluctuations
  in the IGM.  
  The red dashed line shows an unweighted,
  least-squares fit to the values at $z<3$.  We recover 
  $\msigf_{\rm IGM}/\mavgf_{\rm IGM} = -0.068 + 0.077z$.   For
  comparison, we present measurements of the scatter from the BOSS sample of
  \citet[][red circles]{lee+13}, 
  using the sample of 1927 quasars with S/N~$>8$ per pixel. 
  There is excellent agreement between these measurements and 
  our analysis of the QPQ6 dataset.
}
\label{fig:test_igm}
\end{figure}

As described in $\S$~\ref{sec:continuum}, we have generated continua
within the \lya\ forest that are forced to reproduce the mean flux of
the IGM \citep{fpl+08}.  We test the efficacy of this procedure by
measuring the average flux $\mavgf^{2000}$ in a series of contiguous
$\Delta v =2000\mkms$ windows
from $z=1.8$ for all of the QPQ6 spectra. 
The $\Delta v = 2000\mkms$ window is motivated by the analysis that follows
on the regions surrounding quasars; here we assess the behavior 
of this same statistic in the ambient IGM. We restrict
the measurements to the spectral region $\lambda_r = 1041-1185$\AA\ in
the rest-frame of each b/g quasar (to avoid \lyb\ emission and the
proximity zone of the b/g quasar).  
Figure~\ref{fig:test_igm}a shows
the $\mavgf^{2000}$ measurements for the QPQ6 sample, the average
value for the $m$ quasars within each window (\mnaf; equal
weighting), and the standard deviation in the mean,
$s(\mmnaf)/\sqrt{m}$.  
The variation about the mean in each spectrum (the black dots) is 
is caused by a combination of noise and intrinsic fluctuations in the 
forest and continuum errors. 
As expected, we find a decreasing
\mnaf\ value with increasing redshift.  
Overplotted on the
figure is the mean flux of the IGM $\mavgf_{\rm IGM}$, defined by
Equation~\ref{eqn:FIGM}, which was explicitly used in our
MFR continuum procedure.  For $z>2.1$, we find very good agreement
between the mean flux measured from the spectra and the input value
(as expected).  At $z=2$, the $\mavgf^{2000}$ measurements show
systematically higher values which we attribute to poor fluxing of the
BOSS spectra at those wavelengths and to error in extrapolating the
power-law into the bluest spectral region of the BOSS spectra
\citep[see also][]{lee+13}.  At $z<2$, the measurements are made with
follow-up spectroscopy from large-aperture telescopes.  These values
lie slightly below the $\mavgf_{\rm IGM}$ evaluation but are nearly
consistent with Poisson scatter\footnote{Part of the offset may also
  be the results of fluxing errors in these data which are not fully
  corrected by the mean-flux regulation algorithm.}.  
We note that the scatter in the
individual $\mavgf^{2000}$ values is systematically smaller, 
owing to the higher S/N in these spectra. 

\begin{deluxetable*}{lccccccccccccccccc}
\tablewidth{0pc}
\tablecaption{QPQ6 \ion{H}{1} Measurements\label{tab:qpq6_NHI}}
\tabletypesize{\scriptsize}
\tablehead{\colhead{f/g Quasar} &
\colhead{b/g Quasar} & \colhead{\rphys} &
\colhead{Spec.} & 
\colhead{\zfg} &
\colhead{S/N$^a$}  & 
\colhead{$\mavgf^{2000}$} &
\colhead{$\mdeltf^{2000}$} &
\colhead{$\delta v^b$} & \colhead{$\lambda^c$} &
\colhead{\wlya$^d$} & \colhead{log \nhi} & \colhead{flg$_{\rm OT}^e$} \\
&& (kpc) &&&& & & (\kms) & (\AA) & (\AA) &&&}
\startdata
J000211.76$-05$2908.4&J000216.66$-05$3007.6&768&SDSS&2.819& 41&$ 0.66$&$ 0.06$&$ -232$&[4633.4,4647.2]&$ 1.28\pm0.05$&$<18.00$& 0\\
J000426.43$+00$5703.5&J000432.76$+00$5612.5&882&BOSS&2.812& 17&$ 0.58$&$ 0.18$&$ -429$&[4622.4,4632.6]&$ 1.33\pm0.11$&$ $& 0\\
J000536.29$+00$0922.7&J000531.32$+00$0838.9&725&BOSS&2.522& 10&$ 0.79$&$-0.02$&$  317$&[4279.9,4293.7]&$ 1.58\pm0.19$&$<19.00$& 0\\
J000553.32$-03$1200.3&J000551.25$-03$1104.7&533&BOSS&2.547&  6&$ 0.78$&$-0.02$&\\
J000629.92$-00$1559.1&J000633.35$-00$1453.3&711&BOSS&2.333& 11&$ 0.79$&$ 0.03$&$   40$&[4044.8,4060.6]&$ 1.63\pm0.21$&$<19.00$& 0\\
J000839.31$-00$5336.7&J000838.30$-00$5156.7&841&BOSS&2.627& 23&$ 0.59$&$ 0.22$&$  551$&[4411.4,4423.2]&$ 1.55\pm0.08$&$<18.80$& 0\\
J001028.78$-00$5155.7&J001025.73$-00$5155.3&387&BOSS&2.427& 14&$ 0.68$&$ 0.15$&$ -797$&[4149.4,4159.8]&$ 1.40\pm0.11$&$ $& 0\\
J001247.12$+00$1239.4&J001250.49$+00$1204.0&532&BOSS&2.157&  9&$ 0.40$&$ 0.53$&\\
J001351.21$+01$2717.9&J001357.14$+01$2739.2&784&BOSS&2.228&  7&$ 0.79$&$ 0.05$&\\
J001605.88$+00$5654.2&J001607.27$+00$5653.0&176&BOSS&2.402&  7&$ 0.74$&$ 0.07$&\\
\enddata
\tablenotetext{a}{Signal-to-noise per rest-frame Angstrom at the position of \lya, \snlya.  This quantity is measured from the original continuum.}
\tablenotetext{b}{Velocity offset from \zlya\ to \zfg.  Note that absorption lines were assessed only for data with $\msnlya>9.5$.}
\tablenotetext{c}{Wavelength interval for the \lya\ analysis.}
\tablenotetext{d}{Rest equivalent width of the strongest absorption system within 1500\kms\ of \zfg.}
\tablenotetext{e}{Flag indicating an assessment of whether the system is optically thick at the Lyman limit ($-1$=Thin; 0=Ambiguous; 1=Thick).}
\tablecomments{[The complete version of this table is in the electronic edition of the Journal.  The printed edition contains only a sample.]}
\end{deluxetable*}

We may also compute the variance in a set of $m$ measurements of
$\mavgf^{2000}$: 
\begin{equation}
\sigma^2(\mavgf^{2000}) \equiv \frac{\smm_{i=1}^m \ltp \mavgf^{2000} 
- \mmnaf \rtp^2}{m} \;\; .
\end{equation}
As described above, the variance
includes contributions from Poisson
noise, continuum placement, and intrinsic variations in the IGM.
To isolate the latter effect in the following, 
we restrict the evaluation to spectra with
$S/N > 15$ per rest-frame \AA\ at $z_{\rm Ly\alpha}$.  The results for
the QPQ6 data are
shown in Figure~\ref{fig:test_igm}b, using the same set of redshifts as
the upper panel but ending at $z=3.25$ where the sample size is too
small.   In each case, we have
normalized $\msigf$ by the mean-flux at
each redshift $\mavgf_{\rm IGM}$.  
We observe a roughly linear increase in $\msigf/\mavgf_{\rm IGM}$ 
with increasing redshift owing to the evolving relationship between
transmission in the \lya\ forest and overdensity as the universe expands
\citep{hui99}.  The figure also shows a linear, least-squares
fit to the measurements at $z<3$ which yields:
$\msigf/\mavgf_{\rm IGM} = -0.068 + 0.077z$.   Overplotted on the
figure are also a series of $\msigf/\mavgf_{\rm IGM}$ 
measurements drawn from the BOSS dataset of \cite{lee+13}.  
We find excellent agreement and conclude that our linear fit is a good
description for $z<3$.  It will be compared, in the following
sub-section, against the scatter in
$\mavgf^{2000}$ observed in the spectral regions associated with the f/g quasars.

Lastly, we introduce a third statistic which compares a \avgf\
measurement against the average value at that redshift:

\begin{equation}
\mdeltf \equiv \frac{ \mavgf_{\rm IGM} - \mavgf^{\Delta v}}{\mavgf_{\rm IGM}}
  \;\; .
\end{equation}
This quantity is analogous to the standard definition of overdensity
and is defined to be positive in higher opacity (lower flux) regions.
Although it is a relative quantity, it may offer greater physical
significance than the values of \avgf.  Furthermore, by normalizing to
$\mavgf_{\rm IGM}$ we may compare measurements from
sub-samples of QPQ6 having a range of redshifts.

\subsection{\ion{H}{1} Measurements at \llya}
\label{sec:avgf}

Consider first the average flux in 
total intervals $\Delta v = 1000, 2000$ and $3000\,\mkms$ that we
refer to as $\mavgf^{1000}$, $\mavgf^{2000}$,  and $\mavgf^{3000}$. 
The largest interval was chosen to have a high probability ($>90\%$) for containing
\llya\ of the f/g quasar, but it suffers the greatest dilution
from unrelated IGM absorption.  
The smallest velocity window,
meanwhile, does not cover even the $\pm 1\sigma_z$ uncertainty 
interval of \zfg\ for many of the f/g quasars.
We may also report these measurements in terms of the equivalent
width, e.g.\ $\mwlya^{1000} = (1-\mavgf^{1000}) \Delta\lambda^{1000}$ 
where $\Delta\lambda^{1000} = 1215.6701{\rm \AA} \, (1000\,\mkms)/c$.
Table~\ref{tab:qpq6_NHI} lists the \avgf\ and 
\wlya\ values measured in these various windows around each f/g quasar.
The errors listed refer to statistical errors but the uncertainties are
generally dominated by continuum placement.  The latter error
is systematic and scales roughly as the size of the velocity interval; 
 a $10\%$ 
error in the normalization translates to $\sigma(\mwlya) \approx 0.4$\AA\
for $\Delta v = 1000\mkms$.  
This is approximately five times smaller than
the average value observed, but it certainly 
contributes to the scatter in the observed distribution. 
For \snlya=10, the statistical error in a 1000\kms\ window
is $\approx 5-10\%$ depending on the actual $\mavgf^{1000}$ value.

\begin{deluxetable}{lccccccccccccccccc}
\tablewidth{0pc}
\tablecaption{QPQ6 \avgf\  Statistics\label{tab:qpq6_avgf}}
\tabletypesize{\scriptsize}
\tablehead{\colhead{Sample} & 
\colhead{$m_{\rm pair}$} & 
\colhead{$\langle \mzfg \rangle$} & 
\colhead{Median} & 
\colhead{Mean} & 
\colhead{RMS} & 
\colhead{IGM$^a$}  
}
\startdata
\cutinhead{Full QPQ6 with varying velocity window}
1000 km s$^{-1}$& 646&2.415&0.72&0.70&0.21&$0.81$&\\
2000 km s$^{-1}$& 646&2.415&0.73&0.71&0.18&$0.81$&\\
3000 km s$^{-1}$& 646&2.415&0.74&0.73&0.16&$0.80$&\\
\cutinhead{Variations with \rphys\ for a $\Delta v=2000\mkms$ window}
(0,100)  kpc&  20&2.045&0.69&0.61&0.22&$0.78$&\\
(100,200)  kpc&  36&2.137&0.73&0.66&0.22&$0.80$&\\
(200,300)  kpc&  22&2.376&0.72&0.72&0.17&$0.79$&\\
(300,500)  kpc&  70&2.333&0.76&0.74&0.19&$0.81$&\\
(500,1000)  kpc& 451&2.375&0.74&0.73&0.15&$0.82$&\\
(0,300)  kpc&  78&2.181&0.71&0.67&0.21&$0.81$&\\
(300,1000)  kpc& 521&2.369&0.75&0.73&0.16&$0.82$&\\
\enddata
\tablenotetext{a}{Measured from a control sample constructed to match the QPQ6 dataset.}
\end{deluxetable}

Table~\ref{tab:qpq6_avgf} provides statistics for these \avgf\ values
for the full QPQ6 sample.
In every interval,  we find relatively
strong absorption at \lya\ ($\mavgf^{\Delta v} \, \approx 0.7$). 
There is also significant dispersion, which decreases with increasing 
velocity interval. A portion of the scatter is
related to continuum placement and fluctuations in the background
IGM. Nevertheless, a visual inspection of the spectra reveals many
examples with very weak/strong absorption which also implies significant
scatter within the quasar environment.

\begin{figure*}
\includegraphics[width=3.7in,angle=90]{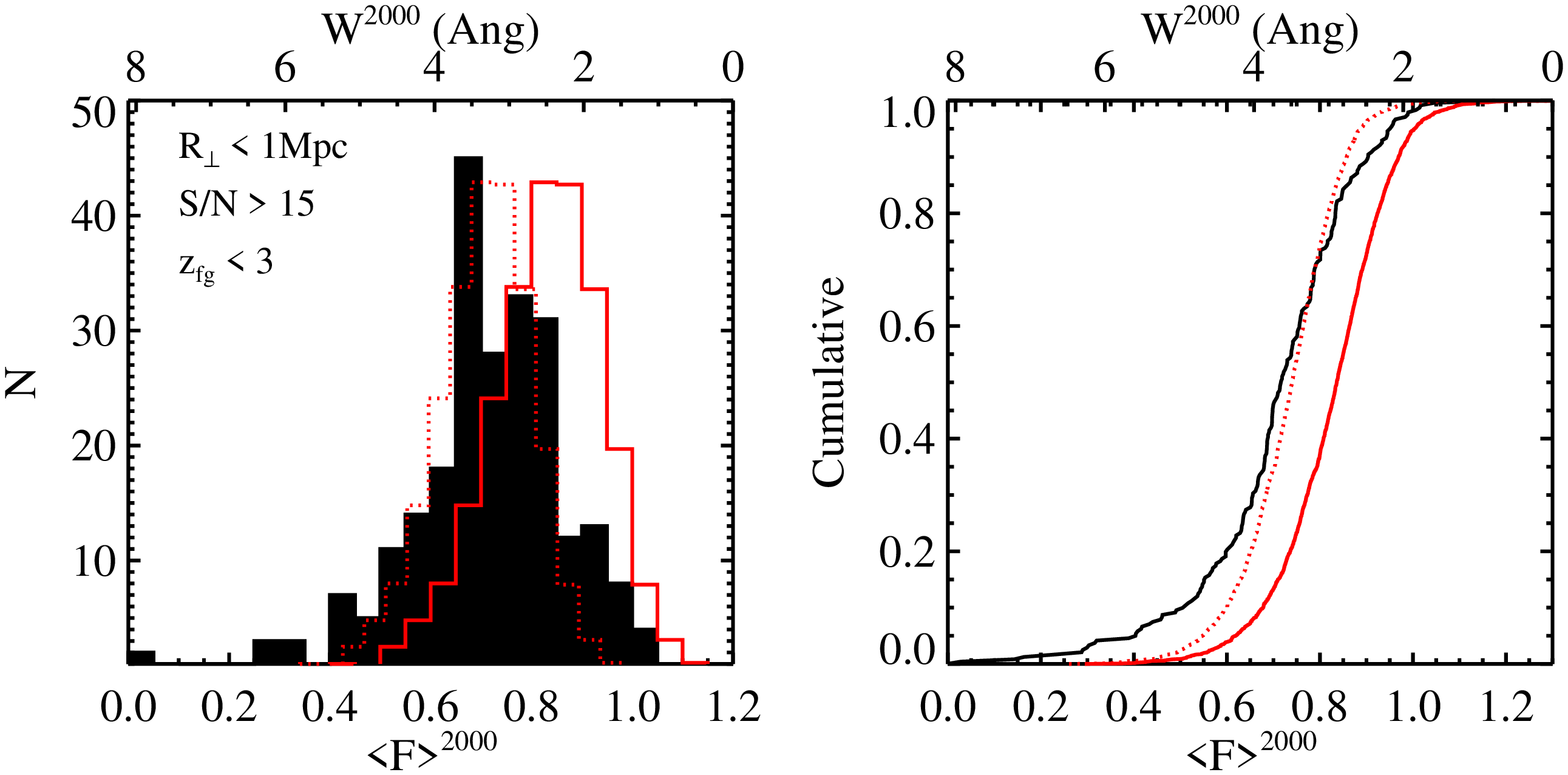}
\caption{(left): Histograms of measured $\mavgf^{2000}$ values for the
 QPQ6 sample restricted to $\msnlya > 15$ and $\mzfg < 3$ (black,
 filled),
 the random control sample with identical restrictions (gray, solid
  line), and
  the control sample with each \avgf\ value scaled by the ratio of the
  means $(1.13)^{-1}$ (gray, dotted).  
  (right): The cumulative $\mavgf^{2000}$
  distributions for the three samples shown in the left-hand panel.
  Even when one scales the control sample to have identical mean as
  the QPQ6 sample it is a poor description of the data because the
  latter exhibits much greater scatter. This
  scatter must be  intrinsic to the quasar environment.
}
\label{fig:hist_avgf}
\end{figure*}

A proper assessment of the \avgf\ values requires placing
them in the context of random regions in the Universe (i.e.\
Figure~\ref{fig:test_igm}).  
Scientifically, we aim to establish whether the quasar
environment has enhanced or reduced \ion{H}{1} absorption relative to
such random regions.  
For each f/g quasar, we randomly chose 10 other
quasar pairs from the QPQ6 sample such that the spectral region
at \llya:
(i) lies within the \lya\ forest of the b/g quasar;  
(ii) lies 1500\kms\ redward of \lyb; 
(iii) has a velocity separation of at least 3000\,\kms\ from the f/g quasar in
that pair;
and
(iv) lies at least 4000\,\kms\ blueward of the b/g quasar (to avoid its
proximity zone).
We then measure \avgf\ and record the
values.  We achieved 10 matches for all quasars in the sample except
for the 9~pairs with $\mzfg >3.5$.
Statistics on the \avgf\ values for this control sample are also given in
Table~\ref{tab:qpq6_avgf}. 
The quasar pair distributions have
lower \avgf\ values at high statistical significance, e.g.\ the mean
$\mavgf^{2000}$ for the full QPQ6 dataset is 0.71 with an error of
0.007 whereas the control sample has a mean of 0.81 with similar
uncertainty.   
Similarly, 
a two-sided Kolmogorov-Smirnov (KS) test rules out the null hypothesis of the QPQ6
and control samples being drawn from the same parent population 
at $>99.99\%$ for any of these velocity intervals.

For the remainder of analysis that follows
we focus on measurements in the 2000\,\kms\ window 
$\mavgf^{2000}$, which we consider to offer the best
compromise between maximizing signal from the quasar environment 
while minimizing IGM dilution. 
This choice is further
motivated by our analysis of individual absorbers ($\S$~\ref{sec:indiv}) and 
composite spectra ($\S$~\ref{sec:stacks}).
Qualitatively, we recover similar
results when using other velocity windows.

Figure~\ref{fig:hist_avgf} shows a comparison of $\mavgf^{2000}$
values for a restricted subset of the QPQ6 sample:
spectra with $\msnlya > 15$, and $\mzfg < 3$. 
We restrict to higher S/N in part to examine the intrinsic
variance of the distribution by minimizing the contribution of Poisson fluctuations
and continuum uncertainty. 
As with the full QPQ6 distribution,
the offset in \avgf\ values between the pair sample and control 
distribution is obvious and the KS test rules out the null hypothesis
at $>99.99\%$ confidence.
We may also compare the dispersion in the \avgf\ measurements.  
We measure $\msigf = 0.17$ for the 245~pairs in this restricted QPQ6
sample. Evaluating our fit to the variance of the 
IGM $\msigf$ ($\S$~\ref{sec:tests}, Figure~\ref{fig:test_igm}b)
at each of the \zfg\ values and averaging the different redshifts, 
we recover 0.087.  An $F$-test
yields a negligible probability that the $\mavgf^{2000}$
values from 
the QPQ6 and control samples have comparable variance.
This result is further illustrated in Figure~\ref{fig:hist_avgf} where we
compare \avgf\ distributions of the quasar pairs and the control
sample.  The dotted line shows the control values scaled by the ratio
of the means of the distributions (0.85).  This scaled
distribution is considerably more narrow than the QPQ6 sample.

The analysis presented above include pairs with a wide
distribution of proper separation and a range in redshifts and
quasar luminosities (Figure~\ref{fig:demograph}).  We now consider
the influence of several of these factors on the \ion{H}{1} absorption
strength. 
We begin with impact parameter \rphys, for which 
we may expect the strongest dependence.
We first restrict the QPQ6 sample 
to pairs with $\mzfg <3$ to produce a sub-sample of pairs
where \zfg\ is less correlated with \rphys.  
Our cut also mitigates against
the likelihood that the properties of the halos hosting quasars evolve significantly with
redshift, as suggested by clustering analysis \citep{shen07}. 
For example, if
higher redshift quasars occur in more massive halos, they might
have systematically distinct associated \ion{H}{1} absorption. 
We caution, however,
that the pairs with smallest \rphys\ do have redshifts that are
a few tenths smaller than those at larger impact parameter.
To further mitigate the effects of IGM evolution, we focus on the \deltf\
statistic instead of \avgf.

\begin{figure}
\includegraphics[width=3.5in]{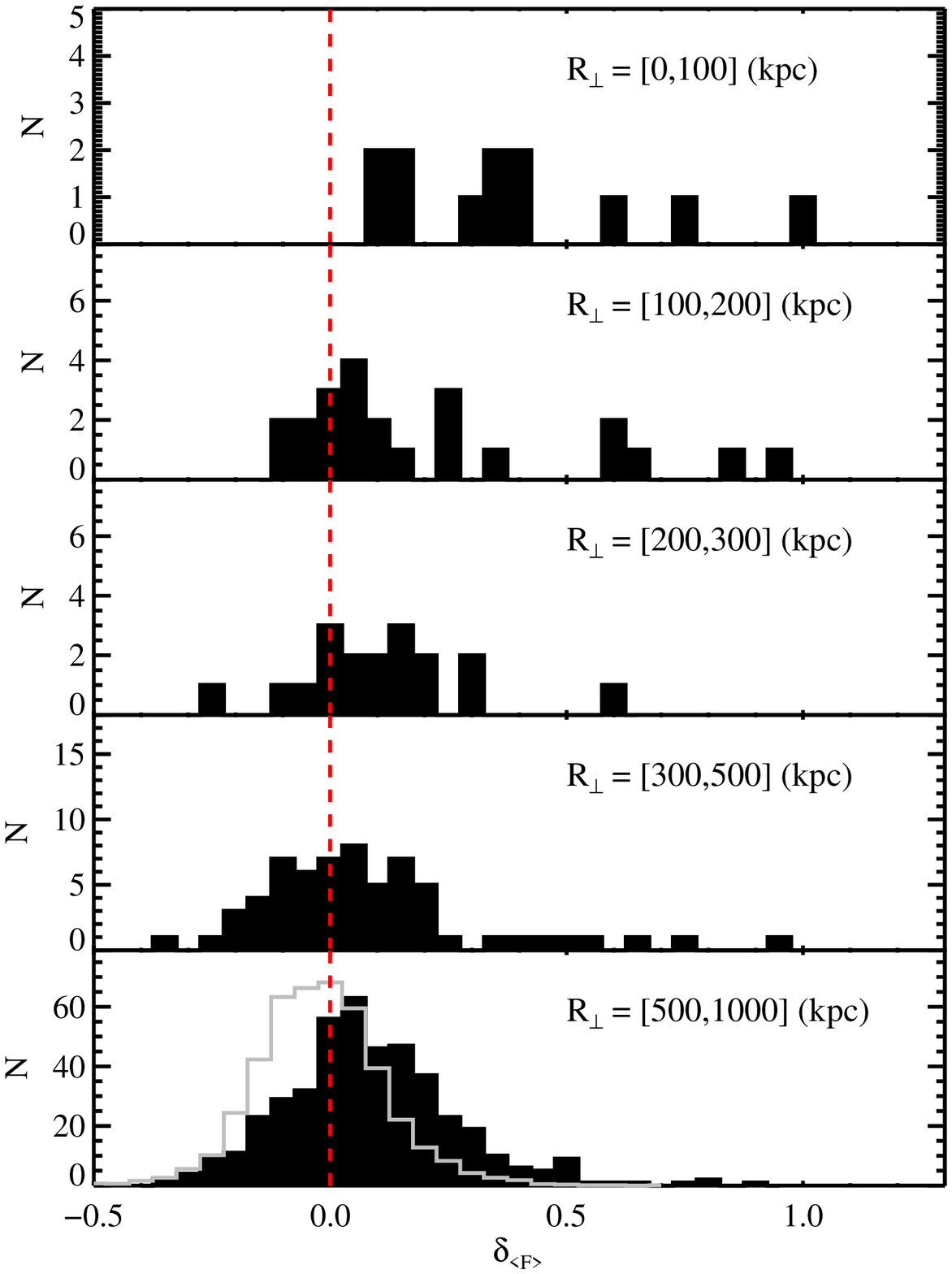}
\caption{These panels present the \deltf\ distributions for
  the QPQ6 quasar pairs (restricted to $\mzfg < 3$) in a series of
  impact parameter intervals.  It is evident that the distributions
  are skewed toward positive values for all of the \rphys\ intervals.
  We conclude: 
  (1) the quasar environment exhibits enhanced \ion{H}{1} absorption at
  all impact parameters $\mrphys < 1$\,Mpc; 
  (2) the excess increases with decreasing \rphys.
  In the last panel, we show the distribution of \deltf\ values for
  our control sample (gray curve).
}
\label{fig:hist_deltf_rho}
\end{figure}

Figure~\ref{fig:hist_deltf_rho} compares the distributions of \deltf\
values in a series of \rphys\ intervals.  Each subset
exhibits a systematic shift towards positive values, and a two-sided
KS test comparison of the values with the control
sample rules out the null hypothesis that the 
f/g quasar distribution is
drawn from the same parent population as the ambient IGM.  
We conclude that there is excess \ion{H}{1} \lya\
absorption at all impact parameters $\mrphys < 1$\,Mpc
from the galaxies hosting $z \sim 2.5$ quasars.  
We also find that the average \deltf\ values increase with decreasing
\rphys\ (see Table~\ref{tab:qpq6_deltf}), indicating the excess is
physically associated to the f/g quasar. 
The Spearman and Kendall tests yield correlation coefficients
implying a correlation at $\approx 98\%$ confidence.  
There is a large dispersion at all \rphys, related to intrinsic variations in the
\ion{H}{1} absorption associated with the f/g quasars, continuum
error, and fluctuations within the IGM. 
Comparing the scatter in these measurements against the
control sample, we find systematically larger scatter in the quasar
pair distributions.
Aside from the $\mrphys =
[200,300]$\,kpc interval (which shows systematically lower \deltf\
values), the $F$-test reports a negligible probability that the
variances are the same.

\begin{figure}
\includegraphics[height=3.5in,angle=90]{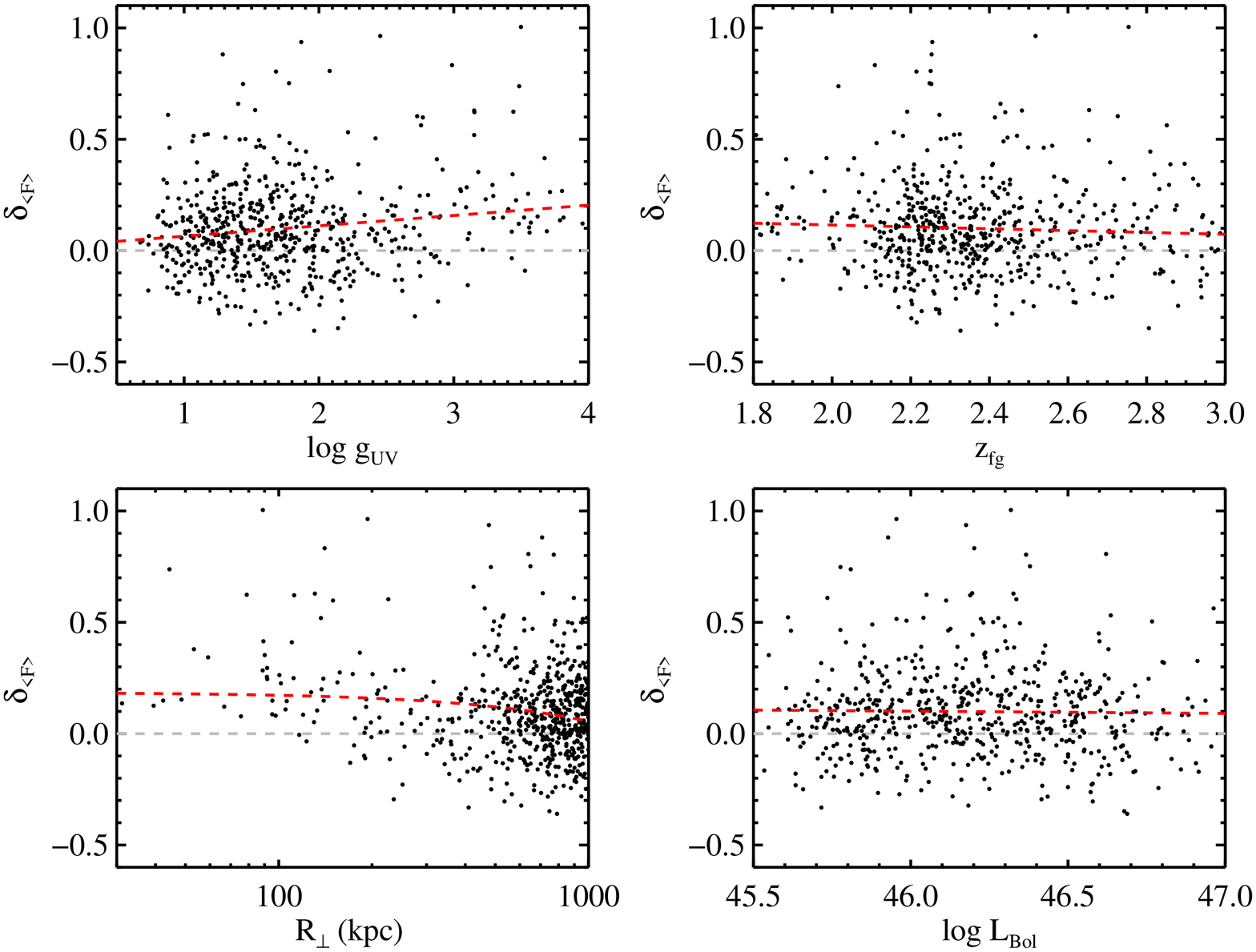}
\caption{Scatter plot of \deltf\ values versus various observables and
  physical
  quantities of the QPQ6 quasar pair sample.  
  For any range of the properties considered, the systematically
  positive \deltf\ values indicate that the quasar pairs
  exceed that of the control sample, i.e.\ enhanced \ion{H}{1}
  absorption occurs independently of $g_{\rm UV}$, \zfg, or quasar luminosity.
  The only apparent
  correlation (the red dotted line shows the least-squares, linear fit)
  is with $\log g_{\rm UV}$ which scales
  inversely with \rphys; therefore, this reflects a result similar to
  that presented in Figure~\ref{fig:hist_deltf_rho}.
}
\label{fig:deltf_scatter}
\end{figure}

In Figure~\ref{fig:deltf_scatter}, we show a series of scatter plots
comparing \deltf\ with various observables and physical quantities of
the quasar pairs: Bolometric luminosity, $g_{\rm UV}$, \zfg, and
\rphys.
We find the \deltf\ values are systematically positive 
indicating that the quasar pairs exhibit
enhanced \ion{H}{1}
absorption independently of $g_{\rm UV}$, \zfg, or quasar luminosity.
Furthermore, 
the only quantity besides \rphys\
where \deltf\ exhibits a correlation is
for $g_{\rm UV}$, and we recognize this as an equivalent result
because $g_{\rm UV} \propto R^{-2}$.

To briefly summarize (see $\S$~\ref{sec:discuss} for further
discussion), the 1\,Mpc environment surrounding quasars exhibit
enhanced \ion{H}{1} \lya\ absorption in their transverse dimension.  The
excess trends inversely with impact parameter indicating a
higher density and/or neutral fraction in gas towards the center of
the potential well.  The distribution of \avgf\ and \deltf\
values also show a larger scatter than the ambient IGM.
The enhanced absorption holds independently of any property of the
quasar or pair configuration. 

\begin{deluxetable}{lcccccccc}
\tablewidth{0pc}
\tablecaption{QPQ6 \deltf\  Statistics\label{tab:qpq6_deltf}}
\tabletypesize{\scriptsize}
\tablehead{\colhead{Sample} & 
\colhead{$m_{\rm pair}$} &
\colhead{$\langle \mzfg \rangle$} & 
\colhead{Median} & 
\colhead{Mean} & 
\colhead{RMS}  
}
\startdata
\cutinhead{Variations with \rphys}
(0,100)  kpc& 12&2.204&0.35&0.38&0.08\\
(100,200)  kpc& 23&2.318&0.09&0.22&0.09\\
(200,300)  kpc& 18&2.494&0.11&0.11&0.03\\
(300,500)  kpc& 63&2.388&0.05&0.08&0.06\\
(500,1000)  kpc&438&2.392&0.07&0.08&0.03\\
(0,300)  kpc& 53&2.352&0.15&0.22&0.08\\
(300,1000)  kpc&501&2.391&0.06&0.08&0.04\\
\enddata
\end{deluxetable}

\clearpage

\section{\ion{H}{1} Absorption from Individual Systems Associated to
  the f/g Quasar}
\label{sec:indiv}

In QPQ5 we demonstrated that a high fraction ($\approx 60\%$) of
the quasar pair sightlines with $\mrphys < 200$\,kpc intersect
optically thick gas surrounding the f/g quasar (see also QPQ1, QPQ2, and QPQ4).  
Furthermore, the
majority of these optically thick systems exhibit strong, metal-line absorption from
lower ionization transitions \citep[QPQ5;][]{farina13}.  
Such absorbers occur relatively rarely in b/g quasar sightlines 
through the intervening IGM (i.e. far from f/g quasars), and 
are thus qualitatively distinct from the canonical \lya\ forest.
In this respect, some of the excess absorption revealed by
Figure~\ref{fig:hist_avgf} must be related
to the individual absorption systems traditionally surveyed by quasar
absorption-line researchers, e.g.\ the Lyman limit systems (LLSs) and
damped \lya\ systems (DLAs).
Motivated by these results, we perform an analysis of the
strongest absorption system associated to each f/g quasar in a $\pm
1500 \mkms$ velocity interval.  We adopt a larger velocity window than
the fiducial 2000\kms\ used for the \avgf\ measurements in the
previous section to increase the
confidence that our analysis includes the strongest absorption related
to the f/g quasar (i.e.\ to more conservatively account for error in
the f/g quasar redshifts).

At $z>2$,  absorption surveys have
tended to focus on strong \lya\ absorbers
\citep[e.g.][]{opb+07,pw09,pow10,noterdaeme+12} and/or gas selected by
metal-line absorption \citep[e.g.][]{ntr05,cooksey+13}.
The definition of these absorption systems is somewhat
arbitrary and are not always physically motivated, 
e.g.\ the velocity window chosen for analysis, the equivalent width
limit adopted.
Similarly, the results derived in the following are not as rigidly
defined as those of the preceding section.  Nevertheless, there
is strong scientific value to this approach and we
again derive a control sample to perform a relative comparison 
to the ``ambient'' IGM.

\subsection{System Definition and Equivalent Widths}
\label{sec:ew}

We have adopted the following methodology to define and characterize
individual absorption systems associated to the f/g quasar.    
First we searched for the
strongest \lya\ absorption feature in the $\pm 1500\mkms$ velocity
interval centered on \llya\ for every QPQ6 pair with $\msnlya > 9.5$. 
This velocity criterion allows for uncertainty in \zfg\ and for
peculiar motions within the halos.
The \snlya\ criterion was imposed to limit the sample to spectra with
better constrained continua and higher quality data for the
\lya\ line assessment and 
associated metal-line absorption. 
A total of \nsys\ pairs were analyzed.  
Second, we set a velocity region for line-analysis based on the
line-profile and the presence of metal-line absorption (rarely
detected).   
The region generally only encompassed the strongest,
Gaussian-like feature at \lya\ but line-blending did
impose a degree of subjectivity.  
Third, we measured the equivalent width across this region, 
estimated the \ion{H}{1}
column density, and assessed the likelihood that the absorption system
is optically thick at the Lyman limit.  In the Appendix, we show a few
examples of this procedure.

\begin{figure}
\includegraphics[height=3.5in,angle=90]{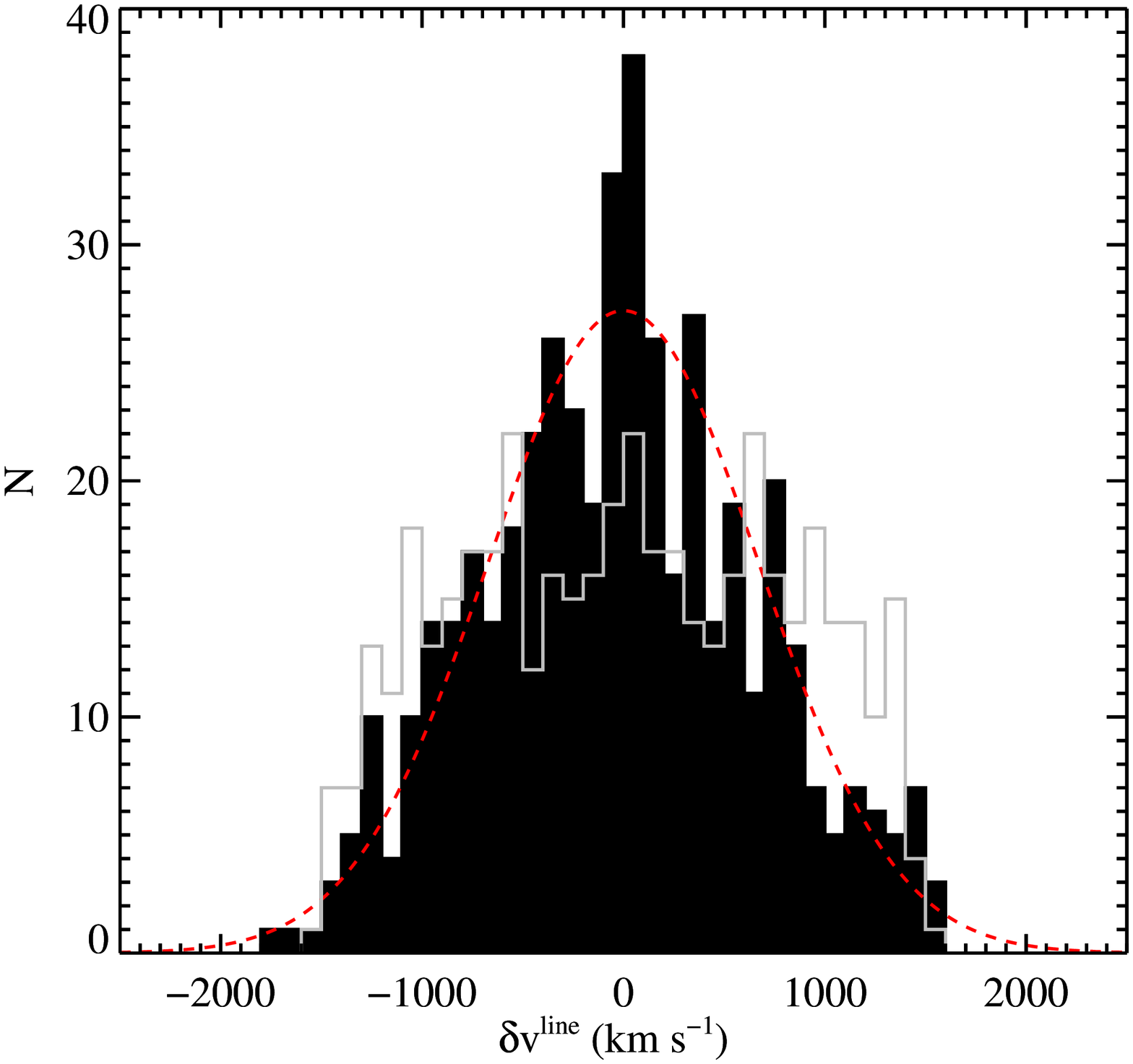}
\caption{ 
  The solid black histogram presents the velocity offsets
  \dvline\ for the strongest absorption lines in a $\pm 1500\mkms$
  interval about \llya\ of the f/g quasars.  The distribution has a
  Gaussian profile with a mean $<\mdvline> \, = -30 \mkms$ and a
  scatter $\sigma(\mdvline) = 670 \mkms$, as traced by the dashed red 
  line.   This distribution is consistent with the estimated error in
  our \zfg\ measurements.
  The gray histogram shows the \dvline\ distribution for a
  control sample.  It is nearly uniform across the $\pm 1500\mkms$
  interval.  
}
\label{fig:both_dv}
\end{figure}

To explore systematic effects associated with the `by-eye' line
identification, we repeated this analysis for a
random control sample of \nsys\ sightlines matched to our pair samples. 
Specifically, this control sample assumes
the same \zfg\ distribution of the
QPQ6 subset but we analyze the spectral region in the \lya\
forest of a randomly chosen spectrum taken from the full set of b/g quasar
data but restricted the b/g quasar as follows: 
we demanded that the spectral region covering \llya\ lies redward
of the \lyb\ emission line, away from the \lya\ emission line, and
away from the known f/g quasar associated to the b/g spectrum.
Figure~\ref{fig:both_dv} presents the velocity offsets \dvline\ between the line
centroid and \llya\ for each of the QPQ6 pairs.  These are centered near
zero\footnote{This implies our \zfg\ measurements have no large,
  systematic offset.}, 
have a Gaussian distribution, and show an RMS of 670\kms\ that is consistent with the
redshift uncertainties of the f/g quasars.
The figure also shows the \dvline\ distribution for the control
sample.  It is nearly uniform, as expected for a random sample. 
We have also examined the velocity offset as a
function of impact parameter.  The scatter is smaller for
$\mrphys < 300$\,kpc, suggesting a more physical association between
the gas and f/g quasar. It may also reflect, however, a somewhat
smaller uncertainty in \zfg\ for that subsample.  

\begin{figure}
\includegraphics[height=3.5in,angle=90]{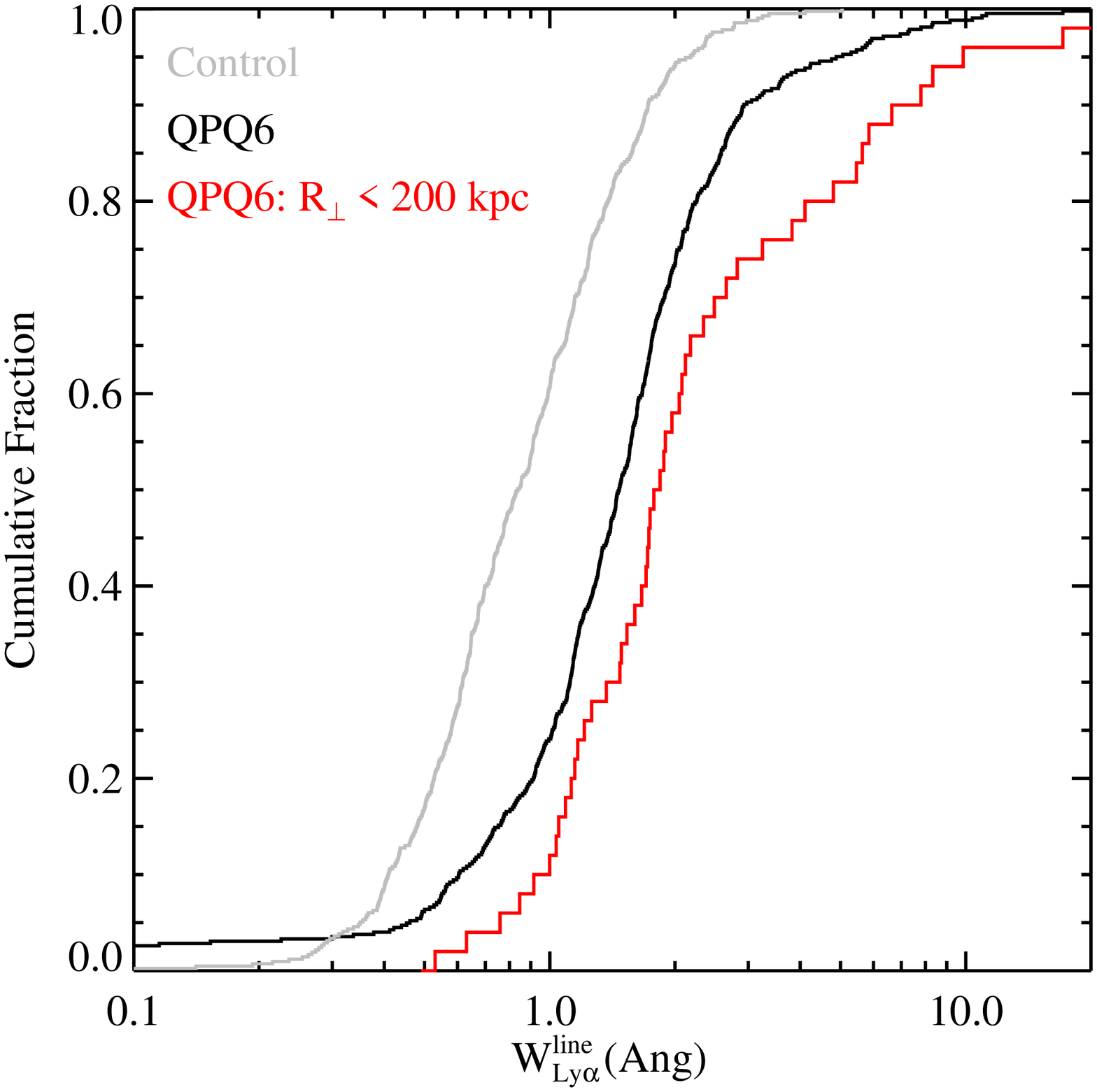}
\caption{The black curve traces the cumulative distribution of the
  rest-frame equivalent widths for the strongest absorption line system
  in a $\pm 1500\mkms$ interval around \llya\ of the f/g quasars from
  QPQ6.  The gray curve is the
  cumulative distribution for a control sample.   It shows
  systematically lower \wsubj\ values.  The red histogram traces the
  cumulative distribution from the subset of quasar pairs with
  $\mrphys < 200$\,kpc.  These pairs show systematically larger
  \wsubj\ values and a two-sided KS test rules out the null hypothesis
  that these are drawn from the same parent sample at high
  confidence ($P_{\rm KS} = 0.005$). 
}
\label{fig:cumul_wline}
\end{figure}

The cumulative distribution of \wsubj\ values
is presented in Figure~\ref{fig:cumul_wline}.  
We report values ranging from a few
0.1\AA\ to over 10\AA, with the majority of the sample having
$\mwsubj \approx 1-2$\AA.  This distribution is compared against the 
control sample, which
shows systematically lower \wsubj\ values 
than the QPQ6 measurements. For example, 60\%\ of the 
random sample have $\mwsubj < 1$\AA\ in comparison to fewer than 25\%\
of the true quasar pairs.
We also show the cumulative
distribution for the 32 pairs with $\mrphys < 200$\,kpc and find that it
is shifted towards even larger \wsubj\ values.  A two-sided
KS test comparing the full distribution (which includes the low
\rphys\ pairs) to the small separation pairs yields a low probability
($P_{\rm KS} = 0.02$) that the two distributions are drawn from the
same parent population.  The probability is even lower if we compare
the low \rphys\ pairs 
with pairs at $\mrphys > 500$\,kpc ($P_{\rm KS} = 0.005$).
We conclude at high confidence that the average strength of
associated \ion{H}{1} absorption lines increase with decreasing \rphys.

\subsection{\nhi\ Measurements}
\label{sec:NHI}

While \wsubj\ is a direct observable that reliably gauges the
\ion{H}{1} absorption strength, it has limited physical significance.
For several scientific pursuits, one would prefer to estimate the
total surface density of \ion{H}{1} gas\footnote{Of course, one would
  prefer to measure $N_{\rm H}$, the column density of total hydrogen
  but that can only be inferred from \nhi\ after estimating an
  ionization correction.}, i.e.\ the \ion{H}{1} column density \nhi.
As Figure~\ref{fig:cumul_wline} indicates, however, the majority of
absorbers exhibit $\mwsubj \approx 1-2$\AA, which places the systems
on the flat portion of the curve-of-growth.  In these cases, the data
have very poor sensitivity to the \nhi\ value; instead
\wsubj\ primarily traces the kinematics of the system.  Nevertheless,
one may resolve the damping wings of \lya\ for
systems with large \nhi\ values ($>10^{19}\cm{-2}$).
There is also a small set of systems showing very
weak absorption ($\mwsubj \ll 1$\AA) which provide upper limits to \nhi.  
As described below, we have also estimated \nhi\
in a broad bin to classify the gas as being optically thick to ionizing
radiation (i.e.\ $\mnhi \gtrsim 10^{17.3} \cm{-2}$).

For each system with $\mwsubj > 1.4$\AA, 
we have performed a Voigt-profile analysis of the \lya\ absorption.
When metal-lines are present, we have set
the \lya\ absorption redshift to correspond to the centroid of these
features.  We then fit the \nhi\ value of
the \lya\ line\footnote{We have assumed a $b$-value of 30\kms\ in this
  analysis.} while simultaneously making minor modifications to the
local continuum as necessary 
\citep[e.g.][]{phw05,opb+07}.
The data and profile fits for all of the systems with measured $\mnhi
\ge 10^{19} \cm{-2}$ are presented in the Appendix.
For those lines without damping wings, we set a conservative upper limit to \nhi\
based on the observed profile.  
Furthermore, systems with $\mwsubj < 1.4$\AA\ are
conservatively assigned to have $\mnhi < 10^{18.5} \cm{-2}$.  
We also analyzed the $\approx 100$~pairs with $\msnlya > 30$ 
where one can place
much tighter upper limits to \nhi\ when the absorption is very weak.
This yielded a set of systems with $\mnhi < 10^{17.3} \cm{-2}$.


The resultant \nhi\ values and upper limits are listed in
Table~\ref{tab:qpq6_NHI}.  Uncertainties in these measurements are
dominated by the systematic errors of continuum placement and
line-blending.  We estimate $1\sigma$ uncertainties of 0.15\,dex for
$\mnhi \ge 10^{20} \cm{-2}$ where the absorption is strongest and
0.25\,dex for systems having $10^{19} \cm{-2} \mnhi < 10^{20} \cm{-2}$ where
line-blending is a particular concern.   We have not attempted to
measure \nhi\ values below $10^{19} \cm{-2}$ but do impose upper
limits below this threshold.
For systems with $\mnhi \approx 10^{19} \cm{-2}$,
the error will not be distributed normally; there will be occasional
catastrophic failures of erroneously classified high-column
  density systems, for which the actual value 
$\mnhi \ll 10^{19} \cm{-2}$ due to unidentified blending.  

In our first pass, we fitted $\approx
60$ systems with $\mnhi \ge 10^{19} \cm{-2}$ and noted that a
significant fraction of these have $\mnhi \approx 10^{19} \cm{-2}$
which produces a damped \lya\ profile that is marginally resolved in
our lower resolution data.  These same classification criteria
resulted in an excess incidence for our random sightlines
over the expectation from previous surveys \citep[e.g.][]{opb+07}.
Therefore, we reexamined each of these systems (the QPQ6 and random
samples) for the presence of associated
low-ion absorption (e.g.\ \ion{C}{2}~1334, \ion{Al}{2}~1670) and
line-blending.  To be conservative, we have set all of the systems
without low-ion absorption or obvious damping wings to have upper
limits to \nhi.  This gave an incidence in the random sample that
is lower than expectation (albeit consistent within Poisson
uncertainty; 3 observed with 5.5 expected) 
and reduced the QPQ6 sample of secure
\nhi\ measurements.  
Given the results on the control sample, we expect if anything that
these conservative criteria have led us to underestimate the incidence
of systems with $10^{19} \cm{-2} < \mnhi < 10^{20} \cm{2}$ associated to
f/g quasars.
We compare the resultant \nhi\ distributions in the Appendix.

We have also examined the data at the Lyman limit for the $\approx 50$
pairs with wavelength coverage.  Most of these data are either
compromised by Lyman limit absorption from a higher redshift system or
poor S/N.  For those with good coverage, 
the presence/absence of strong Lyman
limit absorption is consistent with the \nhi\ values estimated from \lya.  
\begin{figure}
\includegraphics[width=3.5in,angle=90]{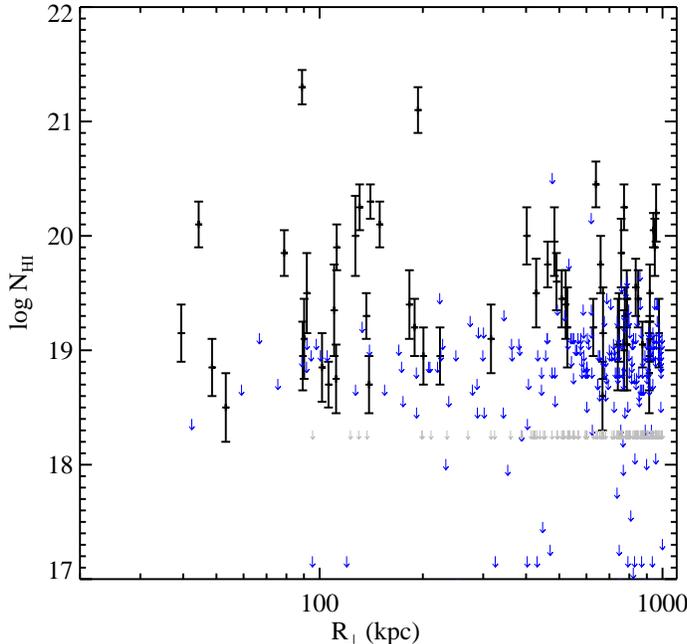}
\caption{Scatter plot of the estimated \nhi\ measurements and upper
  limits against impact parameter for the QPQ6 dataset (restricted to
  $\msnlya > 9.5$).   Perhaps the only discernable trend is a higher
  incidence of measured values at low
  \rphys\ indicating higher \nhi\ values for sightlines penetrating the
  putative CGM of the quasar environment (see also QPQ5).
  The gray arrows refer to the 134 cases with $\mwsubj < 1.4$\AA\ and
  $\msnlya < 30$ where we have automatically assigned an upper limit
  to \nhi\ of $10^{18.3} \cm{-2}$.
}
\label{fig:nhi_vs_rho}
\end{figure}

Figure~\ref{fig:nhi_vs_rho} presents a plot of the \nhi\ values
against impact parameter.  
The result is a complicated scatter plot dominated by upper
limits.   It is somewhat evident, however, that the pairs with
$\mrphys < 200$\,kpc have a much higher incidence of measured \nhi\
values than the pairs at larger impact parameters.  
Furthermore, the 
measured values at all \rphys\ are dominated by systems with $\mnhi
\approx 10^{19} \cm{-2}$ and there are very few systems satisfying the
damped \lya\ (DLA) criterion, $\mnhi^{\rm DLA} \ge 2 \sci{20}
\cm{-2}$ \citep{wgp05}.  In $\S$~\ref{sec:discuss}, we analyze these
measurements to study the clustering and covering fractions of strong
\ion{H}{1} absorbers in the extended, transverse environment of
luminous quasars.

Given the equivalent width for Ly$\alpha$ absorption, our Voigt
profile fits for the $N_{\rm HI}$, the presence/absence of Lyman limit
absorption, and the presence/absence of low-ion metal absorption,
objects were classified into three categories: {\it optically thick,
  ambiguous, or optically thin}. Objects which show obvious damping
wings, Lyman limit absorption, or strong ($W > 0.3$~\AA) low-ion metal
absorption are classified as optically thick. 
For the metals, we focused on the strongest low-ion
transitions commonly observed in DLAs \citep[e.g.][]{pro01}:
\ion{Si}{2}~$\lambda 1260, 1304, 1526$, \ion{O}{1}~$\lambda 1302$,
\ion{C}{2}~$\lambda 1334$ \ion{Mg}{2}~$\lambda 2796,2803$.
A complete description of the metal-line analysis will be presented in
QPQ7 (Prochaska et al., in prep.).
For those cases where
metal-lines are weak, are not covered by our spectral coverage, or are
significantly blended with the Ly$\alpha$ forest of the b/g quasar, a
system is classified as optically thick only if it has $\mwlya \ge
1.8$\,\AA\ in a single, Gaussian-like line.  For a single line with
Doppler parameter $b=40\mkms$, this equivalent width threshold
corresponds to $\mnhi > 10^{18.5} \cm{-2}$.  
There may be a significant number of cases,
however, where unresolved line-blending yields such a high equivalent
width in a system with a total $\mnhi < 10^{18} \cm{-2}$.  
When in doubt, we
designated the systems as ambiguous.  Note that this evaluation
differs slightly from our previous efforts (QPQ1,QPQ2,QPQ4,QPQ5) and the
classifications are not identical but very similar.

The completeness and false positive rate of this analysis are sources of
concern.  Line-blending, in particular, can significantly bias
\wlya\ and the column density high. 
We have assessed this estimate with our control sample, having
evaluated each random sightline for the presence of optically thick gas.  
We detect $\approx 25\%$ more LLS 
(defined to be systems with $\mnhi > 10^{17.3} \cm{-2}$) 
in the control sample 
than expectation from previous surveys \citep{pow10,omeara13}.  
The results are within the Poisson uncertainty (19 detected to 14.7
expected) 
but we allow that the QPQ6 sample may contain a modest set
of false positives.  We stress, however, that a majority of these
systems in the pair sample also exhibit strong low-ion metal
absorption (e.g.\ \ion{C}{2}~1334; QPQ5).

\begin{figure}
\includegraphics[height=3.5in,angle=90]{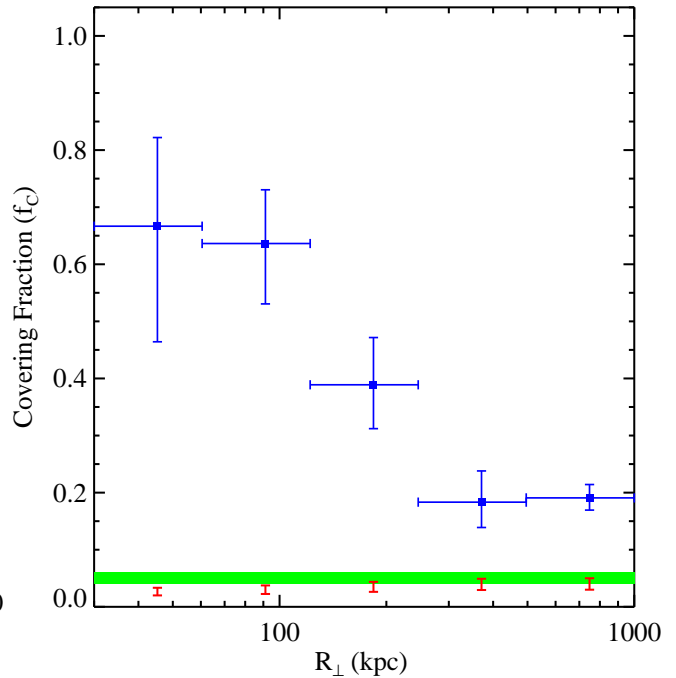}
\caption{
Estimates for the covering fraction \fc\ of optically thick
  absorbers for the QPQ6 sample in logarithmic bins of impact
  parameter.  These are presented formally as lower limits owing to the
  preponderance of ambiguous cases.  The red symbols trace the
  expectation for random quasar sightlines for a $\pm 1500\mkms$
  interval, estimated at \zfg\ and using $\ell(X)$ from
  \cite{ribaudo11}.    The green band shows the measured incidence in
  our control sample which is in good agreement but $\approx 25\%$
  above  random expectation. 
}
\label{fig:othick}
\end{figure}

\begin{figure*}
\includegraphics[height=7.5in,angle=90]{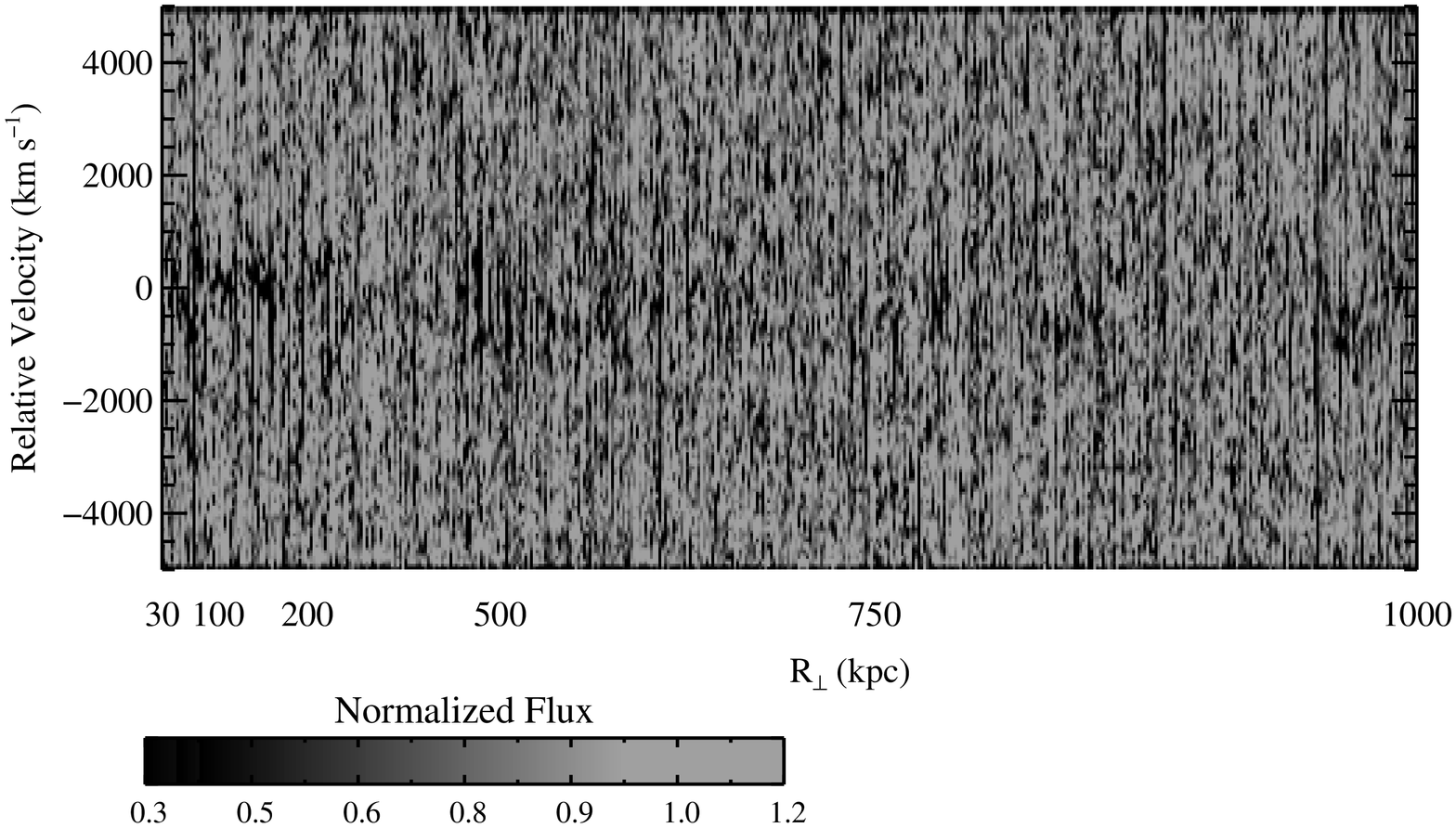}
\caption{Map of the normalized flux, spectrum-by-spectrum ordered by
  impact parameter \rphys, around \lya\ of the f/g quasars
  for the QPQ6 sample (restricted to pairs with $\msnlya > 8$ and
  $\mzfg < 3$).  Each spectrum has been linearly
  interpolated onto a fixed velocity grid with 100\kms\ pixels,
  preserving equivalent width.  One
  notes a likely enhancement in \ion{H}{1} absorption at $\delta v = 0 \mkms$ 
  for nearly all impact parameters.   The few cases which
  appear as nearly a single, black line correspond to DLAs.
}
\label{fig:spec_map}
\end{figure*}

Figure~\ref{fig:othick} presents the covering
fraction $f_C$ of optically thick gas in logarithmic bins of \rphys.
As discussed in our previous work (QPQ1, QPQ2, QPQ4, QPQ5),
$f_C$ exceeds 50\%\ for
$\mrphys < 200$\,kpc, a remarkable result which reveals a massive,
cool CGM surrounding $z \sim 2$ quasars.
With the QPQ6 sample, we extend the \fc\ measurements to 1\,Mpc
(Table~\ref{tab:qpq6_fc}).
We find that \fc\ declines with increasing \rphys, with a
marked decline at $\approx 200$\,kpc which we have interpreted as the
`edge' of the CGM (QPQ5).  
For a halo of mass $M \approx 10^{12.5} \msol$ at $z=2.5$, this modestly exceeds
the the virial radius ($\mrvir \approx 160$\,kpc).
Nevertheless, the covering fraction remains
significant ($\mfc = 19 \pm 2\%$) for $\mrphys > 500$\,kpc.

\begin{deluxetable}{lccccccccccccccccc}
\tablewidth{0pc}
\tablecaption{QPQ6 \fc\ Values\label{tab:qpq6_fc}}
\tabletypesize{\scriptsize}
\tablehead{\colhead{$\mrphys^{\rm min}$} & 
\colhead{$\mrphys^{\rm max}$} & 
\colhead{$m_{\rm pair}$} & 
\colhead{\fc} & 
\colhead{$+1\sigma^a$} & 
\colhead{$-1\sigma^a$} & 
\colhead{$\mfc^{\rm IGM}$} \\
(kpc) & (kpc) &}
\startdata
  30&  60&   6&0.67&0.16&0.20&0.03&\\
  60& 122&  22&0.64&0.09&0.11&0.03&\\
 122& 246&  36&0.39&0.08&0.08&0.03&\\
 246& 496&  60&0.18&0.05&0.04&0.04&\\
 496&1000& 304&0.19&0.02&0.02&0.04&\\
\enddata
\tablenotetext{a}{Confidence limits from Binomial statistics for a 68\%\ interval.}
\end{deluxetable}

The red points in Figure~\ref{fig:othick} show 
estimates for $\mfc^{\rm IGM}$ for the IGM in a random $\pm 1500 \mkms$
interval evaluated at $<\mzfg>$ using the 
\lox\ measurements of \cite{omeara13}.  
These may be compared against the \fc\ value measured from our control
sample (green band).  The two are in fair agreement although, as noted
above, we modestly overpredict the incidence in the control sample.
Nevertheless, even for $\mrphys = [496,1000]$\,kpc we find 
the \fc\ value for the QPQ6 pairs exceeds random by nearly a factor of 5. 
We conclude that the excess \ion{H}{1} absorption inferred from our
statistical measures (e.g.\ Figure~\ref{fig:hist_deltf_rho}),
is also manifest in the strong
\lya\ systems which are generally attributed to the ISM and CGM of
individual galaxies \citep{fumagalli11a,freeke12}.  
Furthermore, there is an enhancement at all scales $\mrphys \le
1$\,Mpc.  We further develop and explore the implications of these results 
in $\S$~\ref{sec:discuss}.


\section{Stacked Spectrum Analysis}
\label{sec:stacks}

The previous sections demonstrated that the environment surrounding
$z \sim 2$ quasars (from $\mrphys = 30$\,kpc to 1\,Mpc)   
exhibits excess \ion{H}{1} \lya\ absorption
relative to random spectral regions of normalized quasar spectra.
We have reached this conclusion through a statistical
comparison of
the distribution of \avgf\ and \deltf\ values and equivalent widths measured about each
f/g quasar compared to the distribution of a 
control sample for the IGM ($\S$~\ref{sec:avgf}, 
Figure~\ref{fig:hist_avgf}; $\S$~\ref{sec:ew}, Figure~\ref{fig:cumul_wline}).
We reached a similar conclusion from the incidence of optically thick
absorption and the observed distributions of \nhi\ values
($\S$~\ref{sec:NHI}, Figure~\ref{fig:nhi_vs_rho},\ref{fig:othick}).
We also presented evidence that the excess \ion{H}{1} absorption 
decreases with increasing impact parameter
(Figure~\ref{fig:hist_deltf_rho}) albeit with substantial scatter from
sightline to sightline.
This scatter 
in the absorption strength (e.g.\ \avgf)
is driven by continuum error, intrinsic scatter in quasar environments,
redshift error for the f/g quasar, and the stochastic nature
of the IGM.  

\begin{figure}
\includegraphics[width=3.5in]{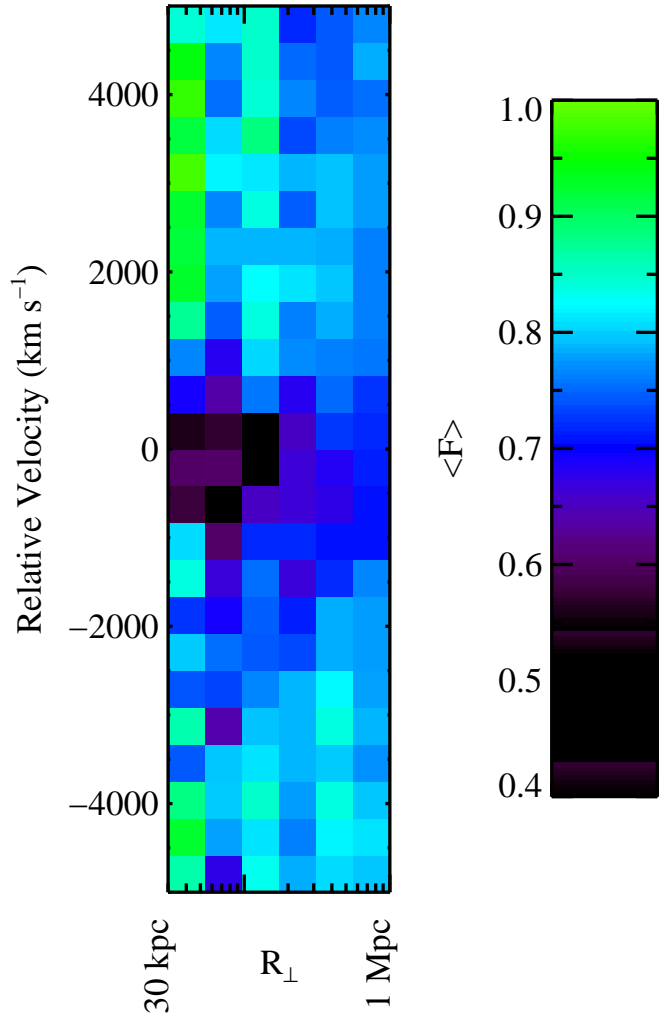}
\caption{A map of the normalized flux around \lya\ of the f/g quasars
  in QPQ6 as presented in Figure~\ref{fig:spec_map} but rebinned onto
  6 logarithmic \rphys\ intervals spanning from 30\,kpc to 1\,Mpc and
  in velocity channels of 400\kms.  One observes a clear enhancement
  in absorption at $\delta v \approx 0 \mkms$ which 
  decreases with increasing \rphys\ and velocity offset.  
  Note that the first column corresponds
  to only 6 quasar pairs and is dominated by sample variance.
}

\label{fig:lya_image}
\end{figure}

There is a complementary approach to assessing the excess (or
deficit) of \ion{H}{1} \lya\ absorption 
which averages over several of 
these sources of uncertainty: the creation of composite spectra.  
A composite spectrum is made by first shifting
the individual spectra to the rest-frame of the f/g quasar
at \lya\ ($\delta v = 0\mkms$ corresponds to \llya).
One then combines them with a statistical measure (e.g.\ average, median) 
in fixed velocity intervals, weighting the individual spectra as desired.
There are several benefits to this approach.  In particular,
one averages down
the stochasticity of the IGM to (ideally) recover a nearly uniform
absorption level in the absence of any other signals.  
Errors in continuum placement are also averaged
down and primarily affect the precision with which one measures
the IGM opacity.  Therefore,
one may then search for excess (deficit) absorption at $\delta v \approx 0\mkms$ 
relative to the IGM level.  
This provides a robust consistency check on results from the previous
sections.  
Quasar redshift error, Hubble flow, and peculiar
motions spread out the absorption, but the total equivalent
width can be preserved by using a straight average.
This technique also generates spectra as a function of
velocity relative to \zfg.  We expect the measured velocity spreads to
be dominated by quasar redshift uncertainty, but one can also
constrain other processes that generate motions of the gas.

This technique was successfully applied in QPQ5 to assess the average
\ion{H}{1} absorption strength in that sample, i.e.\ on scales $\mrphys \le
300$\,kpc.  We observed strong, excess absorption which we concluded
traces the CGM of galaxies hosting $z\sim 2$ galaxies.  In this
section, we extend the analysis to 1\,Mpc and perform an 
assessment of the technique and its uncertainties.
In all of the following, we restrict to the subset of the QPQ6 
sample with $\msnlya > 8$ and $\mzfg < 3$.
These criteria provide a more uniform set of high-quality
input spectra.

Before proceeding to generate composite spectra, 
it is illustrative to first examine maps of the normalized flux.
Figure~\ref{fig:spec_map} presents the 
$\pm 5000 \mkms$ interval surrounding each f/g quasar, 
ordered by impact parameter and restricted to the QPQ6
sample with $\msnlya > 8$ and $\mzfg < 3$.  Each spectrum has been linearly
interpolated (conserving equivalent width), onto a fixed velocity grid
centered at \llya\ with bins of 100\kms.  
For velocity bins of this size, we found it unnecessary to smooth
the data to a common spectral resolution.
A visual inspection reveals obvious
excess \ion{H}{1} \lya\ absorption at
$\delta v = 0 \mkms$ from $\mrphys = 30$\,kpc to $\mrphys =
500$\,kpc and likely beyond.  The absorption scatters about
$\delta v = 0 \mkms$ by many hundreds \kms\  
and an impression is given that the absorption 
declines with increasing \rphys.

Figure~\ref{fig:lya_image} presents a rebinned image,
generated by combining the sightlines in 6 logarithmic intervals in
\rphys, from 30\,kpc to 1\,Mpc and sampling in velocity space with
400\kms\ bins.  We have stretched this image to accentuate the excess
absorption and to illustrate the decreasing absorption strength with
increasing \rphys.
We caution that the first column reflects only 6~pairs and is
dominated by sample variance (the second column corresponds to 
14~pairs).  
Nevertheless, this image illustrates
the primary result of this manuscript: the quasar environment is
characterized by an excess of \ion{H}{1} absorption to 1\,Mpc with
a decreasing enhancement with \rphys\ and $|\delta v|$.

In principle, one could fit a global
model to the full dataset presented in Figure~\ref{fig:spec_map} to
estimate the average \ion{H}{1} absorption as a function of \rphys\
(or any other quantity).
This might maximize the statistical power of QPQ6 but would require a
comprehensive model of the IGM, a proper treatment of the diversity in
spectral resolution,
and complex models for the CGM
surrounding these massive galaxies.  We defer such model comparisons to our
study of the transverse proximity effect (TPE),
using these data and
additional pairs with larger separations.
Here, we instead explore the nature of excess \ion{H}{1} absorption through
the generation of composite spectra.

\begin{figure*}
\includegraphics[width=4.0in,angle=90]{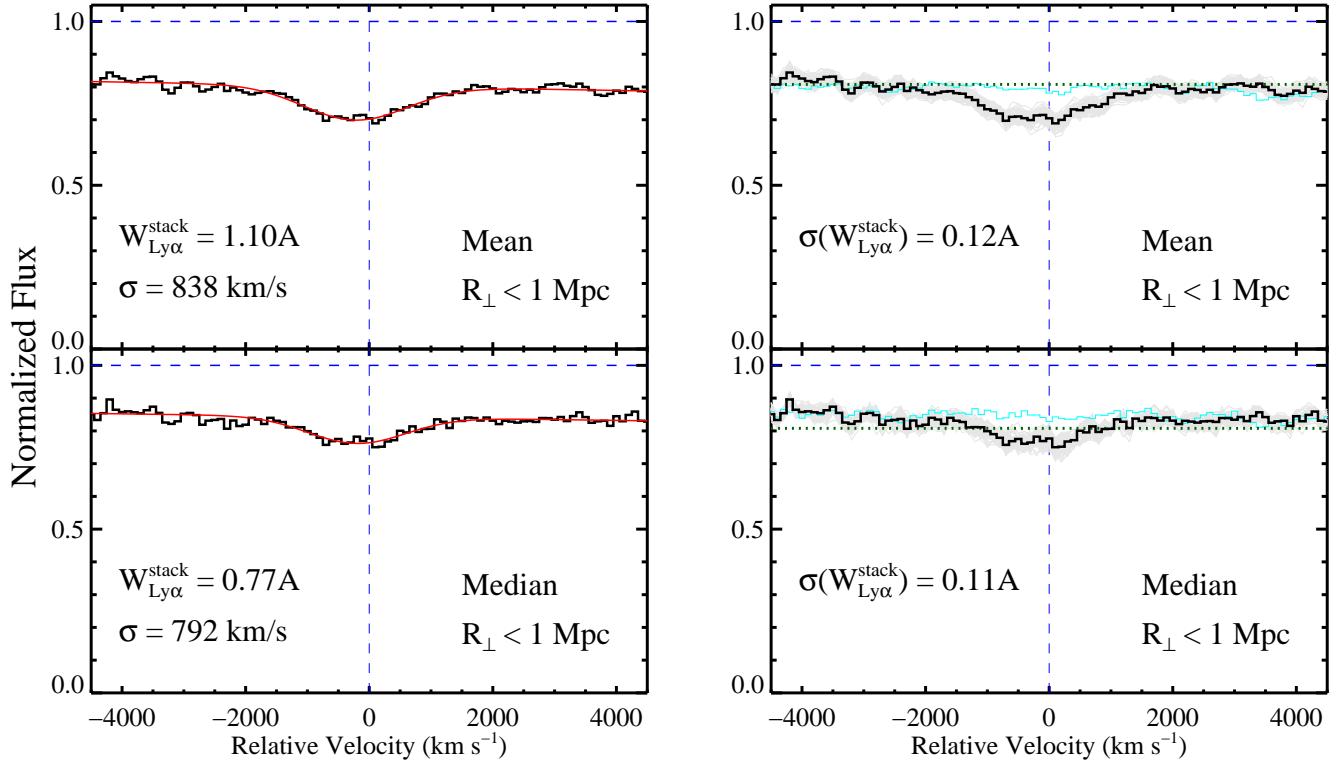}
\caption{Mean and median absorption at \lya\ of the f/g quasars for all 
  QPQ6 pairs with $\msnlya > 8$ and $\mzfg < 3$.  The median impact
  parameter is at $\mrphys = 702$\,kpc.
  The left-hand panels show the composites (black) for the mean
  (upper) and median (lower) statistics.  Overplotted on each these is
  a Gaussian fit (red), normalized to a pseudo-continuum given by the average IGM
  opacity away from $\delta v = 0\mkms$.  The right-hand panels show
  the same composites and also a series of bootstrap realizations
  (gray).  In addition,  a composite spectrum generated from a control sample
  of sightlines is presented (cyan).   
  It shows essentially constant IGM opacity at all
  $\delta v$ (i.e.\ no excess \ion{H}{1} absorption) which matches the
  opacity used to mean-flux regulate the continua
  (green dotted line; $\S$~\ref{sec:continuum}, Equation~\ref{eqn:FIGM}).
}
\label{fig:all_composite}
\end{figure*}

We generate composite spectra
by simply combining the individual spectra shown in
Figure~\ref{fig:spec_map}. 
The results one obtains are somewhat sensitive to the specific
methodology used to generate the composite spectra. 
We have experimented with various approaches and adopt the following.
First, a minimum of 20 individual spectra are required to 
sufficiently reduce the stochasticity of the IGM and thereby yield a robust
estimate of the average opacity at $|\delta v| \gg 0\mkms$.
Second, we found that imposing a \snlya\ threshold of 8 on the
individual spectra offers a good
compromise between maximizing sample size while reducing the
likelihood of severe continuum error and/or Poisson noise.
Regarding the statistic to generate the composite (e.g.\ average,
median) and how to weight the data (e.g.\ \snlya, f/g quasar luminosity), we
consider a few scientifically motivated approaches.

Our primary scientific interest is to measure the excess/decrement of
\ion{H}{1} \lya\ absorption associated with the CGM and extended
environment of the massive galaxies hosting $z \sim 2$ quasars.  
Each quasar pair gives an independent probe of this
region, i.e.\ all pairs have equal weight, at least at a given impact
parameter.   This suggests the composites should be created without any
weighting factor.  
On the other hand, the b/g quasar luminosity should be
independent of the f/g quasar's gaseous environment and one is tempted to
weight the spectra by the measured \snlya\ value.  This would increase,
however, the measured scatter in the IGM absorption because it
effectively reduces the number of sightlines included in the
composite. 
Scatter in the composite spectrum is
dominated by randomness in the IGM instead of other error sources
(i.e.\ continuum error and photon statistics). 
Further, the distribution of \snlya\ is pretty narrowly distributed
around 10 for the majority of the pairs, with a long tail to high
\snlya\ values (Figure~\ref{fig:demograph}).
Therefore, we have proceeded using equal weights\footnote{We have
  repeated the analyses that follow using stacks generated by
  weighting by (\snlya)$^2$.  We find qualitatively similar results,
  but we note that the S/N-weighted values are biased toward the lower
  \rphys\ pairs because many of these were observed with
  large-aperture telescopes.}
  
Regarding the statistics,
an average of the individual spectra yields the best 
estimate of the mean \ion{H}{1} \lya\ absorption and should preserve
the equivalent width. 
This statistic,
however, is more sensitive to outliers, e.g.\ the occasional DLA
system with $\mwlya > 10$\AA.
The median statistic, in contrast, may provide a better
estimate for the `typical' absorption.  
In the following, 
we generate composite spectra using each of these statistics. 

Figure~\ref{fig:all_composite} presents the mean and median composite
spectra, generated from the full set of data 
shown in Figure~\ref{fig:spec_map}.  
One observes significant
absorption at all velocities on the order of $20-25\%$ ($\tau \approx
0.25$).  This
absorption is driven by the IGM and the observed decrement
agrees with previous estimates at $z \gtrsim 2$, as designed 
($\S$~\ref{sec:continuum}).   
At $\delta v \approx 0
\mkms$, one identifies an excess of absorption which we 
associate to \ion{H}{1} gas in the local environments
of the f/g quasars.
The depression is many hundreds \kms\ wide, presumably owing to
quasar redshift error, 
peculiar motions, 'virial motions' and infall in the QSO environment, and Hubble flow. 
To perform a quantitative assessment, we have
fitted Gaussian profiles to 
each composite spectrum.  The fit is performed relative to  
a pseudo-continuum defined by the IGM
absorbed regions (i.e.\ at a level near 0.8 in the normalized flux,
not unity).  This continuum was measured through
a least-squares linear fit 
to each pixel in the composite spectrum with $\delta v < -3000 \mkms$
or $\delta v > 2000\mkms$. We estimate a less than $2\%$ normalization error in
this evaluation.   The Gaussian fits, meanwhile, were limited to
the pixels at $|\delta v| \le 1350 \mkms$ to minimize the effects of 
metal-line blending (see below).  
The width that we recover for the Gaussian ($\sigma=822\mkms$) 
is dominated by redshift error.  The reported equivalent widths are
relative to the pseudo-continuum and are approximately 1\AA\ in
strength. Because these are relative to the mean flux and not the
normalized flux of the quasars, they relate much more
closely to the \deltf\ statistic (Figure~\ref{fig:hist_deltf_rho})
and not the \wsubj\ values measured
for systems associated to each f/g quasar
(Figure~\ref{fig:cumul_wline}).

Regarding the mean vs.\ median composities, the latter has smaller
equivalent width ($\approx 25\%$), as we expected, 
although nearly the same velocity width.  Nevertheless, the strong signal in the median
indicates a majority of the sightlines exhibit excess \ion{H}{1}
absorption.  
Because the composite spectra applied equal weighting, the results in
Figure~\ref{fig:all_composite} are dominated by the hundreds of pairs
in QPQ6 with $\mrphys > 500$\,kpc.  
We may conclude that the quasar environment exhibits excess absorption to
at least 1\,Mpc separation.
In the right-hand panels of Figure~\ref{fig:all_composite}, we present
additional analysis on these composite spectra.  The
cyan curves show composite spectra generated from random \lya\ forest
spectra taken from the same parent sample and using the 
\zfg\ distribution of the QPQ6 sample.
Again, we require that  the random regions lie within the \lya\ forest
of the b/g quasar and away from its known f/g quasar pair.
The resultant composites show uniform absorption
associated to the IGM at all velocities with a magnitude matching 
the mean flux used in the MFR continua, given by Equation~\ref{eqn:FIGM}.  
This gives additional confidence that the observed excess in the QPQ6
composites is solely
related to absorption by gas in the environment of the quasar.  
The gray curves show the composite spectra for 100 bootstrap
realizations of the QPQ6 dataset allowing for duplications.  The RMS
scatter is small -- a few percent -- at all velocities.
Fitting a Gaussian to each of these, we recover a scatter in the
measured equivalent widths of 0.14\AA\ for the mean and
0.12\AA\ for the median composite.

\begin{figure}
\includegraphics[width=3.5in]{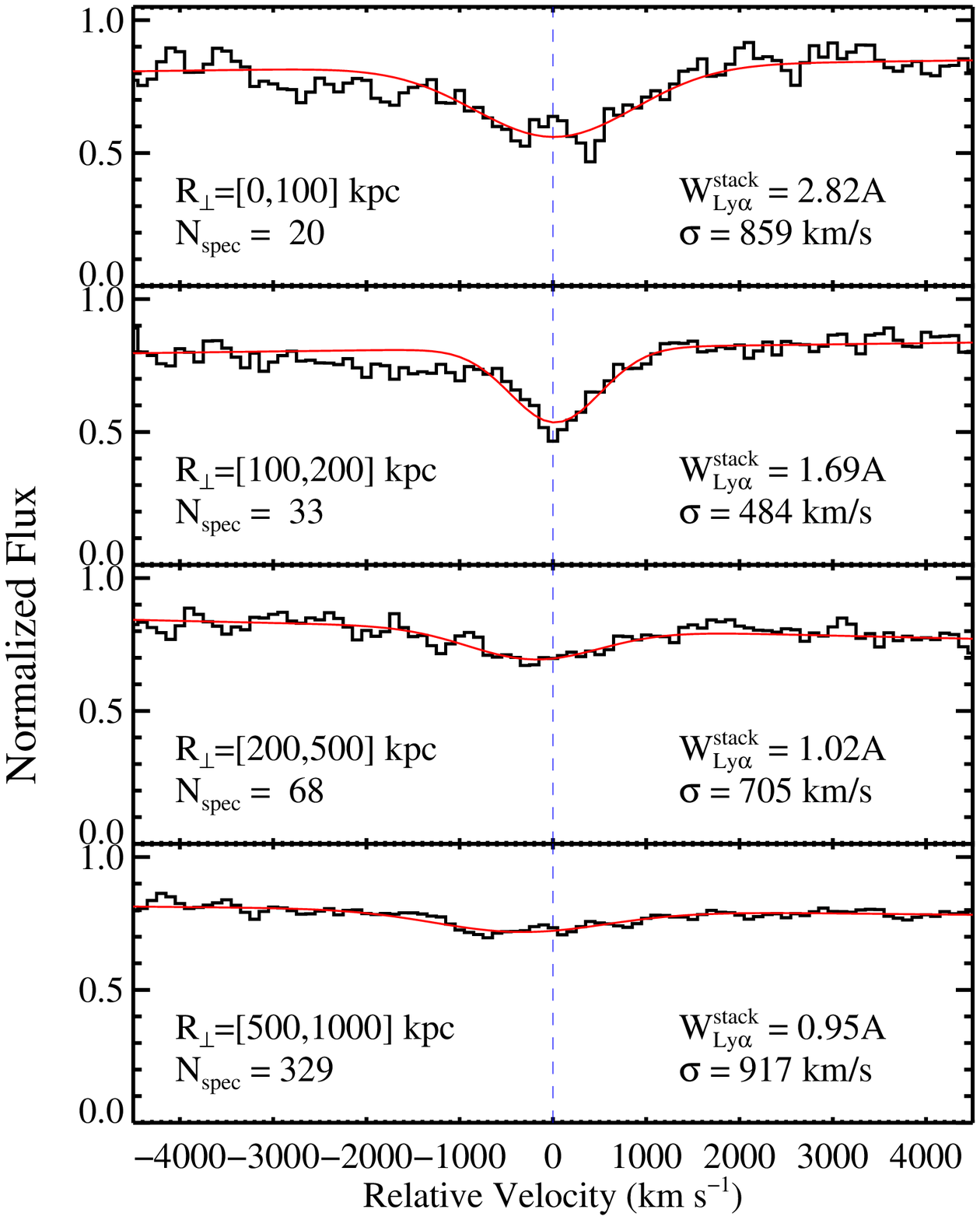}
\caption{Mean composites of the QPQ6 sample centered on \lya\ of the
  f/g quasars in increasing, logarithmic
  intervals of impact parameter \rphys.  The red curves are Gaussian
  fits to the composites adopting a continuum fitted to the average
  opacity (dominated by the IGM) at $|\delta v| \gg 0 \mkms$.  There is excess \ion{H}{1}
  absorption at all impact parameters which decreases with increasing
  \rphys.  
}
\label{fig:lya_stack_avg}
\end{figure}

\begin{deluxetable}{lccccccc}
\tablewidth{0pc}
\tablecaption{\ion{H}{1} \lya\ Equivalent Widths from QPQ6 Stacks\label{tab:qpq6_stacks}}
\tabletypesize{\scriptsize}
\tablehead{\colhead{$\mrphys^{\rm min}$} & 
\colhead{$\mrphys^{\rm max}$} & 
\colhead{Median \rphys} &
\colhead{$m_{\rm pair}$} & 
\colhead{${\mwstack}^a$} &
\colhead{$\sigma(\mwstack)^a$} &
\colhead{${\mwstack}^b$} \\
(kpc) & (kpc) & (kpc) && (\AA) & (\AA) & (\AA)}
\startdata
  31&  98&  89&  20&2.82&0.56&2.19&\\
 100& 199& 139&  33&1.69&0.50&1.05&\\
 201& 495& 387&  68&1.02&0.32&0.53&\\
 503& 999& 786& 329&0.95&0.15&0.72&\\
\enddata
\tablenotetext{a}{Equivalent width measured from the mean stacks, relative to the IGM-absorbed continuum.  The uncertainty is estimated from a bootstrap analysis.  }
\tablenotetext{b}{Equivalent width measured from the median stacks, relative to the IGM-absorbed continuum.}
\tablecomments{The mean equivalent width values are well fitted by a power-law, $\mwstack(\mrphys) = 2.3$\AA ($\mrphys/100\;\rm kpc)^{-0.46}$.}
\end{deluxetable}

The maps shown in
Figure~\ref{fig:spec_map} and \ref{fig:lya_image} 
indicate a significant trend in
absorption strength with \rphys.  
We may study this trend by producing
composite spectra in a series of \rphys\ bins.  
Figure~\ref{fig:lya_stack_avg}
presents the results for four radial bins, in a series of roughly
logarithmic intervals.  
Similar to the full composites, we detect enhanced absorption at
$\delta v \approx 0 \mkms$ in each of the composites.
Additionally, there is an obvious
trend of decreasing absorption with increasing \rphys.
The resultant models are overplotted on the composite
spectra (Figure~\ref{fig:lya_stack_avg}) and the fit parameters are
listed in Table~\ref{tab:qpq6_stacks}.  These single Gaussian models
provide a good description of the data, as follows 
from the central limit theorem.  
The integrated equivalent widths of these Gaussians clearly
decrease with increasing \rphys, decreasing from $\mwstack = 2.8$\AA\
for $\mrphys < 100$\,kpc to $\mwstack = 1.0$\AA\ for $\mrphys >
500$\,kpc.  We note that the values are approximately one half of the
average individual \wsubj\ values that we measured in
$\S$~\ref{sec:indiv} (Figure~\ref{fig:cumul_wline}). 
We have estimated the error in the equivalent width
measurements by bootstrapping each composite 100 times and measuring the
RMS in the resultant \wstack\ values (Table~\ref{tab:qpq6_stacks}).  
We repeated the analysis on a series of median composite spectra.
Similar to the full composites (Figure~\ref{fig:all_composite}), we
detect enhanced absorption at $\delta v \approx 0 \mkms$ in each of
these median composites (Table~\ref{tab:qpq6_stacks}).  

\begin{figure}
\includegraphics[height=3.5in,angle=90]{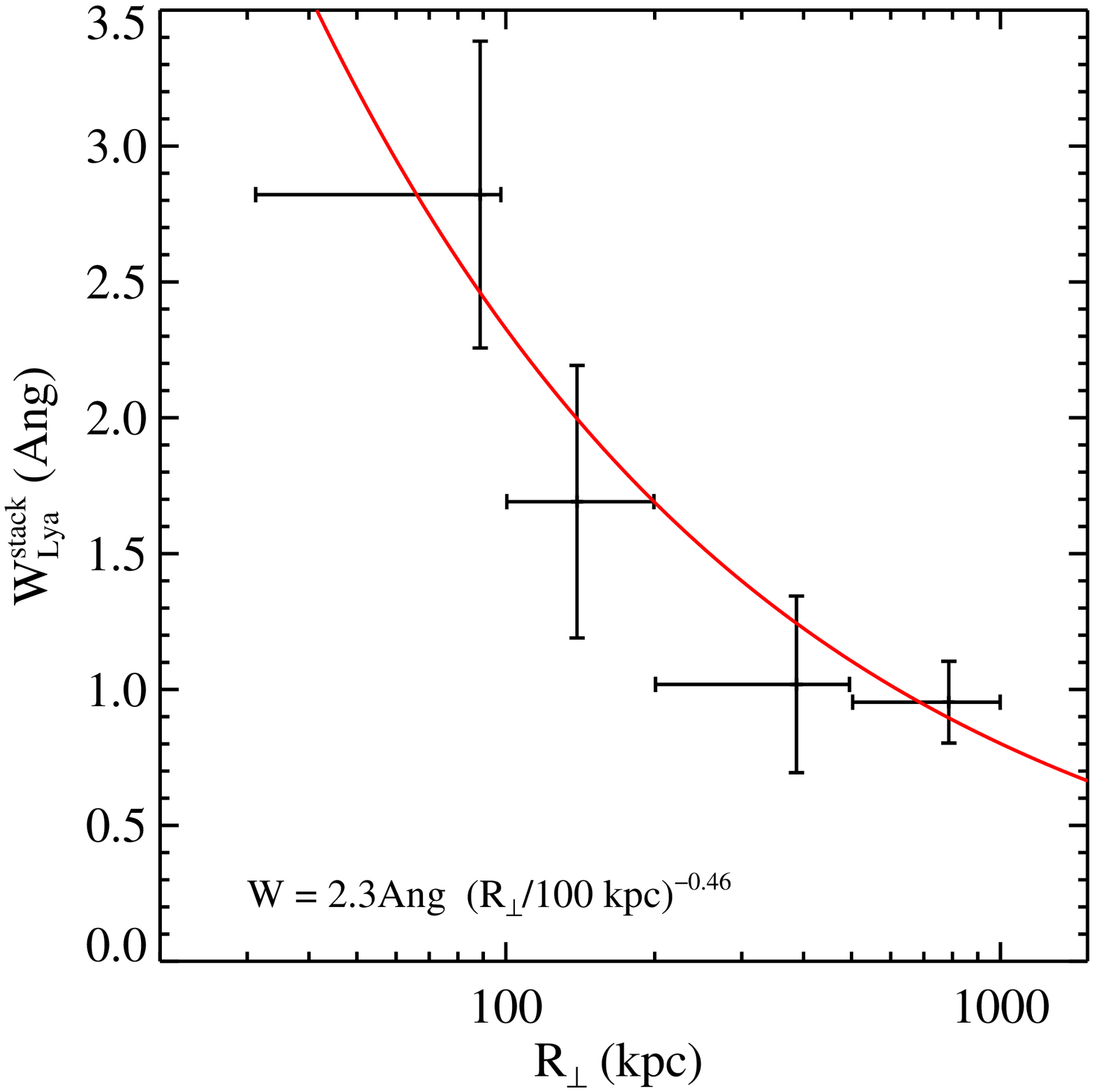}
\caption{Equivalent width measurements \wstack\ for the mean
  composites in Figure~\ref{fig:lya_stack_avg}.  There is a strong,
  non-linear decline in \wstack\ with increasing \rphys. Each bin is
  plotted at the median \rphys\ value of the pairs that contributed to
  the stack (see Table~\ref{tab:qpq6_stacks}).  We have fitted a model of the form 
  $\mwstack = W_0 (\mrphys/{\rm 100\, kpc})^\gamma$ to the \wstack\
  measurements and their errors (estimated from bootstrap
  realizations).  We find $W_0=2.2$\AA\ and $\gamma = -0.45$ and note
  there is significant degeneracy between the two parameters.
}
\label{fig:ewstack_vs_rho}
\end{figure}

Figure~\ref{fig:ewstack_vs_rho} presents the \wstack\ values for the
mean composites versus
\rphys\ for the four intervals, with uncertainties estimated from the
bootstrap analysis.  
There is an obvious, non-linear trend of
decreasing \wstack\ with increasing \rphys. 
We may describe the observed trend with a simple power-law
model: $\mwstack = W_0 (\mrphys/{\rm 100\, kpc})^\beta$. 
By fitting this two-parameter model to the
binned results and assuming Gaussian errors,  we find that $\chi^2$ is
minimized for $W_0 = 2.3$\AA\ and $\beta = -0.46$, with significant
degeneracy between the two parameters.    We emphasize that this
model, despite its relatively good description of the data, 
is not physically motivated.   In fact, we may expect that the observed
trend is the result of two competing and (presumably) unrelated physical
phenomena \citep{fumagalli13b}:  
the CGM of the host galaxy on scales of $\sim 100$\,kpc
and clustering of galaxies and other large-scale structures on Mpc scales.
It is possible that these combine to give this simple model
which describes the observations well. 

\begin{figure}
\includegraphics[width=3.5in]{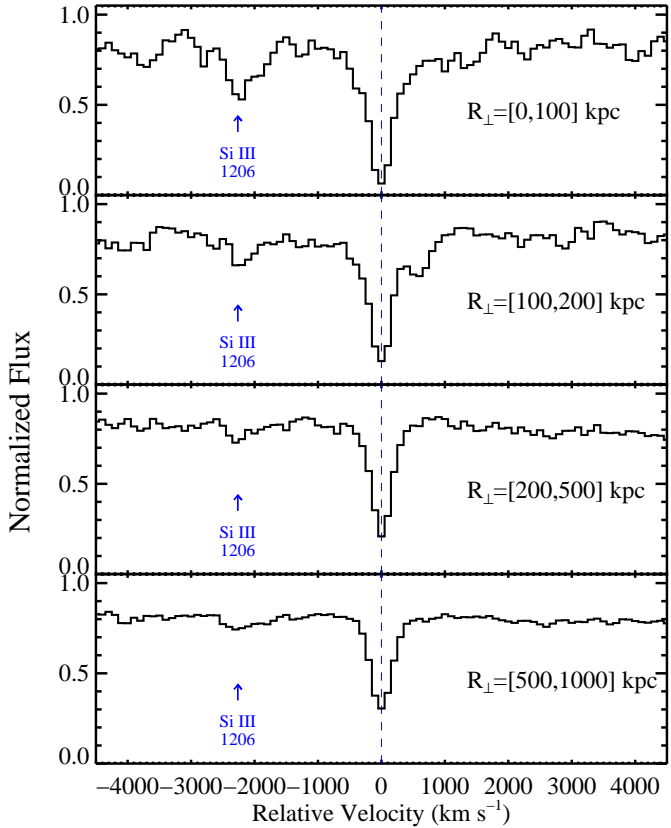}
\caption{Mean composites in a series of \rphys\ intervals using
  individual spectra shifted to the strongest absorption line in the
  $\pm 1500 \mkms$ interval around \zfg\ (see $\S$~\ref{sec:indiv}).
  At $\delta v \approx -2250\mkms$, one identifies significant
  absorption that we associate with the \ion{Si}{3}~1206 transition.
  Its presence affected how we fit the IGM opacity of our
  composites and the Gaussian fits to the excess \ion{H}{1}
  absorption.  
  It may also contribute slightly to the estimated \wstack\ values 
  for the lowest \rphys\ intervals.
  Future work will explore the metal-line absorption
  associated to the quasar environments (QPQ7; see also QPQ3, QPQ5).
}
\label{fig:zlya_stack}
\end{figure}

There is one, somewhat subtle, feature in the composites for the
lowest \rphys\ intervals: 
excess absorption at $\delta v \approx -2000\mkms$ which gives the
impression of greater IGM opacity bluewards of \zfg.
This excess opacity is dominated by \ion{Si}{3}~1206 absorption at $z
\approx \mzfg$ which occurs at $\delta v \approx -2250\mkms$ bluewards
of \lya.  Figure~\ref{fig:zlya_stack}a shows a series of composite
spectra stacked at the flux-weighted centroid of the individual \lya\
lines\footnote{This analysis was
  limited to spectra with $\msnlya > 9.5$, therefore these composites
  do not include all of the data shown in
  Figure~\ref{fig:ewstack_vs_rho}.}
described in $\S$~\ref{sec:indiv}.
We recover similar equivalent widths but the 
\lya\ profiles in these composities are, by design, much more
narrow than the lines in the composites stacked at $z=\mzfg$.
We also recognize significant excess absorption at $\delta v \approx
-2200\mkms$ which we associate to \ion{Si}{3}~1206 absorption.  No
other statistically significant absorption is apparent in this
velocity interval and none is expected.  The presence of strong
\ion{Si}{3}~1206 absorption, especially for the $\mrphys < 200$\,kpc
pairs, motivated our decision to fit the
pseudo-continuum of the composite spectra at $\delta v < -3000 \mkms$.
We perform additional analysis on these composites with emphasis on
the metal-line absorption in QPQ7. 
Future work will also re-examine the composite spectra for these
quasar pairs when near-IR spectroscopy enables a more precise
estimation of \zfg.

\begin{figure}
\includegraphics[width=3.5in]{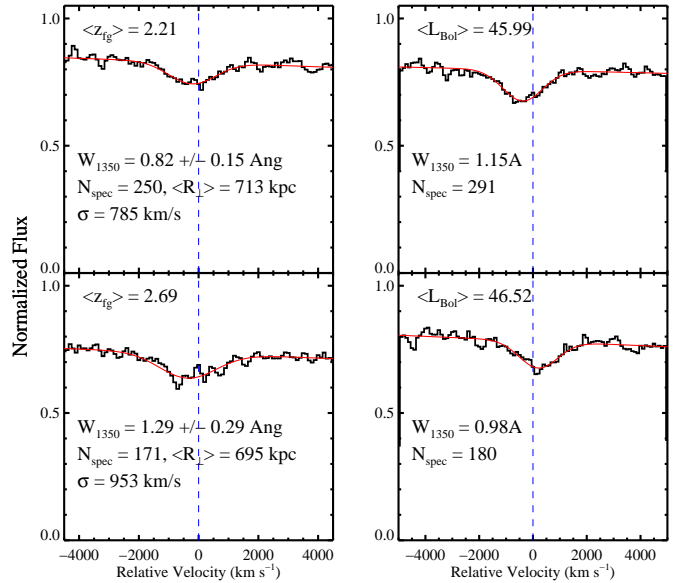}
\caption{
  ({\it left:}) Mean composites cut on f/g quasar redshift (at $\mzfg = 2.4$ and restricted to
  $\mzfg < 3.5$).   The two subsamples have very similar \rphys\
  distributions with $\langle \mrphys \rangle \approx 700$\,kpc.  One observes, as
  expected, that the IGM opacity (evaluated at $\delta v \gg 0 \mkms$)
  is higher for the higher \zfg\ subset.  We also measure a greater
  excess of \ion{H}{1} absorption at $\delta v \approx 0 \mkms$, but
  this is only significant at the $\approx 2 \sigma$ level.  It is
  consistent, however, with the expectation that higher redshift
  quasars are hosted by more massive halos with correspondingly larger
  \ion{H}{1} absorption on these scales.
  ({\it right:)} 
  Mean composites of the QPQ6 sample ($\mzfg < 3$) cut by Bolometric
  luminosity at $\log L_{\rm Bol} = 46.3$.  Each subset has a similar
  mean impact parameter and redshift distribution.  Although the two
  composite spectra have similar equivalent widths, there is an
  obvious systematic shift in the line centroid of over $500\mkms$,
  which we propose is related to a luminosity-dependent shift in the
  \ion{C}{4} emission-line of the f/g quasars (i.e.\ redshift error).
}
\label{fig:zlum_stack}
\end{figure}

We may explore the dependence of \ion{H}{1} absorption on other
aspects of the quasars and/or their environment.  In
Figure~\ref{fig:zlum_stack}a we present composite spectra for two
sub-samples: (1) pairs with $\mzfg = [1.6, 2.4]$ and an average
$<\mzfg> = 2.22$; (2) pairs with $\mzfg = [2.4, 3.5]$ and $<\mzfg> =
2.69$.  Each subsample has an average impact parameter $<\mrphys> \,
\approx 700$\,kpc.  Both show significant IGM opacity, with a 
larger value for the higher \zfg\ sample,
albeit with modest statistical significance ($\approx 2\sigma$).
The results are suggestive that higher \zfg\ quasars exhibit a greater
excess of \ion{H}{1} absorption, consistent with the picture that
these quasars are hosted by more massive halos\footnote{ 
  We also note that the higher \zfg\ quasars have a 0.13\,dex higher
  bolometric luminosity although quasar clustering does not depend
  strongly on quasar luminosity.}
\citep{shen07}.  

We have also generated composite spectra after splitting the QPQ6
stack-sample by f/g quasar luminosity: $L_{\rm Bol}$
(Figure~\ref{fig:zlum_stack}b).  There is no statistically significant
difference in \wstack\ for the $L_{\rm Bol}$ subsets, but we do
measure a systematic offset in the line-center.  We associate this
offset with a luminosity-dependent shift in the \ion{C}{4}
emission-line \citep{hw11}, i.e.\ error in the f/g redshifts we have
measured related to a Baldwin effect.
We have generated additional composite spectra cutting on further
properties of the sample and find no statistically significant
differences.    This follows the results of $\S$~\ref{sec:wlya}
where we noted no strong correlation between the mean fluxes and any
other property (Figure~\ref{fig:deltf_scatter}).

\section{Analysis and Discussion}
\label{sec:discuss}


In this section we perform additional analysis on measurements
from the
previous sections and discuss implications for the properties of
quasars and the nature of gas in their extended environments.

\subsection{Anisotropy in the Quasar Environment}

In previous papers in this series, we have concluded that gas
observed in absorption in b/g sightlines of quasar pairs
is typically not illuminated by the ionizing
flux of the f/g quasars. 
The evidence includes our observation that strong \ion{H}{1}
absorbers are anisotropically distributed about quasars such that 
the line-of-sight
shows a much lower incidence of strong absorbers than background sightlines (QPQ2), 
detailed analysis of the observed ionic ratios of heavy
elements in a single well-studied quasar absorber do not
reflect an intense radiation field (QPQ3), fluorescent \lya\ radiation
from optically thick gas in the quasar CGM is generally
absent (QPQ4), and that there is a high covering fraction of optically
thick, cool gas with metals predominantly in low-ionization
states (QPQ5).  These results are impossible to reconcile with an
isotropically emitting source unless quasars vary on very
short timescales $\sim 10^{5}$\,yr.\footnote{This timescale is the
transverse light crossing time.}
Furthermore, this anisotropic emission follows naturally from 
the obscuration frequently invoked in unification models of AGN
which center the accreting black hole within an obscuring torus of
gas and dust \citep[e.g.][]{Anton93,elvis00,gaskell09}.

The results presented in this manuscript offer further evidence for
anisotropic emission.  First, we have demonstrated that the environment
surrounding $z \sim 2$ quasars exhibits enhanced \ion{H}{1} absorption
relative to the average IGM (Figure~\ref{fig:all_composite}).  Because
the quasar's ionizing flux exceeds the UV background to 1\,Mpc and
beyond (Figure~\ref{fig:demograph}), 
one expects a smaller neutral fraction for illuminated gas, and a corresponding
decrease \ion{H}{1} absorption, absent other effects. 
Clearly, any reduced absorption caused by ionization effects has been compensated 
for
by the increased absorption caused by the overdensity of gas on scales $\mrphys < 1$\,Mpc.
Second, the enhanced \ion{H}{1} absorption actually 
{\it increases} with decreasing impact
parameter (Figure~\ref{fig:hist_deltf_rho},\ref{fig:ewstack_vs_rho}), 
opposite to one's
expectation if ionization effects were playing a dominant
role, where the $R^{-2}$ increase in flux
would imply a stronger reduction in absorption at smaller impact
parameters \citep[see also][]{font13}. 
Again,
the observed signal implies an increasing density and covering fraction
of \ion{H}{1} gas towards the quasar.
Third, we have established that the high covering fraction of optically thick
gas exceeds random expectation even at $\mrphys \approx 1$\,Mpc
(Figure~\ref{fig:othick}). 

All of these results run contrary to the characterization of
\ion{H}{1} \lya\ absorption near quasars but along our line-of-sight,
where one observes comparable or suppressed absorption relative to the 
ambient IGM \citep[e.g.][]{sbd+00,KT08,dww08} known 
as the line-of-sight proximity effect \citep{bdo88,sbd+00}. 
We conclude that the environments surrounding $z \sim 2$
are anisotropically illuminated. 

This conclusion may be further quantified through a comparison of the
clustering signal measured transverse to the quasars with that along
the line-of-sight, as estimated from the incidence of proximate
absorption systems.  Such analysis was first presented in QPQ2.  
We do not repeat those estimates here, but 
instead, we consider a different but analogous argument based on covering
fractions.   We have measured that the covering fraction of
optically thick gas at $\mrphys < 200$\,kpc exceeds 50\%\ 
and that \fc\ increases with decreasing \rphys\
(Figure~\ref{fig:othick}).  
Because radiation directed towards us must travel along these scales
(and at even smaller separations),
we may reasonably assume that \fc\ exceeds 0.5 along 
the direct line-of-sight i.e.\ 
the majority of quasars should exhibit strong Lyman
limit absorption near their emission redshifts. 

\cite{pow10} examined the
incidence of so-called proximate LLS (PLLS) in their survey of Lyman
limit absorption, using $z>3.5$ quasars drawn from the SDSS.
They found a modest {\it deficit} in the incidence of PLLS
(measured within 3000\kms\ of \zem)
relative to intervening LLS, discovering only 35 PLLS in over
450 quasars surveyed.  Similarly, we note that there are
very few PLLS in the
\cite{omeara11} dataset of $z \approx 2.5$ quasars. 
We rule out at very high confidence that $z \sim 2-3$
quasars exhibit a
covering fraction of $\approx 50\%$ along their sightlines.

Of course, the quasar flux is sufficiently intense to photoionize
optically thick gas to distrances of hundreds kpc
corresponding to $\approx 100\mkms$
\citep[QPQ2,][]{cmb+08}.   
This should, however, have a negligible effect on the incidence within
a 3000\kms\ window, corresponding to $\approx 12$\,Mpc (proper) at 
$z=2.5$. 
Indeed, the clustering results described in
$\S$~\ref{sec:clustering} 
imply the radiation field must affect LLS on Mpc scales.  
Adopting $r_0^{\rm LLS} = 15 \mcMpc$, 
we calculate that the incidence of LLS from $1-10$\,Mpc
is boosted by a factor of $\approx 2$.  
In contrast, \cite{pow10}
measured an approximately $25\%$ reduction in $\ell(X)$ for gas 
with $\delta v \le 3000\mkms$ relative to the quasar.  
Therefore, the
quasar must be ionizing gas to well beyond 1\,Mpc 
to match the observed incidence of PLLS.  At a proper distance of
1\,Mpc, the radiation field of a $g=20$\,mag quasar exceeds the EUVB
by a factor of $g_{\rm UV} \approx 25$.  Provided the LLS gas is sufficiently
diffuse ($n_{\rm H} < 10^{-1} \cm{-3}$; QPQ2), 
this could conceivably be achieved.  The primary conclusion is that
quasars impose a radiative feedback to gas on very large scales 
that may otherwise fuel
star formation.  It is important to study the consequences of such a  
feedback process on galaxy formation 
and properties of gas in the IGM
\citep[e.g.][]{gh12,os13b}. 

\subsection{Comments on the Proximity Effects}
\label{sec:proximity}

Throughout this manuscript we have made reference to the proximity
effect of quasars: the expectation that ionizing radiation emitted
by these bright sources over-ionizes their surrounding media, thereby
reducing the \ion{H}{1} \lya\ opacity.  There are several scientific
motivations for exploring the proximity effect, both along the
line-of-sight (the line-of-sight proximity effect; LPE) and in the transverse
direction (TPE).  These include assessing the
intensity of the UV background \citep{bdo88} and constraining the
astrophysics of quasar emission \citep{croft04}.  The QPQ dataset
affords a new opportunity to perform LPE and TPE measurements but
we defer such analysis to later papers in the series and here offer
a few comments that may guide future work.

First,  we have not detected the signature of a `classical' TPE, i.e.\
lower \ion{H}{1} \lya\ opacity analogous to the LPE, but instead find 
enhanced absorption presumably driven by the overdensity of gas on
1\,Mpc scales.  Similar results were
reported previously by \cite{croft04} and \cite{font13} on much larger
scales (where the quasar flux is low) and by \cite{KT08} from a
modest sample of pairs with $\mrphys \approx 1$\,Mpc.  
If the quasar radiation field is illuminating gas transverse to the
sightline, one must search for an alternate manifestation of the TPE.
In principle, one could predict {\it ab initio} the overdensity of gas
surrounding quasars on $\sim 1$\,Mpc scales and compare against our
observations.  This would require, however modeling the non-linear 
structure of the IGM around massive quasar halos, as well as a characterization
of the host halo mass distribution for quasars. If small scales characteristic
of the CGM $\mrphys < 300\,$kpc are included in the analysis, this model must
also reproduce the properties of the frequently optically thick quasar CGM; 
hydrodynamic simulations
and radiative transfer would also be required.  
Alternatively, one may empirically compare
the LPE and TPE in the same quasars to test whether an (assumed)
isotropic overdensity implies radiation emitted in the transverse
dimension.  

Another approach is to search for trends in the \ion{H}{1}
absorption strength along the transverse dimension 
which might reflect
episodic/anisotropic emission \citep{croft04}.
Indeed, \cite{KT08} reported on a probable excess of \ion{H}{1}
absorption transverse to quasars but with an offset in velocity from \zfg.
If the measured velocity offset is interpreted as Hubble flow, then the excess gas lies
behind the quasar and this might be interpreted as a transverse
proximity effect that results from 
short quasar lifetimes ($t_{\rm QSO} < 1$\,Myr).
Our observations have not confirmed the \cite{KT08} result; we do find
excess \ion{H}{1} \lya\ absorption but it is roughly centered on \zfg\
and it spans several hundred \kms\ to both positive and
negative velocities.
We did identify an offset in \ion{H}{1} absorption that is dependent on
quasar luminosity (Figure~\ref{fig:zlum_stack}), but we interpret this
result as a luminosity-dependent shift in quasar emission
lines\footnote{We also note that \cite{KT08} measured quasar 
redshifts from their own dataset, primarily from \ion{C}{4} emission
lines to which they applied a $+753\mkms$ offset.  Our analysis assumes a shift of
+851\kms\ when only \ion{C}{4} is measured.}
\citep[e.g.][]{hw11}. 
We suspect that this also explains the \cite{KT08} result although we
cannot strictly rule out a TPE effect in that data. 
A proper assessment awaits more
precise measurements of f/g quasar redshifts (Hennawi et al.\, in
prep.).

Regarding the LPE, our results indicate -- beyond a shadow of a doubt
--
that the environments of luminous quasars at $z>2$ are overdense with
respect to the ambient IGM \citep[see also][]{font13}.  
Therefore, any attempt to infer the
intensity of the EUVB with the LPE must properly model the underlying
density field \citep[e.g.][]{fg08_prox}.  Indeed, all previous estimates
of the EUVB intensity have likely overestimated the
true value.  This may well reconcile the perceived offset in
EUVB measurements between those derived from the LPE technique and
estimates from analysis of the flux decrement in the IGM \citep[e.g.][]{flh+08}.
In turn, one concludes that star-forming galaxies need not contribute
as significantly to the extragalactic UV background as demanded by LPE
measurements of the EUVB, especially at redshifts $z<4$.

\subsection{Excess \ion{H}{1} Absorption in the 1\,Mpc
  Environments of Massive Halos}
\label{sec:mass}

We have concluded that the gas transverse to quasar sightlines
is generally not illuminated by the observed f/g quasar. 
Instead, we observe an excess of \ion{H}{1}
absorption relative to the ambient IGM
which we interpret as the overdensity of gas in the
environment of quasars.
This interpretation follows from the current paradigm of the IGM, i.e.\ the
fluctuating Gunn-Peterson approximation.  In this model, which is well
substantiated by numerical simulation \citep[e.g.][]{mco+96}, the
\ion{H}{1} \lya\ opacity traces the local over-density
\citep[e.g.][]{GnedHui98}.  
Given that quasars mark
extreme over-densities, one expects that their environments have
densities much greater than the mean density, on scales of 1\,Mpc and
beyond.  Indeed, \cite{kim08} proposed and proceeded to estimate the
mass of the dark matter halos hosting $z\sim 2$
quasars through analysis of the observed excess of \ion{H}{1}
absorption on scales of several to tens $h^{-1}$~Mpc.

The masses of dark matter halos hosting quasars have also been
constrained from clustering measurements. 
The auto-correlation analysis yields
characteristic masses $M_{\rm halo}^q \approx 10^{12.5} \msol$ at $z \sim 2$
\citep[][see also \citet{croom05,pmn04}]{white12}. 
Similarly, \cite{ts12} have assessed
the cross-correlation between quasars and star-forming galaxies to
estimate $M_{\rm halo}^q = 10^{12.3 \pm 0.5} \msol$.
\cite{font13} have studied the cross-correlation of
quasars with the \lya\ forest and measure a similar quasar bias
factor.

\begin{figure}
\includegraphics[height=3.5in,angle=90]{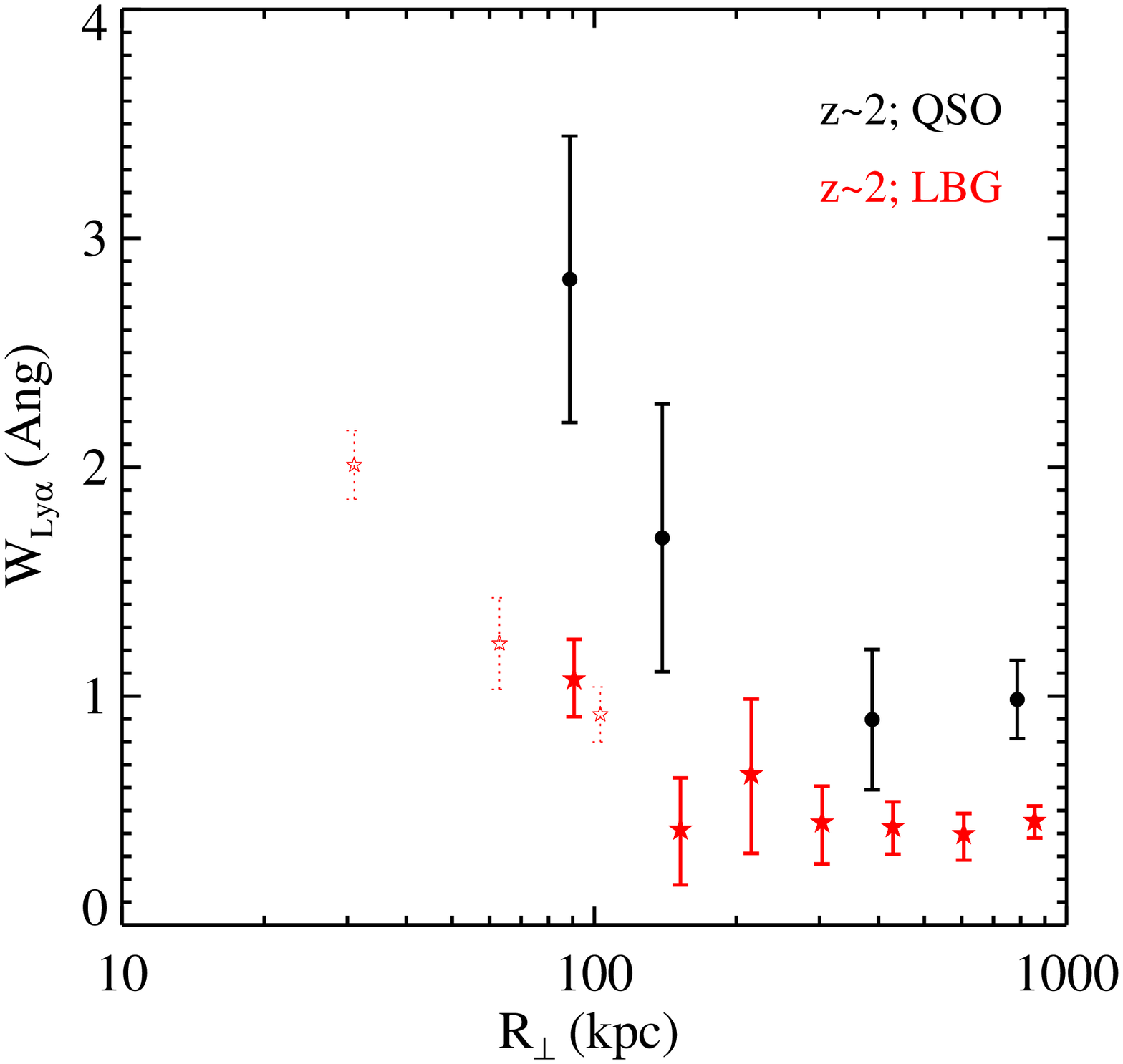}
\caption{
Comparison of the excess equivalent width of \ion{H}{1} \lya\ absorption
relative to the ambient IGM for the environments surrounding $z \sim
2$ quasars (black; \wstack\ values from this paper) and 
the $z \sim 2$ Lyman break galaxies
\citep[red;][]{steidel+10,rakic12}.  At all proper impact parameters \rphys,
the \ion{H}{1} absorption associated to the galaxies hosting quasars
is stronger.  This is consistent with the result that quasars reside
in more massive halos than those hosting LBGs at $z \sim 2$.
}
\label{fig:qpq_vs_lbg}
\end{figure}

These quasar host halo masses  
exceed the halo masses estimated for the $z
\sim 2-3$ star-forming galaxy population termed Lyman break galaxies
\citep[LBGs;][]{sgp+96}.  Clustering analyses and halo occupation
models for the LBGs yield mass estimates and limits ranging from
 $M_{\rm halo}^{\rm LBG} \approx 10^{11.4} - 10^{12} \msol$
\citep[e.g.][]{adel05,bielby+13}.   Within $\Lambda$CDM,
therefore, one predicts that 
$z \sim 2-3$ quasars will exhibit greater \ion{H}{1} \lya\
absorption than LBGs at the same proper separation.
Figure~\ref{fig:qpq_vs_lbg} compares the equivalent width of
\ion{H}{1} \lya\ from our stacked spectra against comparable
measurements derived from LBGs \citep{steidel+10,rakic12}.  At all
proper impact parameters \rphys, the environments surrounding $z\sim
2$ quasars show greater \ion{H}{1} \lya\ 
absorption than that of the LBGs (see also QPQ5).  
The immediate conclusion is that quasar environments have greater
overdensities on these scales.  As emphasized in QPQ5, this includes
$\mrphys < 100$\,kpc where one likely intersects the virial radius and
current numerical simulations tend to predict less
cool gas in massive halos \citep[e.g.][]{kkd+09,stewart11a,freeke12,fumagalli13b}.

To further explore these results in the context of $\Lambda$CDM, 
we have compared our observations
against outputs from the cosmological simulation of a massive dark
matter halo at $z>2$.
Specifically,
we have produced simulated mock spectra from a high-resolution simulation 
of the formation of a massive halo at $z\sim2.4$ obtained with
the \texttt{RAMSES} code \citep{teyssier02} as described in detail 
in appendix B of \cite{cantalupo12}. The simulation has a box
size of 40 comoving Mpc (corresponding to about 2800\kms\ at 
at $z\sim2.5$) and a maximum spatial (mass) resolution of
about 180 proper pc ($1.8\times 10^6 \msol$) at $z\sim2.4$ using
adaptive mesh refinement. The
box has been centered on the most massive halo at $z=2.4$,
$M_{\mathrm{halo}}\sim6\times10^{12} \msol$, as a representative
host of a luminous quasar. We include in the hydro-simulation the
photo-ionization from the cosmic UV background (Haardt \& Madau 2012),
star formation, metal enrichment, metal cooling and supernova feedback
\citep{cantalupo12}. 

\begin{figure}
\includegraphics[height=3.5in,angle=90]{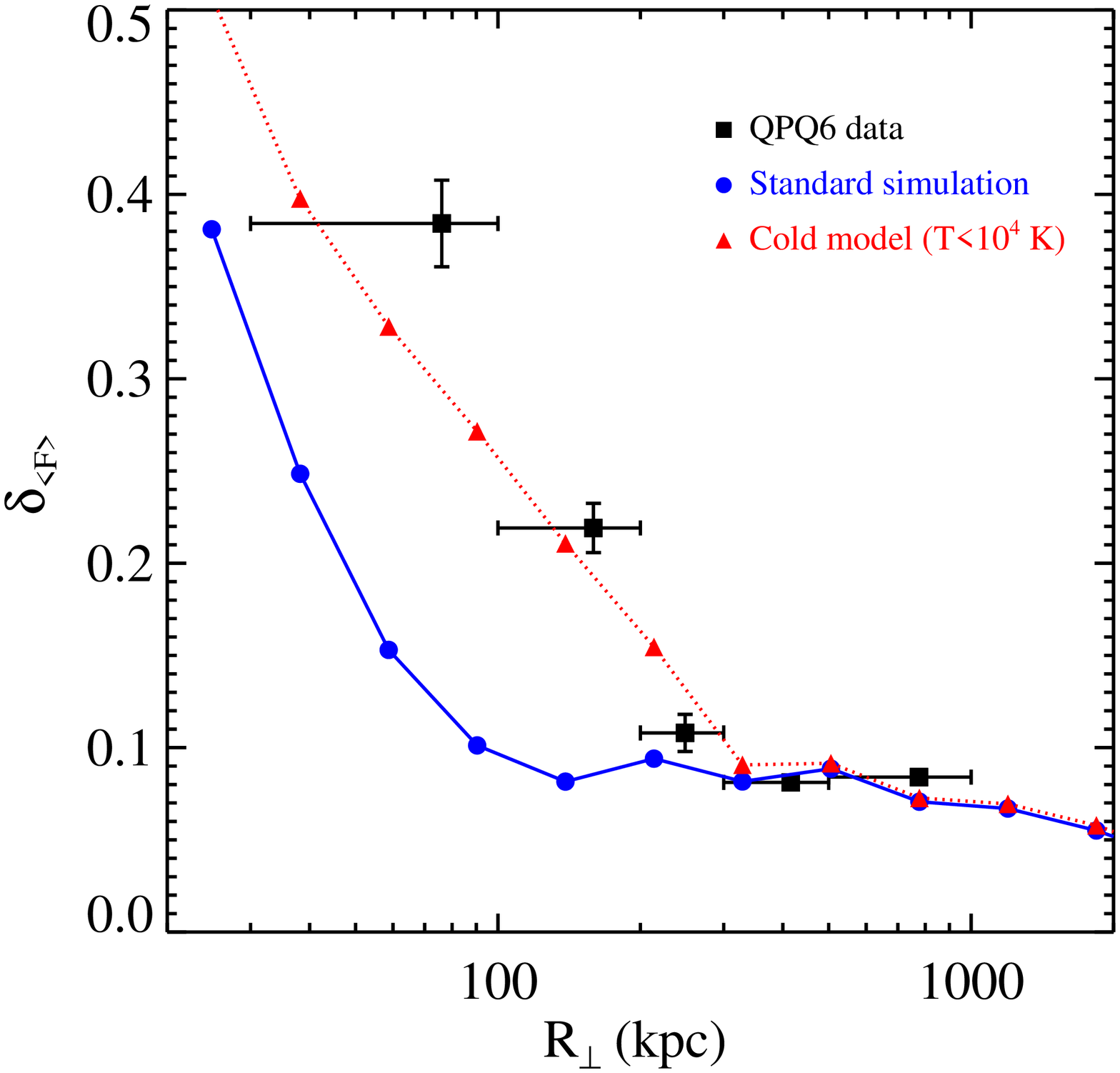}
\caption{
Comparison of \deltf\ values measured from the QPQ6 dataset (black
points) with predictions from simulations of a massive 
halo at $z=2.4$ \citep[$M_{\rm halo} \approx
10^{12.8} \msol$;][]{cantalupo12}.  The
solid blue curve shows the standard output where we have neglected any
radiation from a central quasar.  This model matches the data well at
large impact parameters $(\mrphys \gtrsim 200$\,kpc) but
under-predicts the absorption on scales comparable to the virial
radius.   We conclude that such simulations underpredict the amount of
cool gas within the CGM of massive galaxies \citep[see
also][]{fumagalli13b}.  
As a simple attempt to reproduce the data, we generated a second, toy model
where the gas within 200\,kpc is restricted to $T \le 10^4$\,K (red, dotted curve).
This provides a better match to the observations on small scales.
Future work will explore the astrophysical processes within the CGM of
massive galaxies that may reproduce the implied reservoir of cool gas.
}
\label{fig:sc}
\end{figure}

The hydro-simulation has been post-processed with the radiative transfer code
\texttt{RADAMESH} \citep{cp11} to model the effect of
the cosmic UV background.  We do not discuss here models that include
radiation from the f/g quasar.
%
We have simulated 7500 mock spectra with randomly 
distributed orientations and impact parameters with respect to the quasar host. 
The simulated spectra include the effect of gas peculiar velocities
and the errors on the quasar systemic redshifts (randomly generated
from a Gaussian distribution with $\sigma_{z}=520 \mkms$, the
typical value associated with our quasar sample). Note, however, that
these effects are practically negligible given the large size of the
velocity window used for the flux averaging (2000\kms).
The simulated
absorption shows good agreement with the observations on large scales
($\mrphys \gtrsim 200$\,kpc), as shown in Figure~\ref{fig:sc}.  
In this respect the observations are consistent with quasars tracing
massive halos at $z \sim 2.5$.  On the other hand, it is apparent that
our `standard' simulation shows
too little absorption on small scales, especially within 200 proper
kpc from the quasar.  Even without quasar radiation,
the model does not show the observed
steep increase in $\delta_{<F_{2000}>}$ 
on these scales.\footnote{Coincidentally,
the standard model does provide a good match to the 
LBG values shown in Figure~\ref{fig:qpq_vs_lbg} \citep{rakic12}.}

We note that $\mrphys \sim 200$\,kpc corresponds to about one virial radius for
the selected, massive halo. Simulated gas within this radius is
shock-heated to very high temperatures ($\sim10^7$ K) and therefore
highly ionized by collisional ionizations. 
The significant lack of absorption in the simulated spectra may be an indication that
the simulations are missing a population of small, cold clumps within the hot gas,
due to resolution or other effects possibly related to multi-phase gas physics or
metal mixing. 
We have therefore considered a toy model in which, as an extreme case, we have
forced all the gas within the halo to have an initial, maximum temperature of $10^{4}$ K
(without changing its density), before applying the radiative transfer.
 In this case, the gas is photo-ionized and the 
contribution from collisional excitations is neglected. The resulting
\deltf\ values
indicated by the red dotted line in Figure~\ref{fig:sc}, 
are now in much better agreement with the QPQ6 data.
Although this `cold' model is clearly an oversimplification and it does
not represent a univocal solution to the discrepancy with data, it
suggests that the halo contains a sufficient reservoir of gas but that
a non-negligible fraction should be much cooler than predicted by our
standard model. This cold gas may be present in form of small, dense
clumps that are not resolved by our simulation. Higher resolution and
additional astrophysical processes, e.g. a proper treatment of metal
mixing and cooling in a multi-phase halo gas, may be required 
to resolve the discrepancy with the data. We will explore these
processes in future works. 


\subsection{Clustering of \ion{H}{1} Absorbers with $z \sim 2$
  Quasars in the Transverse Dimension} 
\label{sec:clustering}

In this sub-section, we measure the cross-correlation signal of strong
\ion{H}{1} absorbers with quasars to explore the physical nature of
these systems. 
The formalism was introduced and applied in QPQ2 to measure the
quasar-absorber correlation function $\xi_{\rm QA}$ 
of \ion{H}{1} absorbers transverse to quasar sightlines. 
In that paper, we measured a large correlation length $r_0 =
9.2^{+1.5}_{-1.7} \mcMpc$ for systems with $\mnhi \ge 10^{19} \cm{-2}$,
assuming $\xi_{\rm QA} \propto (r/r_0)^\gamma$ with $\gamma = 1.6$. 
We compared this result with measurements along
the line-of-sight and concluded that these absorption 
systems were anisotropically clustered around quasars. 
We further surmised that the gas detected in b/g sightlines is
typically not illuminated by the quasar, 
either because of anisotropic emission or episodic variability (we
currently favor the former).
Using the QPQ6 dataset, we 
update this calculation with a much larger pair sample and also
extend the analysis to absorbers with lower \ion{H}{1} column
densities.  A description of the \nhi\ measurements was given in
$\S$~\ref{sec:NHI} (see also the Appendix).   

\subsubsection{Clustering Analysis}

Define $\xi_{\rm QA}(r)$ to be the cross-correlation function that describes
clustering of absorbers to quasars.
%
Given the small sample size in QPQ2, we parameterized
the correlation function as a power-law $\xi_{\rm QA} = (r/r_0)^{-\gamma}$
with a fixed exponent, $\gamma=1.6$, following standard results from
galaxy-galaxy clustering \citep[e.g.][]{adel05}.
With QPQ6, we have analyzed $N_{\rm pair} = 393$ quasar 
pairs for strong \ion{H}{1} absorbers
at comoving impact parameters ranging from $\mrcom \approx 0.1 - 3
\mcMpc$ and $\mzfg < 3$.  
We begin by fixing $\gamma = 1.6$, test whether this provides a good
description of the observations, and then allow $\gamma$ to vary.

In practice, we have surveyed absorption systems in spectral windows
$\pm 1500\mkms$ around \zfg\ at a range of comoving impact parameters
\rcom.  Following the formalism in QPQ2, we evaluate $\xi_{\rm
  QA}(r)$, by maximizing the likelihood of the data for observing
$N_{\rm sys}$ absorption systems amongst the $N_{\rm pair}$ pairs,
where the probability of finding an absorber in the redshift interval
$\Delta z = 2(1+z)/ (\Delta v/c)$ at separation \rcom\ is $P(\mrcom,z)
= \ell_q \, \Delta z$ with $\Delta v = 1500\mkms$ and $\ell_q$ 
evaluated along the sightline.
Note that $P(\mrcom,z) = \ell_q \, \Delta z$ is equivalent to the
covering fraction $f_C$ which we evaluated in $\S$~\ref{sec:NHI} for
optically thick gas (and in proper units).
Specifically, we have

\begin{equation}
\ell_q(z,\mrcom,\Delta v) = \ell_{\rm IGM}(z) \; [1 +
  \chi_\perp(\mrcom,\Delta v)]
\label{eqn:ellq_proj}
\end{equation}
where $\ell_{\rm IGM}(z)$ is the incidence of absorbers in a random
region of the universe,

\begin{equation}
\chi_{\perp} (\mrphys, \Delta v) \approx \frac{a H(z)}{2 \Delta v} 
\intl_{-\Delta v / [a H(z)]}^{\Delta v / [a H(z)]} \, dZ \, \xi_{\rm
  QA} (\sqrt{\mrphys^2 + Z^2})  \;\;\;, 
\label{eqn:tcorr}
\end{equation}
$a \equiv 1/(1+z)$, and $Z$ is the comoving distance along the
sightline relative to \zfg.

To compute $\ell_{\rm IGM}(z)$, we adopt
the $f(\mnhi,X)$ distribution of \cite{omeara13} at $z \approx 2.5$
assuming a redshift evolution $(1+z)^{3/2}$
and evaluate $\ell_{\rm IGM}(z) = \ell(X)_{\rm IGM} dX/dz$ with

\begin{equation}
\frac{dX}{dz} = \frac{H_0}{H(z)}(1+z)^2  \;\;\; ,
\label{eqn:dX}
\end{equation}
and 

\begin{equation}
\ell_{\rm IGM}(X) = \intl_{N_{\rm HI}^{\rm min}}^{N_{\rm HI}^{\rm max}}
  f(\mnhi,X) \, d\mnhi \;\;\; .
\label{eqn:lox}
\end{equation}
We perform this analysis for three absorption-line samples: 
(1) the damped \lya\ systems (or DLAs) with $\mnhi \ge 10^{20.3}
\cm{-2}$.  The QPQ6 dataset is complete for
such systems.  The \cite{omeara13} results give 
$\ell^{\rm DLA}_{\rm IGM} (z) \approx 0.2[(1+z)/(1+2.5)]^{2.1}$, in good
agreement with \cite{phh08};
(2) super Lyman limit systems\footnote{ Note that this
differs from QPQ2 which analyzed the combined incidence of SLLS and
DLAs together (i.e.\ $\mnhi \ge 10^{19} \cm{-2}$).}  
(or SLLS) with 
$10^{19} \cm{-2} \le \mnhi \le 10^{20.3} \cm{-2}$.  
An evaluation of equation~\ref{eqn:lox} gives 
$\ell^{\rm SLLS}_{\rm IGM} (z) \approx 0.44 [(1+z)/(1+2.5)]^{2.1}$.
As discussed in $\S$~\ref{sec:NHI}, we have taken a conservative
approach towards identifying SLLS in these lower resolution data.  Therefore,
our estimate of the clustering signal may be
an underestimate;
(3) optically thick systems (or LLS) with $\mnhi \ge 10^{17.3}
\cm{-2}$, corresponding to $\tau \ge 2$ at the Lyman limit.  
As described in $\S$~\ref{sec:NHI}, we do not observe
directly the Lyman limit of the gas but infer that systems are optically thick
based on several criteria.  
Evaluating the \cite{omeara13} distribution function, we find
$\ell^{\rm LLS}_{\rm IGM} (z) \approx 1.05 [(1+z)/(1+2.5)]^{2.1}$, in good
agreement with \cite{ribaudo11}.

\begin{figure}
\includegraphics[width=3.5in]{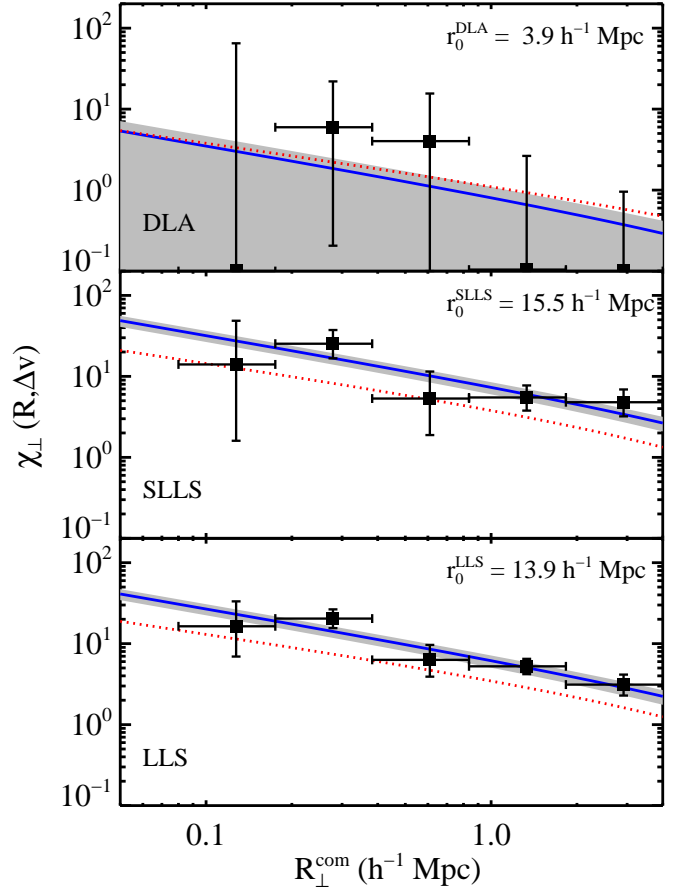}
\caption{The black data points show binned evaluations of the
  transverse cross-correlation function $\chi_\perp$ as a function of
  comoving impact parameter for DLA, SLLS, and LLS absorption systems
  with quasars.  The solid blue curves show the best-fit model for
  $\xi_{\rm QA} = (r/r_0)^{-\gamma}$ with $\gamma=1.6$,
  evaluated along the sightline at that impact
  parameter (Equation~\ref{eqn:tcorr}).  These models, derived from a
  maximum likelihood analysis, offer an excellent description of the
  observations. The gray shaded region indicates uncertainty in the
  models (we have adopted a 25\%\ systematic error).  The clustering
  amplitude for DLAs matches previous estimates and is consistent with
  gas tracing the ISM of individual galaxies.  The very large
  amplitudes for the LLS/SLLS imply a bias in the environments of
  massive galaxies that enhances the formation of optically thick gas.
  We propose that this gas arises in the overdense but uncollapsed
  large-scale structures (e.g.\ filaments) of the cosmic web around
  massive galaxies.  For comparison, we show the galaxy-quasar
  cross-correlation function from \cite{ts12} as a dotted red line.
}
\label{fig:correlation}
\end{figure}

The results from these analyses are presented in
Figure~\ref{fig:correlation} 
which shows binned evaluations of the transverse correlation function
$\chi_\perp$ evaluated in logarithmic bins of separation by comparing
the observed incidence of systems relative to random expectation (see QPQ2).
We overplot on these evaluations the best-fit models for
$\xi_{\rm QA}$, converted into $\chi_\perp$ (Equation~\ref{eqn:tcorr}).
These models were derived using our maximum likelihood estimator and the
uncertainties were determined using a Monte Carlo method (see QPQ2 for details).  
For $\gamma=1.6$, we recover clustering amplitudes
$r^{\rm DLA}_0 = \sixrdla$, 
$r^{\rm SLLS}_0 = \sixrslls$,  and
$r^{\rm LLS}_0 = \sixrlls$.  All of the observations are well described
by these models;  a Kolmogorov-Smirnov (KS) test comparing the observed
\rcom\ distribution of the \ion{H}{1} systems with the distribution
predicted from the best-fit models (adopting the QPQ6 distributions of
\rcom\ and \zfg) gives
$P_{\rm KS} \approx 0.25$ for the DLAs and $P_{\rm KS} > 0.9$ for the
SLLS and LLS. 

Our results confirm the primary conclusion of QPQ2, that strong
\ion{H}{1} absorbers are highly correlated with quasars
in the transverse dimension.  In fact, for the LLS and SLLS
populations (which overlap significantly) we derive clustering
amplitudes higher than those reported in QPQ2\footnote{Most of
  the difference is due to the higher $\ell_{\rm IGM}(z)$ estimation
  used for the
  background incidence of absorbers in QPQ2.}, even exceeding
quasar-quasar and quasar-galaxy clustering at $z \sim 2.5$
\citep{shen07,ts12}.    
To emphasize the difference, we overplot the projected
cross-correlation function of quasar-galaxy clustering 
in Figure~\ref{fig:correlation} using the results\footnote{Note that
  \cite{ts12} focused on hyper-luminous quasars but quasar clustering
  does not show a strong dependence on luminosity
  \citep{croom04,shen07}.}
of \cite{ts12}. 

\begin{figure}
\includegraphics[width=3.5in]{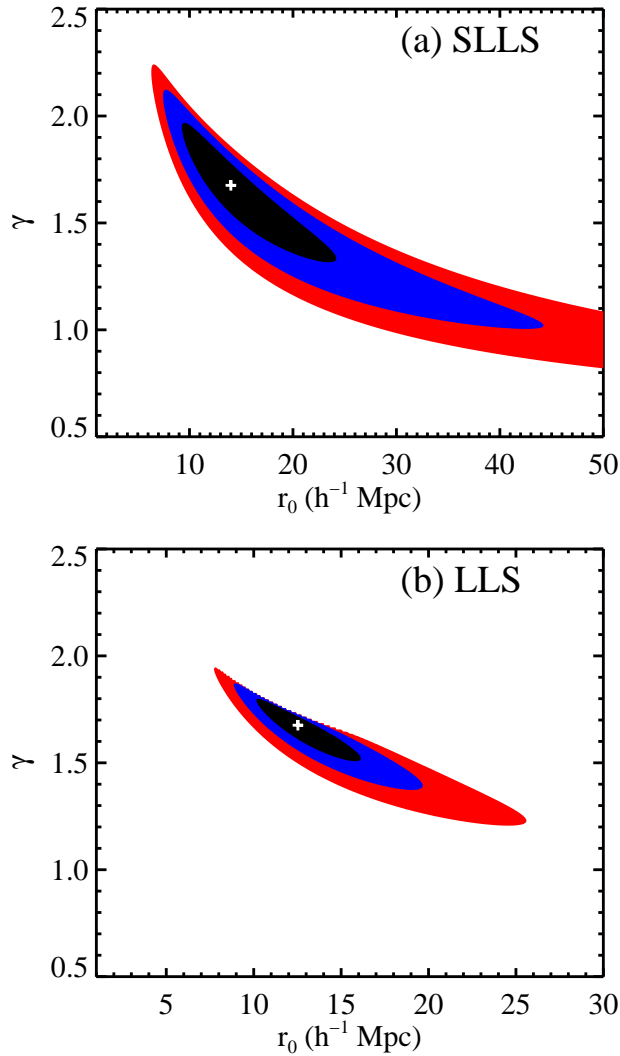}
\caption{Confidence contours (black, blue, red for 68.3\%, 95.4\%, and
  99.7\% respectively) for clustering fits 
  $\xi_{\rm QA} = (r/r_0)^{-\gamma}$ to the (a) SLLS and (b)
  LLS around quasars.  There is an obvious degeneracy between the $r_0$ and
  $\gamma$, although the results are reasonably well
  constrained.  Integrating over each parameter we recover
  $r_0^{\rm SLLS} = \rslls$ with $\gamma^{\rm SLLS} = \gslls$
  for the SLLS and 
  $r_0^{\rm LLS} = \rlls$ with $\gamma^{\rm LLS} = \glls$ for the LLS.
}
\label{fig:2parm_cluster}
\end{figure}

The formal error bars for the SLLS/LLS are very small, owing to the
large clustering signal (see QPQ2) and our having fixed $\gamma = 1.6$.  
Indeed $\gamma$ and $r_0$ are highly correlated.  
In Figure~\ref{fig:2parm_cluster}, we present the confidence contours
for $\gamma$ and $r_0$ for the LLS and SLLS having allowed each
parameter to vary freely.  For the SLLS, the results give 
$r_0^{\rm SLLS} = \rslls$ with $\gamma^{\rm SLLS} = \gslls$
at 68\%\ c.l. 
Similarly, the analysis gives 
$r_0^{\rm LLS} = \rlls$ with $\gamma^{\rm LLS} = \glls$ for the LLS.
These results are more reflective of the
uncertainty in the clustering amplitude.
Furthermore, the Monte Carlo error estimate does not reflect uncertainty
in the \nhi\ measurements, neither statistical nor systematic (see
$\S$~\ref{sec:NHI}).  
We also estimate an uncertainty of $\approx 20\%$ in the 
background incidence which affects the clustering measurements. 
Considering
these effects together, we recommend adopting an $\approx 25\%$ 
uncertainty for
our best-fit $r_0$ values\footnote{Note that $r_0$ scales as the
  incidence to the $\gamma^{-1}$ power.}.

Of greatest importance to the discussion that follows is the enhancement in
SLLS (and LLS) at large separations.  As such, we have carefully
scrutinized this result (QPQ5 focused on small scales). 
For the SLLS, we discovered 14 systems\footnote{We note that one of
  these SLLS occurs twice in the clustering analysis because the field
  (J1421+5727) shows two f/g quasars at nearly the same redshift.
  This is the only example were a b/g spectrum is analyzed twice at
  essentially the same \zfg.} 
in the QPQ6 dataset at $\mrcom
> 1 \mcMpc$ where only 2.2 were
predicted.  Of these 14, at least 10 show strong low-ion absorption
indicative of SLLS and we consider them secure. 
Restricting to the 10 systems, we recover $r_0^{\rm SLLS} = 14.7
\mcMpc$ for $\gamma = 1.6$.
Similar conclusions are drawn for the LLS.
We conclude that $r_0 > 10 \mcMpc$ is a robust result for the
clustering of SLLS and LLS for $\gamma < 1.8$.

\subsubsection{Implications for \ion{H}{1} Absorbers and the CGM}
\label{sec:cluster_discuss}

The previous sub-section presented new measurements on the clustering
of strong \ion{H}{1} systems to quasars, as measured from the QPQ6
survey, corresponding to comoving separations of $\mrcom < 2.5
\mcMpc$.  We remind the reader that quasars inhabit massive halos
inferred (in part) from the large quasar-quasar auto-correlation length 
\citep[$r_0 = 8.4 \hMpc$ for $\gamma = 2$;][]{white12}.  This sets a
reference for which to consider our results.  
In principle, one may combine the quasar-absorber cross-correlation with the
auto-correlation function to estimate the mass of the dark matter
halos hosting the absorbers.  This requires, however, that the
absorbers trace the dark matter density field deterministically and
that the analysis be performed on the same scales. 
In general, absorbers are believed to be hosted
by dark matter halos that are less biased than quasars, 
such that the quasar-absorber cross-correlation length 
is expected to be smaller than the quasar auto-correlation. 

Previous studies on absorber clustering
have focused on galaxies, in part because the latter have a much higher 
comoving number density than quasars.
At low redshift, such measurements have been used to examine the
origin of strong \ion{Mg}{2} absorption
\citep[e.g.][]{bmp04,tc08,padilla+09}
and the nature of the \lya\ forest \citep{cpw+05,tejos+12}.
At $z \sim 2$, studies have examined the link between \ion{C}{4} gas
and LBGs \citep{ass+05,martin+10} and have placed constrains 
on the halo masses
for DLAs \citep{Bouche04,cwg+06}.
Regarding quasar-absorber clustering, in addition to QPQ2,
\cite{wkw+08} measured the \ion{Mg}{2} and \ion{C}{4} clustering with
quasars at $z \sim 1$ and 2 respectively on large, transverse 
scales ($\mrcom \gg 1 \mcMpc$).  
The clustering lengths
($r_0 \approx 5 \mcMpc$) were used to infer that quasars are hosted by
halos with masses $M > 10^{12.5} \msol$ at $z \sim 2$ and over
$10^{13} \msol$ at $z \sim 1$.  
The quasar-absorber cross-correlation amplitude also offers insight
into the physical nature and origin of strong \ion{H}{1} absorbers and
on astrophysical processes for cool gas in massive halos.

Consider first our results for the DLAs.  The estimated clustering
length is consistent with previous estimates from 
LBG-DLA cross-correlation analyses performed on similar scales
\citep{cwg+06}.  
Formally, $r_0^{\rm DLA} = 3.8 \mcMpc$ is preferred and the
correlation length is
restricted to $r_0^{\rm DLA} <9 \mcMpc$ (95\%\ c.l.) for $\gamma = 1.6$.
Our results are consistent with models where DLAs trace dark
matter halos \citep[i.e.\ galaxies;][]{pgp+08} 
and the clustering amplitude implies
characteristic masses of $M_{\rm halo}^{\rm DLA} \sim 10^{11} -
10^{12} \msol$ \citep{lnh+11,font12}.  
There is no signature of 
the so-called one-halo term where $\xi_{\rm QA}(r)$ shows excess
small-scale clustering due to
satellites of the central halo, but we caution that this conclusion
is tempered by the small sample size. 
In summary, the results for DLAs follow one's expectation for gas
tracing galaxies within common dark matter halos, i.e.\ the ISM of
star-forming galaxies.

In contrast to the DLAs, the
LLS and SLLS exhibit very large clustering lengths, e.g.\ 
$r_0^{\rm SLLS} > 10\mcMpc$,  which even exceeds the correlation
lengths for quasar-quasar and quasar-galaxy clustering at $z \sim 2$.
Naively, following the discussion above,
one might infer that these optically thick
absorbers would need to trace
amongst the most massive dark matter halos in the Universe 
($M > 10^{13} \msol$). This assertion is easily
dismissed, however, because such halos are so exceedingly rare at $z
\sim 2$ that they could not possibly account for the observed incidence
of optically thick gas.  
For example, given the low number density of halos with $M > 10^{13}
\msol$, these 
would need to have an effective area $A_{\rm eff} > 2\sci{6} \, {\rm
  kpc^2}$ 
corresponding to a typical radius of 800 kpc 
to match the LLS incidence (see $\S$~\ref{sec:qal} for further details).
This would require a 100\%\ covering factor of SLLS to well beyond the
virial radius in these massive halos.  
We conclude that at the small scales probed here ($\mrphys < 2.5
\mcMpc$) from luminous QSOs, LLSs are no longer simple 
biased tracers of the underlying dark matter. 

Instead, we infer that the large
$r_0$ values are driven by astrophysical processes which bias the
$\approx 1$\,Mpc environments surrounding quasars to preferentially
exhibit optically thick gas.  
On small scales, i.e.\ within the virial radius, the clustering signal
could be driven by gas within the CGM of the host halo. 
The clustering observed on larger scales, however, implies the presence of a
dense, self-shielding medium giving rise to LLS/SLLS which lies between
dark matter halos.  We hypothesize that this gas is located
within the large-scale structures (e.g.\ filaments) that connect the
massive halo to its neighbors.
Because we do not observe such clustering in the DLAs, the structures
must have surface densities preferentially below 
the $\mnhi = 10^{20.3} \cm{-2}$ threshold.

We now discuss several such scenarios that may explain the
large
clustering amplitude, 
focusing first on the inner regions $r \lesssim \mrvir$ ($\approx
160$\,kpc for a halo with $M \approx 10^{12.5} \msol$ at $z=2.5$).
Might the large clustering length (and correspondingly high covering
fraction) of optically thick gas result from satellite galaxies in the
central halo?  Indeed, quasars reside in massive halos and
$\Lambda$CDM clustering predicts an abundance of such satellite
galaxies \citep[e.g.][]{mps+01}.  If we confine the gas from such
satellites to lie within their halos, however, one finds a small covering
fraction and a correspondingly small clustering amplitude
\citep[e.g.\ QPQ3;][]{tumlinson+13}.  
For example, one predicts an average of 5
satellites with $M > 10^{10} \msol$ within the virial radius of
halos with $M_{\rm halo} = 10^{12} - 10^{13} \msol$ at $z=2.5$
\citep{behroozi13}.  Even if we assumed that optically thick gas
extended uniformly to a
radius of 20\,kpc within each satellite, these would
cover $\lesssim 10\%$ of the projected area of the central halo.
We reached the same conclusion in QPQ3 based on analysis of the
observed quasar-galaxy cross-correlation function.
In addition, satellite galaxies should also contribute to DLA
absorption and significant quasar-DLA clustering which is not observed.  

If gas within satellite galaxies is insufficient, one must
conclude that this optically thick gas comprises the
ambient circumgalactic medium (CGM) of the central halo.  
But from where did this material originate?    
As discussed in QPQ5, the high incidence of strong
metal-line absorption indicates the gas is chemically enriched.  This
requires that a non-negligible fraction of the material has previously
cycled through a galaxy.  The total gas mass implied 
is large ($M > 10^{10} \msol$), and it may thus be
unlikely that the majority of this CGM was stripped/expelled
from the
halo's satellite galaxies. 
Indeed, numerical simulations of galaxy
formation currently predict that the majority of dark matter and
baryons accrete onto halos in diffuse streams 
\citep{kkw+05,dbe+08,brooks09}.
On the other hand, such streams alone are unlikely to reproduce the
clustering signal \citep[$\S$~\ref{sec:mass},][]{fumagalli13b}.  
Material driven from the galaxy hosting the f/g quasar, e.g.\ via
supernovae or AGN feedback, may also inject gas and metals into the
surrounding halo.  A fraction of this material may be cool and could
be optically thick, but we question whether it can travel to large
radii given the implied energetics (QPQ3).
We suspect that none of these effects (streams, satellites,
expelled gas) alone can reproduce the observations.  Absent any other
obvious effects (e.g.\ the condensation of cool gas out of a hot halo), we
contend that they all must contribute to match the observations.
We await the output of new `zoom-in' simulations of massive halos at
$z \sim 2$ to further explore these issues
\citep[e.g.][]{cantalupo12,fumagalli13b}.  

Another question that follows relates to the survival of this CGM
gas: Is it ephemeral, i.e.\ does it require constant replenishment?
If the gas exists outside satellites and their dark matter
halos, we may assume it is not self-gravitating.  For these `clouds'
to avoid rapid dissipation through adiabatic expansion, we
propose that it be embedded within a warm/hot
medium that offers pressure confinement (e.g.\ QPQ3).  Previous work has
described a range of physical processes that can and should destroy
cool gas clouds within a hot halo
\citep[e.g.][]{mm96,mcd99,mb04,sck07}.  
These would
apply to the CGM of our massive halos, albeit with a presumably hotter
halo (previous work considered lower mas halos with virial
temperatures $\sim 10^6$\,K)
and higher mean density than average. 
In QPQ3, we demonstrated that such clouds could exist in pressure
equilibrium with the hot virialized tenuous plasma predicted to be present in the massive
dark matter halos hosting quasars. 
Evaluating the survival time of such clouds and the processes that may
generate a new population deserves focused, numerical studies that
include radiative-transfer coupled to the hydrodynamics.

While one may evoke various astrophysical mechanisms 
within the CGM
of the central halo to explain the high clustering signal on small scales ($r
< \mrvir$), we
reemphasize that the observations show substantial clustering even at
$\mrcom \approx 3 \mcMpc$.  And again, the measured amplitude likely
exceeds anything that could be 
produced by the LLSs tracing the underlying dark matter, 
because in order to reproduce this amplitude
the LLSs would have to exist in extremely massive halos implying 
unphysically large cross-sections. 
Instead, we 
hypothesize that a baryonic bias exists for optically thick gas
on proper scales of $\sim 1$\,Mpc.  We further
speculate that this gas arises in the
large-scale structures (e.g.\ filaments) surrounding massive
halos.  The gas within these overdense regions 
self-shields and gives rise to SLLS and LLS
\citep[e.g.][]{mcquinn11,altay11}.
While lower mass halos should also exhibit extended structures (albeit
on smaller scales), the amplitude of their overdensities are lower
and may be insufficient to produce optically thick gas.  
We encourage future observational and theoretical work to 
(a) confirm
the clustering signal at $\mrcom \approx 3 \mcMpc$ and extend it to
larger impact parameters; and
(b) explore the likelihood that SLLS/LLS preferentially arise in the
large-scale structures surrounding massive halos at $z \sim 2$.

There are a few additional points to emphasize.  First, the strong
clustering signal implies a bias of SLLS/LLS to massive halos.  Under
the expectation that such environments exhibit preferentially higher
metallicity, this may bias the 
absorbers to have higher enrichment. 
Indeed, one observes a remarkable number of SLLS with solar or
super-solar abundances \citep[e.g.][]{peroux06,poh+06}, perhaps exceeding
the incidence of high-metallicity DLAs
\citep{pdd+07,dz09,kph+10}.
Indeed, this has led to speculations that SLLS trace more
massive galaxies \citep{kks+10}.
Second, there is an
apparent mass dependence to the presence of optically thick gas.
Although the LLS cross-correlation with LBGs has not yet been
measured, the observed covering factor of LLS is
considerably lower for LBGs at the same proper scales than for
quasars (QPQ5).
Similarly, \cite{ass+05} measured the cross-correlation of strong
\ion{C}{4} systems (which may preferentially trace strong \ion{H}{1}
absorbers) and measured $r_0 \sim 4 \mcMpc$ on scales of a few
\cMpc.  
This is much smaller
than what we record for optically thick gas and
quasars, although we do caution that our SLLS show relatively modest
\ion{C}{4} equivalent widths (QPQ5). 

Third, the significant clustering of optically thick gas has
implications for the attenuation of quasar ionizing radiation and the
resultant intensity of the UV background.
The most important effect of LLS
clustering would be to underestimate the mean free path 
of photons (if one were to neglect it).
Lastly, it is worth considering whether any of these
results are biased by the presence of a luminous quasar.  
For example, if the quasar's
radiation field illuminates gas throughout the environment, it
should suppress optically thick gas and thereby reduce the
clustering amplitude (and covering fraction).  Such a bias would only
accentuate the conclusions drawn above.  On the other hand, one could
speculate that quasars shine only during episodes of intense galaxy
formation, e.g.\ during galaxy-galaxy mergers, which could enhance the
presence of cool gas in the environment.   We consider this unlikely 
because
the signal we observe extends to Mpc whereas 
the fueling of AGN occurs on pc scales.  It would be
remarkable for such processes to be so tightly coupled.
Furthermore, estimates of the star-formation rates of galaxies hosting
luminous quasars do not indicate extreme activity
\citep{santini+12,rosario+13}.

\subsection{The Contribution of Massive Halos to QAL Systems}
\label{sec:qal}

Quasar absorption line (QAL) systems with the largest \ion{H}{1}
column densities have been associated with galaxies and the dark
matter halos within which they reside.  This includes the damped \lya\
systems \citep[DLAs, $\mnhi \ge 10^{20.3} \cm{-2}$;][]{wgp05}, the
super Lyman limit systems \citep[SLLS, $\mnhi \ge 10^{19}
\cm{-2}$;][]{peroux05,opb+07}, and the optically thick Lyman limit
systems \citep[LLS, $\mnhi \gtrsim 10^{17}
\cm{-2}$;][]{pow10,fumagalli13a}.
Their association to galaxies is motivated by the nearly ubiquitous
detection of heavy elements \citep{rafelski+12,pow10,peroux05,fop11}
and the expectation that such high \nhi\ values may only be achieved
within collapsed structures.  Such inferences are supported by
analysis of cosmological simulations
\citep[e.g.][]{pgp+08,fg11,fumagalli11a,erkal12,rahmati+13}.

It is of great interest is to establish the masses of the halos
hosting these
absorption systems.  
In $\S$~\ref{sec:clustering}, we measured the clustering of these absorbers
with quasars in the transverse dimension.
Our results require that these massive halos contribute
a non-zero fraction of strong \ion{H}{1} absorption systems. 
On the other hand, massive halos are sufficiently rare that their
integrated contribution to the incidence of these absorbers may be small.
By comparing the contribution of massive halos to the cosmological incidence of
these QAL systems, one may in turn constrain their host halo masses.
For example, if massive halos have insufficient cross-section to LLS
absorption, then this gas must manifest in alternate regions of the
universe.  

\begin{figure}
\includegraphics[width=3.5in]{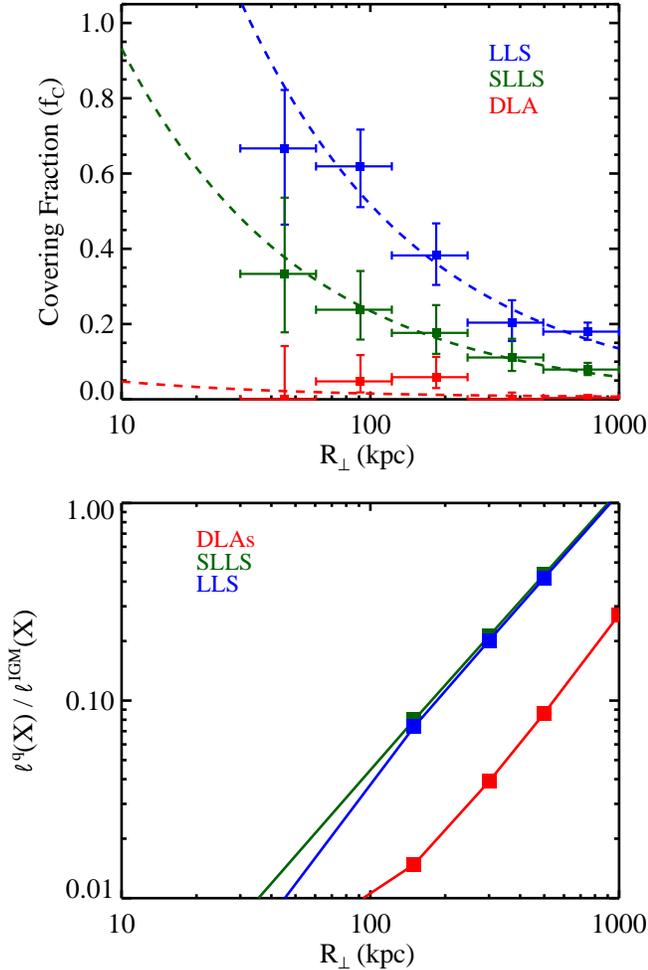}
\caption{
  (top): Binned evaluations of the covering fraction \fc\ for strong
  \ion{H}{1} absorption systems (LLS, SLLS, DLAs) as a function of
  proper impact parameter \rphys\ from a luminous $z \sim 2$
  quasar.  Specifically, \fc\ is defined as the fraction of sightlines
  at \rphys\ that show at least one such absorber in a velocity window
  $\Delta v = \pm 1500\mkms$ around the f/g quasar redshift.  
  The dashed curves show estimations of \fc\ from the evaluation of
  the cross-correlation functions derived in
  $\S$~\ref{sec:clustering} and shown in Figure~\ref{fig:correlation}. 
  (bottom): The solid curves show the fractional contribution of gas in the
  environments of $z \sim 2$ luminous quasars to the total incidence
  of strong \ion{H}{1} absorbers (LLS, SLLS, DLAs). These assume the \fc\
  evaluations shown above and that quasars
  trace dark matter halos with $\mmmin \ge 10^{12} \msol$.   The
  curves also assume that $\mfc = 1$ for $\mrphys^{\rm min} \le
  10$\,kpc.  DLAs rarely occur within a few hundred kpc
  of massive halos.  In contrast, quasar environments contribute
  significantly to optically thick gas (LLS, and especially the SLLS).
  For an assumed virial radius $r_{\rm vir} = 150$\,kpc, massive halos
  contribute approximately 10\%\ of the incidence of these optically
  thick absorbers.  
  The contribution of the extended environment ($\sim 1$\,Mpc)
  approaches 100\%.  This
  suggests that all such gas lies within $\approx 1$\,Mpc of a
  massive halo, e.g.\ neighboring and lower mass dark matter halos and/or
  the large-scale structures that connect these halos.
}
\label{fig:qal}
\end{figure}

For a population of halos with a comoving number density \ncom\
and an effective, physical cross-sectional area \aeff\  
to a given \ion{H}{1} column density, the incidence per unit
absorption length $dX$ (Equation~\ref{eqn:dX})
is given by

\begin{equation}
\ell(X) = \frac{c}{H_0} \, \mncom \, \maeff \perd
\end{equation}
We will assume that quasars occupy all halos with masses
greater than a minimum value $M_{\rm halo}^{\rm min}$.
Assuming a $\Lambda$CDM cosmology and standard Press-Schecter analysis
\citep[e.g.][]{bardeen86,db06}, 
this sets \ncom\ at a given redshift.
For the effective area, we adopt $\maeff = f_C \, \pi \mrphys^2$ with
\fc\ the measured covering factor within \rphys.

Figure~\ref{fig:qal}a presents the evaluation of \fc\ for LLS,
SLLS, and DLAs from our QPQ6 dataset (see also
Figure~\ref{fig:othick}).  
For an evaluation of \fc\ at any impact parameter, we may recast
the cross-correlation functions measured in the previous sub-section
in terms of a covering fraction.  In comoving coordinates, \fc\ at
comoving impact parameter \rcom\ is given by 
$f_C(\mrcom) = \ell_q(\mrcom,\Delta v) \Delta z$ with $\ell_q$ given
by Equation~\ref{eqn:ellq_proj} and we take $\Delta v = \pm 1500\mkms$ to
match our approach to assessing strong \ion{H}{1} systems
($\S$~\ref{sec:indiv}).
The resultant \fc\ curves assuming $\gamma=1.6$
are overplotted on Figure~\ref{fig:qal}a,
assuming $z=2.4$ to convert \rcom\ to \rphys\ only to match the x-axis. 
These provide a good description of the observations (as expected).

We then calculate 
\begin{equation}
\maeff(\mrphys) = \intl_0^{R_\perp} \, f_C(\mrphys') \, 2 \pi \mrphys' \,
d\mrphys'  \;\; .
\end{equation}
Note that we restrict $f_C \le 1$ and we have set $f_C = 1$ at impact
parameters $\mrphys^{\rm min} \le 10$\,kpc for all absorbers to account for the
ISM of a presumed host galaxy.  This primarily impacts the results for the
DLAs.

Figure~\ref{fig:qal}b illustrates the fractional contribution of massive halos to
the observed incidence of strong \ion{H}{1} absorbers evaluated at
several characteristic impact parameters: 30\,kpc, 150\,kpc
(corresponding to approximately one virial radius \rvir),  
300\,kpc ($\approx 2 r_{\rm vir}$), 500\,kpc, and 1\,Mpc.
All of the curves assume $M_{\rm halo}^{\rm min} = 10^{12.5}
\msol$, giving $n=1.2\sci{-4}$ halos Mpc$^{-3}$ (comoving) at $z=2.4$.
For the incidence of LLS, SLLS, and DLAs in the ambient 
IGM, we have adopted
$\ell^{\rm LLS}_{\rm IGM}(X) = 0.3$,
$\ell^{\rm SLLS}_{\rm IGM}(X) = 0.13$,
and
$\ell^{\rm DLA}_{\rm IGM}(X) = 0.055$ 
respectively.  These values are drawn from traditional surveys of quasar
absorption systems at $z \approx 2.4$ \citep{pw09,opb+07,omeara13},
increase by $\approx 50\%$ from $z=2$ to 3,  
and have approximately 20\%\ uncertainty.

Consider first the DLAs.  Figure~\ref{fig:qal}b reveals that gas within
the virial radius of massive
halos ($M_{\rm halo} \ge 10^{12.5} \msol$) does not contribute significantly
to the total incidence of these strongest \ion{H}{1} absorption
systems.  Even if we set $f_C=1$ for $\mrphys^{\rm min} \le 20$\,kpc, we find
that $\ell^{\rm DLA}_{q} (X)$ is less than 5\%\ of the total
incidence at $\mrphys = 150$\,kpc, which matches or exceeds \rvir\ 
for the central halo.
Unless the halos of quasars have suppressed DLA gas (contrary to all
other inferences we have drawn), 
we conclude that the majority of DLAs must arise in halos with masses
$M_{\rm halo}^{\rm DLA} \le 10^{12.5} \msol$.  
\cite{font12} have measured the cross-correlation between the
\lya\ forest and DLAs and concluded that the latter may be hosted by
more massive halos than previously derived from numerical simulations.  
In their fiducial model, which matches
the clustering and incidence of DLAs (we caution
this model is not unique),
they assert that $10^{12} \msol$ halos must have an effective area
for DLAs of 
$A_{\rm eff}^{\rm DLA} = 1400 \, {\rm kpc^2}$. 
For the massive halos traced by quasars, we have identified 2
DLAs from the 38 pairs with $\mrphys < 150$\,kpc.
This incidence is consistent with gas in the outer halo contributing a
significant fraction of the $A_{\rm eff}$ value suggested by
\cite{font12}, but clearly a much
larger sample is required to offer a robust measurement.

In contrast to the DLAs, gas associated to massive galaxies
contributes significantly to the SLLS and LLS populations.  
For $M_{\rm halo}^{\rm min} = 10^{12.5} \msol$, gas within $r \approx
\mrvir$ of such massive halos yields
$\approx 10\%$ of the SLLS in the Universe
(Figure~\ref{fig:qal}b). 
This is a remarkable result given that $M_{\rm halo} > 10^{12.5} \msol$ halos represent
a small fraction of all dark matter halos at $z=2.4$, e.g.\ only $\sim \sci{-3}$ 
of all halos with $M _{\rm halo}> 10^{10} \msol$.
Despite being rare, their large physical size and covering fractions
imply a substantial contribution to LLS.
Even more remarkable, the results in Figure~\ref{fig:qal}b indicate
that the extended environment
of massive halos (gas to $\mrphys \sim 1$\,Mpc) yields an incidence
of optically thick gas that roughly matches the IGM average\footnote{ 
  The number density of $10^{12.5} \msol$ halos is sufficiently small that
  one proper Mpc is much less than their mean separation. 
  Therefore, the extended environments as we have defined them do not
  overlap (i.e.\ no double counting).}.
While there are substantial uncertainties in this estimate 
(e.g.\ $\ell_q(X)$ is sensitive to $M_{\rm halo}^{\rm min}$, \fc\ has associated
uncertainty), our quasar pair analysis implies that a substantial
fraction of all optically thick gas occurs within the structures
extended around massive galaxies at $z \sim 2$.
This assumes, of course, that the condition of having a luminous
quasar within these massive galaxies does not imply special
characteristics for the gas on scales of 1\,Mpc. 
If confirmed by future work, this result has significant consequences
for our understanding of the LLS population, their contributed  
opacity of the IGM, and the attenuation of ionizing sources
that generate the EUVB.

\section{Summary}
\label{sec:summary}

We have constructed a sample of \npairs\ projected quasar pairs
  where spectra of the b/g quasar covers the \ion{H}{1} \lya\ line at
  the redshift of the f/g quasar.  We have restricted the sample to
  data with $\msnlya > 5.5$, proper separations at \zfg\
  of $\mrphys \le 1$\,Mpc, and redshift separation such that
  \ion{H}{1} \lya\ of the f/g quasar lies in the rest wavelength
  interval $1030 \, {\rm \AA} < \lambda_r < 1200 \,\rm \AA$ of the b/g
  quasar. 
The f/g quasars have redshifts $\mzfg = 1.6 - 4.5$ with a median
  value of 2.34 and Bolometric luminosities ranging
  from $\approx 10^{45}-10^{47} \, {\rm erg \, s^{-1}}$.
  If these sources emit isotropically, their UV fluxes exceed
  the UV background by factors of 10 to 10,000.
The quasar spectroscopy comprises a heterogeneous dataset drawn from
  SDSS, BOSS/DR9, and a diverse set of instruments on large-aperture
  telescopes.  We have used these data to re-measure the f/g quasar
  redshifts and we have continuum normalized the b/g quasar spectra with
  an automated routine that mean flux regulates the data
  to the average \ion{H}{1} \lya\ opacity of the IGM \citep{lee+12}.

We then proceeded to analyze these spectra to study the \ion{H}{1}
\lya\ absorption in the $\mrphys \le 1$\,Mpc environment of $z \sim 2$
luminous quasars.  The primary findings are:

\begin{itemize}
\item 
  The 1\,Mpc environments surrounding the massive
  galaxies which host luminous $z\sim 2$ quasars have enhanced \ion{H}{1}
  \lya\ opacity.  For the complete sample, which has a median impact
  parameter of 725\,Mpc, we find that $\mdeltf \equiv (\mavgf_{\rm
    IGM} - \mavgf^{2000})/\mavgf_{\rm IGM} = 0.09$.  
  \item The excess \ion{H}{1} absorption increases with
  decreasing \rphys\ consistent with the gas tracing a massive
  overdensity and not being illuminated by the f/g quasar.
  Analysis of composite spectra
  binned in intervals of \rphys\ yields an excess \ion{H}{1} \lya\
  equivalent width: 
  $\mwstack = 2.3 \, {\rm \AA} (\mrphys/{\rm 100\,kpc})^{-0.46}$.

\item Comparing the data against a numerical simulation of a massive
  halo \citep{cantalupo12}, we find good agreement on scales $\mrphys
  \gtrsim 200$\,kpc but that the model shows much less 
  absorption than observed on smaller scales.   Current models of structure
  formation appear to underpredict the distribution of cool gas in the
  CGM of massive galaxies.

\item The \ion{H}{1} \lya\ opacity around quasars exceeds
  that observed for LBGs \citep{rakic12}, consistent with the latter
  galaxies occupying systematically lower mass halos at $z \sim 2-3$.


\item We analyzed the quasar-absorber cross-correlation function
  $\xi_{\rm QA}(r)$ to comoving impact parameter $\mrcom \approx 2.5
  \mcMpc$.  Parameterizing $\xi_{\rm QA}(r)$ as a power-law
  $(r/r_0)^\gamma$, we find 
  $r^{\rm DLA}_0 = \sixrdla$ for a fixed $\gamma=1.6$ for DLAs, 
  $r_0^{\rm SLLS} = \rslls$ with $\gamma^{\rm SLLS} = \gslls$
  for the SLLS, and 
  $r_0^{\rm LLS} = \rlls$ with $\gamma^{\rm LLS} = \glls$ for the LLS.
  We estimate a systematic
  uncertainty of $\approx 20\%$ in these values.

\item The amplitude for DLAs is consistent with
  previous galaxy-DLA cross-correlation measurements and 
  follows expectation for gas tracing the ISM of galaxies. The very
  large clustering amplitude for optically thick gas (LLS, SLLS)
  indicates a strong bias toward such material in the environments of
  massive halos.  We speculate that this gas arises predominantly in
  large-scale structures (e.g.\ filaments) that connect the central
  halo to neighboring dark matter halos on $\sim 1$\,Mpc scales.

\item We estimate that gas within the virial radius of 
  massive halos $(M_{\rm halo} >
  10^{12.5} \msol$) contribute $\sim 10\%$ of the observed optically
  thick gas at $z \sim 2.5$.  In contrast, these halos yield less than
  5\% of DLAs.
  Extending to 1\,Mpc, the environments of
  these massive halos may dominate the Universe's Lyman limit opacity.

\item Our observations of large enhancements in absorption in b/g sightlines provide
compelling evidence for a large overdensity around the f/g quasar. 
Previous analyses of the line-of-sight proximity effect that have neglected this density enhancement
will have systematically overestimated the EUVB intensity.

\end{itemize}

Future work will focus on 
(1) the metal-line absorption of gas in the quasar environment on both
small ($\mrphys < \mrvir$) and large scales;
(2) kinematics of the gas using a subset of the sample with precisely
measured \zfg\ values from near-IR spectra;
(3) an assessment of the transverse proximity effect;
and 
(4) detailed analysis of echelle and echellette spectra of b/g spectra to study and analyze
the CGM of quasar hosts. 
In parallel, we are pursuing and encouraging numerical simulations of
massive galaxies at $z \sim 2$ to further explore the assembly of the
most massive structures at early times.

\acknowledgements
We thank A. Elvin and J. Primack for their analysis of the Bolshoi
simulation.   We thank M. Fumagalli for valuable comments and
criticism and his software to construct mass functions. 
JFH acknowledges generous support from the Alexander von Humboldt
foundation in the context of the Sofja Kovalevskaja Award. The
Humboldt foundation is funded by the German Federal Ministry for
Education and Research.  JXP and SC acknowledge support from the National
Science Foundation (NSF) grant AST-1010004. JXP and AM
thank the Alexander
von Humboldt foundation for a visitor fellowship to the MPIA where
part of this work was performed, as well as the MPIA for hospitality
during his visits.
SGD acknowledges a partial support from the NSF grants AST-0407448 and
AST-0909182, and the Ajax Foundation.
CM acknowledges support from NSF grant AST-1109288.

Much of the data presented herein were obtained at the W.M. Keck
Observatory, which is operated as a scientific partnership among the
California Institute of Technology, the University of California, and
the National Aeronautics and Space Administration. The Observatory was
made possible by the generous financial support of the W.M. Keck
Foundation.  Some of the Keck data were obtained through the NSF
Telescope System Instrumentation Program (TSIP), supported by AURA
through the NSF under AURA Cooperative Agreement AST 01-32798 as
amended.

Some of the data herein were obtained at the Gemini Observatory, which
is operated by the Association of Universities for Research in
Astronomy, Inc., under a cooperative agreement with the NSF on behalf
of the Gemini partnership: the NSF (United
States), the Science and Technology Facilities Council (United
Kingdom), the National Research Council (Canada), CONICYT (Chile), the
Australian Research Council (Australia), Minist\'{e}rio da
Ci\^{e}ncia, Tecnologia e Inova\c{c}\~{a}o (Brazil) and Ministerio de
Ciencia, Tecnolog\'{i}a e Innovaci\'{o}n Productiva (Argentina). 

The authors wish to recognize and acknowledge the very
significant cultural role and reverence that the summit of Mauna Kea
has always had within the indigenous Hawaiian community. We are most
fortunate to have the opportunity to conduct observations from this
mountain.

Funding for the SDSS and SDSS-II has been provided by the Alfred P. Sloan 
Foundation, the Participating Institutions, the National Science Foundation, 
the U.S. Department of Energy, the National Aeronautics and Space 
Administration, the Japanese Monbukagakusho, the Max Planck Society, 
and the Higher Education Funding Council for England. The SDSS Web Site 
is http://www.sdss.org/.

The SDSS is managed by the Astrophysical Research Consortium for the
Participating Institutions. The Participating Institutions are the
American Museum of Natural History, Astrophysical Institute Potsdam,
University of Basel, University of Cambridge, Case Western Reserve
University, University of Chicago, Drexel University, Fermilab, the
Institute for Advanced Study, the Japan Participation Group, Johns
Hopkins University, the Joint Institute for Nuclear Astrophysics, the
Kavli Institute for Particle Astrophysics and Cosmology, the Korean
Scientist Group, the Chinese Academy of Sciences (LAMOST), Los Alamos
National Laboratory, the Max-Planck-Institute for Astronomy (MPIA),
the Max-Planck-Institute for Astrophysics (MPA), New Mexico State
University, Ohio State University, University of Pittsburgh,
University of Portsmouth, Princeton University, the United States
Naval Observatory, and the University of Washington. 

Funding for SDSS-III has been provided by the Alfred P. Sloan
Foundation, the Participating Institutions, the National Science
Foundation, and the U.S. Department of Energy Office of Science. The
SDSS-III web site is http://www.sdss3.org/.l 

SDSS-III is managed by the Astrophysical Research Consortium for the
Participating Institutions of the SDSS-III Collaboration including the
University of Arizona, the Brazilian Participation Group, Brookhaven
National Laboratory, University of Cambridge, Carnegie Mellon
University, University of Florida, the French Participation Group, the
German Participation Group, Harvard University, the Instituto de
Astrofisica de Canarias, the Michigan State/Notre Dame/JINA
Participation Group, Johns Hopkins University, Lawrence Berkeley
National Laboratory, Max Planck Institute for Astrophysics, Max Planck
Institute for Extraterrestrial Physics, New Mexico State University,
New York University, Ohio State University, Pennsylvania State
University, University of Portsmouth, Princeton University, the
Spanish Participation Group, University of Tokyo, University of Utah,
Vanderbilt University, University of Virginia, University of
Washington, and Yale University.


%

\clearpage




\appendix

{\bf APPENDIX: System Analysis}

\begin{figure}
\includegraphics[width=5in]{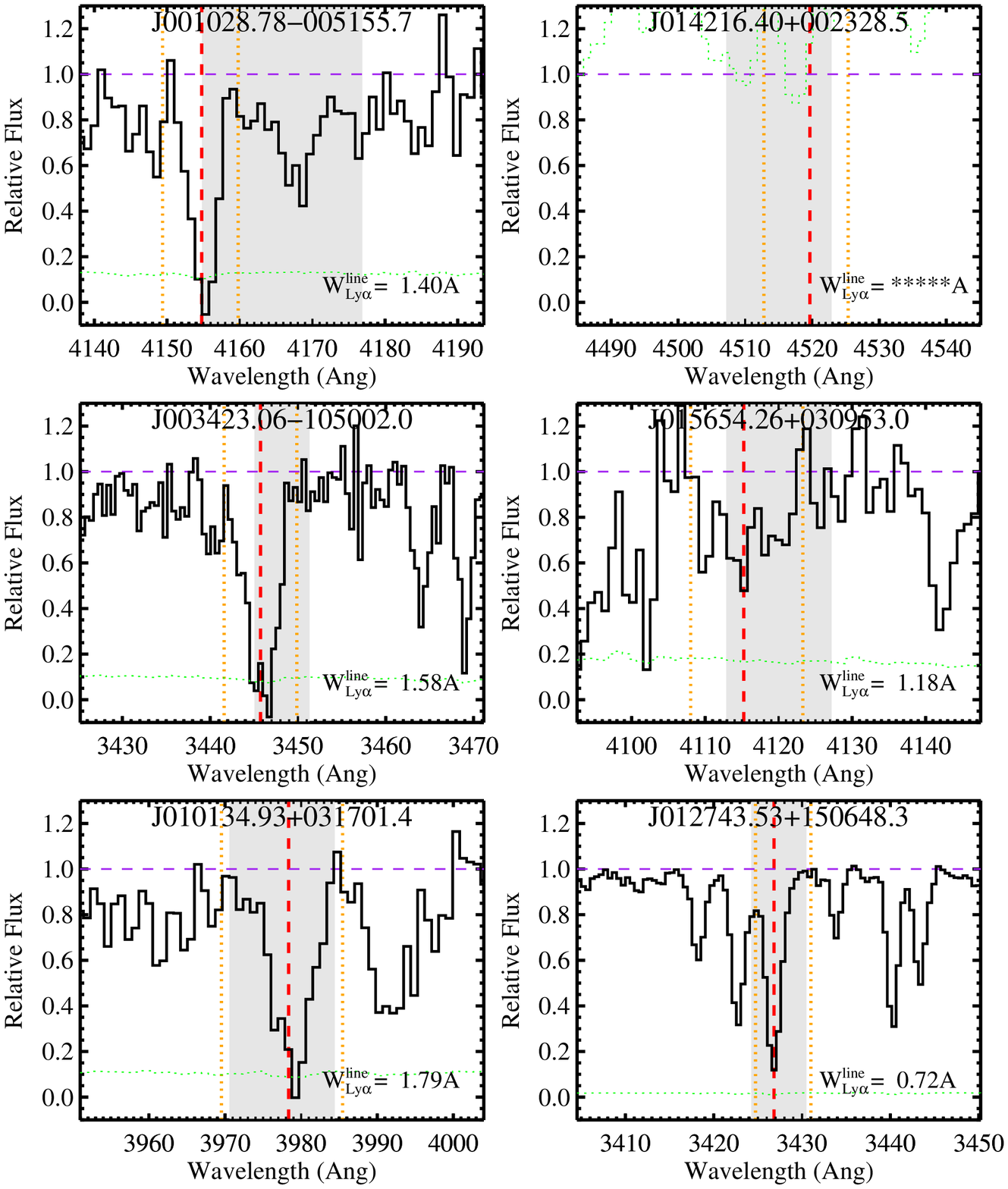}
\caption{The panels show snippets of spectra centered on \lya\ of the
  f/g quasar \llya, in a window of $\pm 2000\mkms$.  
  The purple dashed line traces our estimates for the quasar continua
  (original; $\S$~\ref{sec:continuum})
  and the green dotted line is the estimate of the $1\sigma$
  uncertainty in the relative flux. The gray shaded region
  specifies the $\pm 1\sigma_z$ uncertainty in \zfg.  The vertical,
  red-dashed line marks the strongest absorption line within a
  $\pm 1500\mkms$ interval about \llya.  We measure the rest-frame
  equivalent width \wsubj\ through simple boxcar summation of the
  region designated by the vertical orange dotted lines.  
  These data describe some of the complexity and uncertainty in
  identifying the strongest \lya\ line associated to the f/g quasar
  and assessing its \ion{H}{1} \lya\ absorption strength.
}
\label{fig:exmpl_lya}
\end{figure}

Figure~\ref{fig:exmpl_lya} presents spectral regions 
centered on \llya\ for a representative set of the sample, 
with the shaded region indicating the $1\sigma$ uncertainty in \zfg.
The figure describes the diversity of absorption apparent in the
sample, both in terms of equivalent widths and velocity offsets of the
absorbers from \zfg.  It also illustrates some of the subjectivity
involved in defining systems. 
One occasionally has multiple absorption lines within the $\pm 1500\mkms$
interval and the system nearest \zfg\ need not be the strongest.
Nevertheless, the assignment of the `strongest line'
was generally unambiguous and we were
relatively confident in defining the integration windows for the
\wlya\ measurements.  All of 
the velocity regions and rest-frame \lya\ equivalent widths \wsubj\ for the 
\ion{H}{1} systems are provided in Table~\ref{tab:qpq6_NHI}.

\begin{figure}
\includegraphics[width=5in,angle=90]{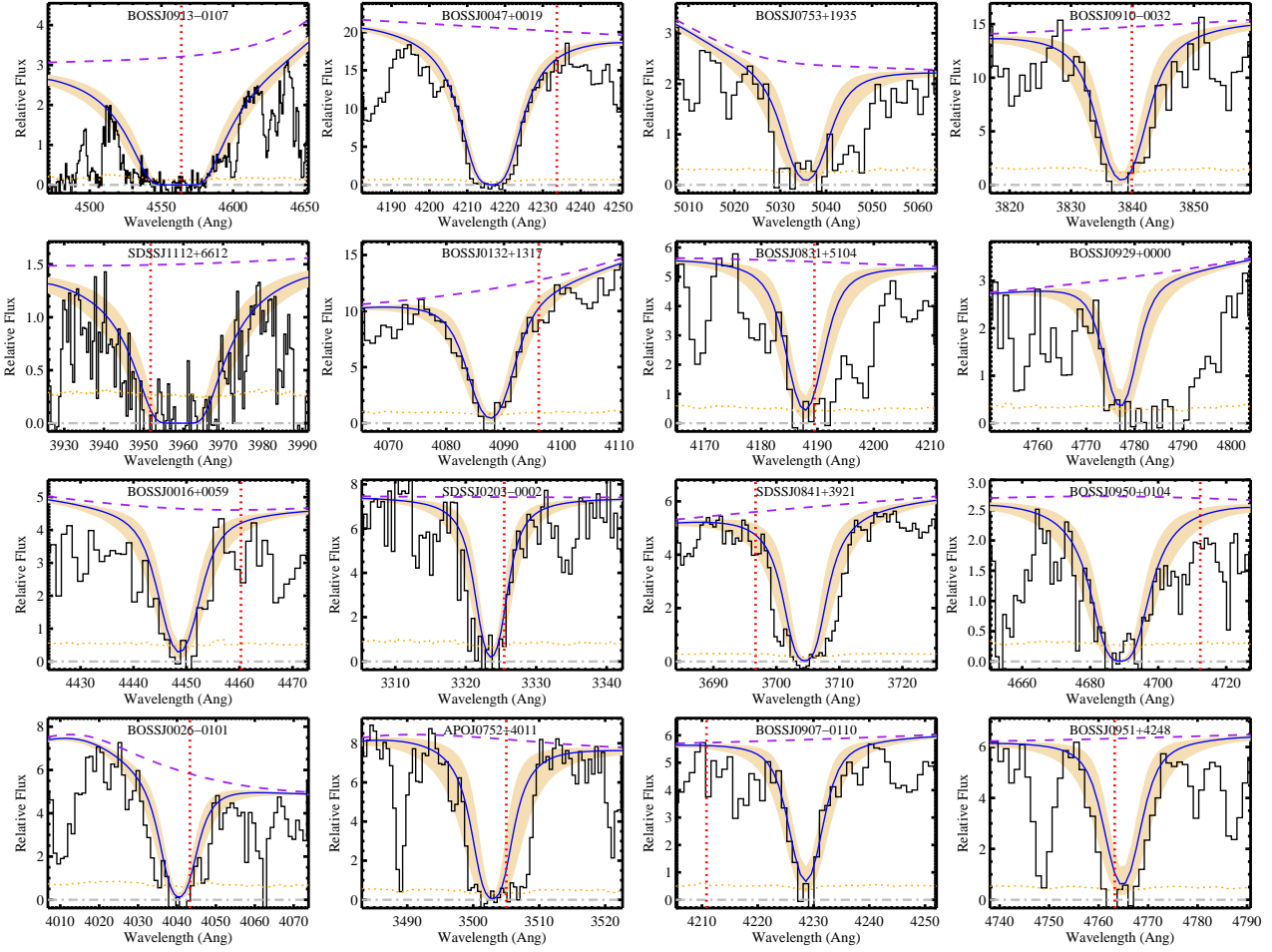}
\caption{ Voigt profile models for the strongest \ion{H}{1} absorbers
  in the QPQ6 dataset (solid blue curves with shaded regions
  showing the error estimate; Table~\ref{tab:qpq6_NHI}).  The purple dotted
  lines trace our estimate of the quasar continuum, the red dotted
  line marks \lya\ of the f/g quasar (\llya), and the dotted curve
  traces the error array.
}
\label{fig:fits}
\end{figure}

\begin{figure}
\includegraphics[width=5in,angle=90]{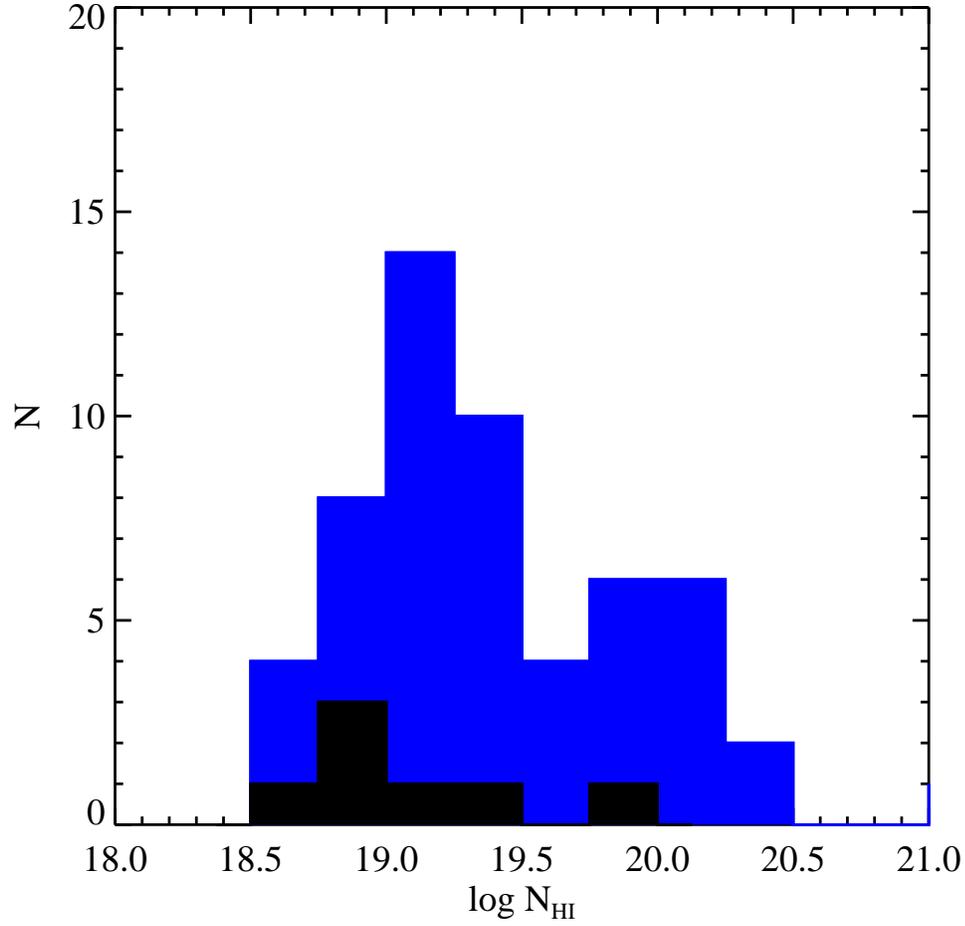}
\caption{ Distribution of measured \nhi\ values along the line of
  sight towards quasar pairs (blue) compared to the distribution from
  our control sample 
  (black).  We find the shape of the two distributions are similar.   
  The cutoff at $\mnhi \approx 10^{19} \cm{-2}$ follows
  from our sensitivity to damping wings in the \lya\ transition. 
}
\label{fig:histogram_nhi}
\end{figure}

The data at \llya\ for those f/g quasars where we have measured an
associated absorption system with $\mnhi \ge 10^{19} \cm{-2}$ are
presented in Figure~\ref{fig:fits} (see also $\S$~\ref{sec:NHI}).  
Overplotted on these data are the
Voigt profile fits and the shaded regions show our estimate of the
$1\sigma$ uncertainty.  Most of the systems with $\mnhi \approx
10^{19} \cm{-2}$ exhibit low-ion absorption (e.g.\ QPQ5).
All of the measurements are given in Table~\ref{tab:qpq6_NHI}.
Figure~\ref{fig:histogram_nhi} presents a histogram of the measured
\nhi\ values for the quasar pairs and compares them to the \nhi\
measurements for systems identified in the control sample.  The
distributions are similar but the control sample is much smaller owing
to the large clustering signal of SLLS around quasars.

\end{document}